\documentclass[12pt]{article}
\usepackage{showlabels}
\usepackage{cite}
\usepackage[english]{babel}

\newlength{\abstractwidth}
\usepackage{makeidx}
\usepackage{indentfirst,hyperref,color,psfrag,verbatim}
\usepackage{amsmath,amssymb,amscd,mathrsfs,graphicx,dsfont,caption,
subcaption,mathtools,empheq,mathbbol}

\usepackage{listings}
\usepackage{slashed}
\usepackage{ulem}
\usepackage{scalerel}

\newcommand{\dslash}{d \hspace{-0.8ex}\rule[1.2ex]{0.9ex}{.1ex}}
\newcommand{\deltaslash}{\delta \hspace{-0.8ex}\rule[1.2ex]{0.9ex}{.1ex}}

\newcommand{\sumint}{\sum \hspace{-4ex}\int}

\newcommand{\dfn}{\vcentcolon=}
\newcommand*\widefbox[1]{\fbox{\hspace{0.5em}#1\hspace{2.3em}}}

\newcommand{\ssk}{\smallskip}
\newcommand{\equalhat}{\widehat{=}}

\renewcommand{\thefootnote}{\fnsymbol{footnote}}
\renewcommand{\thanks}[1]{\footnote{#1}}
\newcommand{\starttext}{
\setcounter{footnote}{0}
\renewcommand{\thefootnote}{\arabic{footnote}}}

\begin{document}
\starttext
\setcounter{footnote}{0}

\begin{flushright}
KCL-PH-TH/2016-60\\
\end{flushright}

\baselineskip=18pt
\rightline{{\fontsize{0.40cm}{5.5cm}\selectfont{}}}
\vskip 0.3in

\begin{center}

{\LARGE {\bf Highly Excited Strings I:}}\\
\vskip 0.2cm
{\LARGE {\bf Generating Function}}

\vskip 1cm

{\large \bf Dimitri P.~Skliros$^{(a,b)}$, Edmund J.~Copeland$^{(a)}$ \\and Paul M.~Saffin$^{(a)}$}
\vskip 0.2cm

{ \sl (a) School of Physics and Astronomy}\\
{\sl University of Nottingham, Nottingham, NG7 2RD, UK} 

\vskip 0.08in

{\sl (b) Theoretical Particle Physics and Cosmology Group, Department of Physics,\\ 
       King's College London, Strand, London WC2R 2LS, UK}
       
\vskip 0.1in

{\tt \small d.p.skliros@gmail.com; edmund.copeland@nottingham.ac.uk; paul.saffin@nottingham.ac.uk}

\end{center}

\vskip -1 in

\begin{abstract}
\begin{center}
\begin{minipage}[c]{0.85\textwidth}
This is the first of a series of detailed papers on string amplitudes with highly excited strings (HES). In the present paper we construct a generating function for string amplitudes with generic HES vertex operators using a fixed-loop momentum formalism. We generalise the proof of the chiral splitting theorem  of D'Hoker and Phong to string amplitudes with arbitrary HES vertex operators (with generic KK and winding charges, polarisation tensors and oscillators) in general toroidal compactifications $\mathcal{E}=\mathbb{R}^{D-1,1}\times \mathbb{T}^{D_{\rm cr}-D}$ (with generic constant K\"ahler and complex structure target space moduli, background Kaluza-Klein (KK) gauge fields and torsion). We adopt a novel approach that does not rely on a ``reverse engineering'' method to make explicit the loop momenta, thus avoiding a certain ambiguity pointed out in a recent paper by Sen, while also keeping the genus of the worldsheet generic. This approach will also be useful in discussions of quantum gravity and in particular in relation to black holes in string theory, non-locality and breakdown of local effective field theory, as well as in discussions of cosmic superstrings and their phenomenological relevance. We also discuss the manifestation of wave/particle (or rather wave/string) duality in string theory.
\end{minipage} 
\end{center}
\end{abstract}

\newpage
\setcounter{tocdepth}{1}
{\footnotesize\tableofcontents}

\section{Introduction}

Highly excited strings (HES) are responsible for numerous miraculous properties of string perturbation theory, they play a central role in ensuring the absence of ultraviolet (UV) divergences and are a crucial ingredient in regarding string theory as a theory of quantum gravity. A fundamental driving force underlying much of the recent flourish of interest \cite{GiveonItzhaki12,ShenkerStanford14b,ShenkerStanford14a,GiveonItzhaki13,Silverstein14,MertensVerscheldeZakharov15,ShenkerStanford15,Martinec15,GiveonItzhakiKutasov15,MertensVerscheldeZakharov16,DodelsonSilverstein15b,DodelsonSilverstein15a,Ben-IsraelGiveonItzhakiLiram15,Ben-IsraelGiveonItzhakiLiram16,BianchiMoralesPieri16,GiveonItzhakiKutasov15b,MertensVerscheldeZakharov16b,PuhmRojasUgajin16} is that a detailed and careful study of HES (in addition possibly to branes and other solitons in string theory) will likely reveal previously unforeseen features of quantum gravity. For example, in reference to quantum black holes (BH), both in terms of microstates out of which the black hole interior is constructed \cite{Mathur05,Mathur09,MathurTurton14,BenaMartinecTurtonWarner16,BenaGiustoMartinecRussoShigemoriTurtonWarner16} (usually with some additional ingredients), and in terms of (possibly non-local) structure at the black hole horizon due to string effects\cite{Silverstein14,DodelsonSilverstein15a,DodelsonSilverstein15b,GiveonItzhakiKutasov15,MertensVerscheldeZakharov15,MertensVerscheldeZakharov16,PuhmRojasUgajin16}, building on earlier ideas \cite{'tHooft90,SusskindThorlaciusUglum93,Susskind93a,Susskind93,LowePolchinskiSusskindThorlaciusUglum95,HorowitzPolchinski97,HorowitzPolchinski98,AmatiRusso99,DamourVeneziano00,Mathur05,CornalbaCostaPenedonesVieira06}, with possible implications for the black hole information paradox.  See \cite{Polchinski16} for a recent review of various traditional ideas in this direction. 

Contrary to field theoretic intuition, it has also recently been shown  \cite{FlorakisRizos16} that in semi-realistic heterotic string compactifications with spontaneously broken supersymmetry and exponentially small values for the cosmological constant, the global structure (or ``shape'') of the effective potential (around certain self-dual points in moduli space, corresponding to extrema of the effective potential) is strongly influenced by contributions from massive string modes (as well as non-level matched string states), thus further highlighting the importance of the HES contributions even in low energy effective field theories. 

\ssk
On a parallel note, highly excited strings may have also been produced in the early universe \cite{CopelandMyersPolchinski04,DvaliVilenkin04,Polchinski04,BeckerBeckerKrause06,CopelandPogosianVachaspati11,CopelandKibble10,Hindmarsh11}, possibly during one or more symmetry-breaking phase transitions \cite{SarangiTye02,JonesStoicaTye02,MajumdarDavis02,DvaliVilenkin04} (but see also \cite{BarnabyBerndsenCLineStoica05}), providing an observational signature for superstrings \cite{CopelandPogosianVachaspati11,Hindmarsh11,AvgoustidisCopelandMossSkliros12,CharnockAvgoustidisCopelandMoss16}. In this context HES are referred to as {\it cosmic superstrings}, and if produced they can have a wide variety of signatures, most notably {\it gravitational wave signatures} \cite{BinetruyBoheCapriniDufaux12,Dufaux12b}, including gravitational wave bursts produced from cusps and kinks \cite{Burden85,VachaspatiVilenkin85,DamourVilenkin00,DamourVilenkin01,DamourVilenkin05,SklirosCopelandSaffin13}, covering a wide frequency range that can be probed by pulsar timing arrays, ground-based interferometers (such as LIGO) and the much anticipated eLISA \cite{BinetruyBoheCapriniDufaux12,Dufaux12b}, to mention a handful. In addition, even though the Planck satellite has placed strong constraints from the temperature data of the cosmic microwave background (CMB), the CMB still offers observational prospects via polarisation and non-Gaussianity \cite{CharnockAvgoustidisCopelandMoss16}, see also \cite{CopelandPogosianVachaspati11,Hindmarsh11} and references therein for a more complete list of observational signals. One major uncertainty \cite{Hindmarsh11} is the eventual destination of the energy of a string network, gravitational and possibly massive radiation (and the associated backreaction) which is believed to play a major role in determining the average size of the produced string loops. Furthermore, incorporating backreaction in theoretical predictions for gravitational wave bursts from individual loops is still an unresolved issue that is suspected may play a major role in their observational prospects. 

\ssk
Partly motivated by the above developments, let us now zoom in further on HES in the context of string perturbation theory in particular. From this viewpoint, the first step will be to set-up an efficient construction that will directly yield string amplitudes in the presence of HES vertex operators, a complete set of which (in a coherent state basis) was first constructed in \cite{HindmarshSkliros10,SklirosHindmarsh11}. In \cite{SklirosCopelandSaffin13} we used these vertex operators to compute decay rates and power associated to massless radiation, while also making contact with low energy effective theory (in a certain IR limit). The tools we have constructed are powerful enough to capture a wide range of phenomena, including, e.g., radiative backreaction corrections to the classical gravitational wave results \cite{Burden85,VachaspatiVilenkin85,DamourVilenkin00,DamourVilenkin01,DamourVilenkin05,SklirosCopelandSaffin13}, cross sections and decay rates associated to HES, including loop corrections, etc. The current document is the first of a series of technical papers on string amplitudes with HES \cite{SklirosCopelandSaffin16bb,SklirosCopelandSaffin16cc,SklirosCopelandSaffin16dd,SklirosCopelandHindmarshSaffin16dd2,SklirosCopelandSaffin16ee}. 

\ssk
A fundamental tool that we make use of is the chiral splitting theorem of D'Hoker and Phong \cite{D'HokerPhong89}, whereby string amplitudes at fixed-loop momenta chirally factorise.\footnote{This chiral factorisation is closely related to that of the classical theory \cite{SklirosCopelandSaffin13,Burden85}, and hence making use of the chiral splitting theorem immediately draws the string perturbation theory approach closer to the classical effective theory while retaining the full set of stringy corrections.} We will make use of and generalise this framework in a number of ways, but to motivate further our approach in this document let us begin with some introductory technical comments. 

\ssk
In a series of recent papers \cite{PiusSen16,Sen16,Sen16b,Sen16c,Sen16d,PiusRudraSen14,PiusRudraSen14b,PiusRudraSen14c} Sen and collaborators  have revisited various aspects of superstring theory (unitarity of string amplitudes, mass and wavefunction renormalisation \cite{PiusRudraSen14b,PiusRudraSen14c,Sen16b}, perturbation theory around dynamically shifted string vacua \cite{PiusRudraSen14}, offshell string amplitudes \cite{Sen15b}, Wick rotations and analytic continuations \cite{PiusSen16,Sen16,Sen16b}, one-particle irreducible (1PI) quantum effective actions, etc.). In a very careful and complete study \cite{PiusSen16} Pius and Sen derived Cutkosky rules for superstring field theory amplitudes to all orders in perturbation theory by providing a prescription for taking integration contours of loop energies in the complex plane (which would naively otherwise yield divergent results for the corresponding S-matrix elements). In \cite{Sen16} Sen then showed that this prescription of introducing loop momenta and appropriately deforming the integration contours is equivalent to that of Berera \cite{Berera94} and Witten \cite{Witten13b}, (see also \cite{Polchinski88b,DaiPolchinski89,MitchellTurokWilkinsonJetzer89,WilkinsonTurokMitchell90,IengoRusso02} and in particular \cite{DHokerPhong95} for related discussions on analyticity of string amplitudes), in the worldsheet approach to superstring theory to all orders in perturbation theory.  In the approach of Witten \cite{Witten13b} one is to deform the integration cycles over moduli space of punctured Riemann surfaces into a complexified moduli space, and this establishes consistency of the former fixed-loop momenta approach with S-matrix unitarity. 
Sen then recently also discussed \cite{Sen16b} an application of the fixed-loop momenta approach, building on earlier work \cite{PiusRudraSen14b,PiusRudraSen14c} and in particular \cite{PiusSen16}, namely mass renormalisation of unstable massive string states (where a naive computation yields divergent results for the two-point one-loop amplitude), explaining how to obtain finite results that are consistent with unitarity. The basic reason for the aforementioned divergences are ultimately due to the fact that the analogue of the `$i\epsilon$' prescription of quantum field theory is somewhat subtle in string theory \cite{Witten13b} because string amplitudes are most naturally defined in Euclidean space where they are real \cite{DHokerPhong95}. Therefore, e.g., potential imaginary parts  (that are required by unitarity) in Lorentzian signature string amplitudes show up as divergences in the corresponding Euclidean space amplitudes. Motivated by string field theory, Sen \cite{Sen16} and Pius and Sen \cite{PiusSen16}  have provided a well-defined prescription for dealing with such analytic continuations by reformulating string amplitudes in terms of {\it fixed-loop momenta} and deforming their integration cycles into the complex plane following a specific prescription (whereby loop energy contours are pinned down at $\pm i\infty$ following a non-trivial but well-defined path in between, leading to Lorentzian signature amplitudes).

Drawing from analogies with string field theory \cite{PiusSen16}, the introduction of fixed-loop momenta is central to Sen's analytic continuation approach to string amplitudes (which are traditionally given as integrals over moduli space, in the ``Schwinger parametrisation'' with implicitly integrated loop momenta). Fixed-loop momenta amplitudes have a long history in string theory that dates back to the old dual models, see e.g.~\cite{Schwarz82} and references therein, but it was not until Dijkgraaf, E.~Verlinde and H.~Verlinde \cite{DijkgraafVerlindeVerlinde88} (building on \cite{VerlindeVerlinde87}) that fixed-loop momenta appeared in the path integral formulation of string theory, where various interesting properties were also noted, one such property being that (taking into account also the Belavin-Knizhnik theorem \cite{BelavinKnizhnik86}) bosonic amplitudes with tachyonic external vertex operators in simple toroidal compactifications {\it chirally factorise}. This observation was later explored in a much more complete manner and in the full superstring context by D'Hoker and Phong \cite{D'HokerPhong89} (although even here the explicit results were derived for massless external states, and also non-compact flat spacetime). This study led to a {\it chiral splitting theorem}: when string amplitudes are written in terms of integrals over loop momenta\footnote{In particular, $A_I$-cycle loop momenta, with $I=1,\dots,\mathfrak{g}$, with $\mathfrak{g}$ labelling the genus of the Riemann surface and $\{A_I,B_I\}$ a canonical intersection basis for the homology cycles.} and fermion zero modes (when present), the corresponding integrands chirally factorise (in terms of their supermoduli, abelian differentials, worldsheet coordinates of vertex operators, polarisation tensors and momenta). 

One thing we would like to highlight is that in the approach of D'Hoker and Phong \cite{D'HokerPhong89} and Sen \cite{Sen16b}, the fixed-loop momenta amplitudes are constructed by a ``reverse engineering'' method, whereby string amplitudes are computed using the conventional approach \cite{Polchinski_v1}, and it is only at a later stage of the computation (after integrating out the path integral fields and hence obtaining the ``Schwinger form'' of amplitudes) that it is noted that explicit string amplitudes under consideration can be written as integrals over loop momenta. Unfortunately, such an approach is almost hopeless when considering amplitudes with arbitrarily massive HES vertex operators, because the fixed-loop momenta amplitudes will typically have quite a complicated form that one is to somehow guess, hence the name `reverse engineering' mentioned above. (A systematic approach to ``guessing'' the correct loop momentum integrand given the Schwinger parametrisation was given by Sen at one-string loop in \cite{Sen16b}, but this is somewhat tedious and messy and requires a case-by-case study.) D'Hoker and Phong \cite{D'HokerPhong89} made fundamental further progress by making the remarkable observation that the correct fixed-loop momenta result can be obtained directly from the path integral if the path integral fields of vertex operators,\footnote{This is the prescription for bosonic fields; the corresponding correlation functions for worldsheet fermions are (up to zero modes) already chirally split, as are the corresponding (Grassmann-even and odd) ghost insertions and supermoduli \cite{D'HokerPhong89}.} $x^{M}(z,\bar{z})$, are replaced by (anti-)chiral fields $x_{+}^M(z)$ (and $x_-^M(\bar{z})$) as appropriate for left- and right-moving degrees of freedom, while inserting exponential factors:\footnote{Our worldsheet conventions are presented in Appendix \ref{sec:RS}.}
$$
e^{i\mathbb{Q}_I\cdot \oint_{B_I}\partial x_+}, \qquad e^{i\bar{\mathbb{Q}}_I\cdot \oint_{B_I}\bar{\partial} x_-},
$$
into the integrands, where $Q_{IM},\bar{\mathbb{Q}}_{IM}$ are loop momenta spanning all (non-compact in their derivation) spacetime directions, the (anti-)chiral {\it effective} fields, $x_{\pm}$, being defined by standard Wick contractions using `chiral propagators':
$$
\langle x_+^M(z)x_+^N(w)\rangle_+ =-\frac{\alpha'}{2}G^{MN}\ln E(z,w),\qquad \langle x_-^M(\bar{z})x_-^N(\bar{w})\rangle_- =-\frac{\alpha'}{2}G^{MN}\ln \bar{E}(\bar{z},\bar{w}).
$$
Given that these effective rules were derived explicitly by making use of the `reverse engineering' method, it is important to show that the prescription for the effective rules does not depend on it.\footnote{Here $E(z,w)$ is the usual (worldsheet moduli dependent) prime form of Fay \cite{Fay,Mumford_v12} (that has a compact representation in terms of Riemann theta functions, their derivatives and abelian differentials \cite{Mumford_v12}). $E(z,w)$, although multiple-valued \cite{DHokerPhong}, generalises the notion of distance, $z-w$, on $\mathbb{C}$ to arbitrary genus Riemann surfaces, having the unique property that it vanishes only at $z=w$ as $E(z,w)\simeq z-w+\mathcal{O}((z-w)^3)$. }

Of course, one expects the chiral splitting theorem to also hold for generic vertex operators and in generic toroidally compactified string theories, but as the derivation of D'Hoker and Phong \cite{D'HokerPhong89} was carried out explicitly in non-compact Minkowski (or Euclidean) spacetime (and in \cite{DijkgraafVerlindeVerlinde88,AokiD'HokerPhong04} in the context on $\mathbb{Z}_2$ orbifold compactifications) and for massless vertex operator insertions it would be desirable to discuss amplitudes with generic vertex operators more explicitly.\footnote{For instance, in D'Hoker and Phong's approach it is not obvious (and in fact it is a highly non-trivial statement) that shifting the target space embedding of the string (in both the action and vertex operators) by instanton or soliton contributions (that would be present in compactified spacetimes) would yield a similar result with appropriately identified loop momenta.} Furthermore, it would be desirable to adopt an approach that does not rely on the reverse engineering method outlined above, and hence show in particular that the potential ambiguities discussed by Sen \cite{Sen16b} are absent for generic vertex operators when using the effective chiral splitting rules of D'Hoker and Phong. The ambiguity we are referring to is the following. In \cite{Sen16b} Sen has shown that the same Schwinger parameter representation of a given string amplitude (which arises directly from the usual path integral formulation of string theory) can be represented in {\it more than one way} from a fixed-loop momenta representation, see the discussion associated to equation (2.17) in \cite{Sen16b} and also the footnote below (3.20) there, leading one to question whether this reverse engineering method that has been adopted to-date could potentially be ambiguous for generic vertex operators. Sen then went on to argue that this ambiguity will actually not be visible in the result for the full amplitude {\it after} having integrated out the loop momenta using the prescription for avoiding  singularities \cite{PiusSen16} in the loop momenta integrations. 

Although this is a very important step forward (given in particular the subtleties of the loop momentum contour integrations in Lorentzian-signature target spacetimes), there are situations where it is {\it not} desirable to integrate out the loop momenta completely, and identify the loop momentum {\it integrand} with a physical observable. This at first sight may seem unphysical, not least because the integrand of the loop momentum integrals are {\it not} modular invariant (defining loop momenta requires specifying a homology basis \cite{DHokerPhong}, which transforms non-trivially under the mapping class group), but nevertheless there do exist physical observables that are sensitive to this integrand. Put simply, strings in loops can go onshell and can therefore also appear in the detector of a given observer. Their momenta can thus be measured, and so it does make sense to consider the integrands of loop momentum integrands of string amplitudes as being physical. One can also make this argument by appealing to the optical theorem and unitarity. A good example is the following. 

Consider the power emitted from a massive string per unit solid angle. Here we imagine a generic highly excited string that is unstable and emitting radiation while it decays, usually anisotropically, with massless radiation often being the dominant decay channel. We place a detector far from the interaction region and measure the power absorbed by the detector as a function of energy or frequency of the radiation, and so extract a spectrum which will contain information about the radiating string state (allowing one {\it in principle} to reconstruct the quantum numbers of the decaying string). The orientation of the emitting string can also be determined in this manner, due to the anisotropy of the decay; this is particularly relevant for gravitational wave emission from strings with cusps where the associated burst of radiation is highly anisotropic, and this ultimately provides one of the strongest signals in cosmic string phenomenology; this has a long history, see  \cite{Burden85,VachaspatiVilenkin85,DamourVilenkin00,DamourVilenkin01,DamourVilenkin05} for an effective theory computation and \cite{SklirosCopelandSaffin13} for a corresponding analytical string theory computation, the two being in precise agreement when backreaction is neglected and one is confined to low energies.\footnote{There are a number of other approaches in the literature to decay rate computations of highly excited strings, but these typically rely on numerical approximations or saddle-point evaluations of {\it integrated}-loop momentum two-point one-loop amplitudes (but also tree-level amplitudes), in order to obtain order of magnitude estimates, and they also consider leading Regge trajectory states only, see \cite{MitchellTurokWilkinsonJetzer89,WilkinsonTurokMitchell90,DabholkarMandalRamadevi98} and more recently \cite{IengoKalkkinen00,IengoRusso02,IengoRusso03} and \cite{ChialvaIengoRusso03,ChialvaIengo04,ChialvaIengoRusso05,GutperleKrym06}, and also \cite{IengoRusso06} for a review on string decay.} The observable in this thought experiment can, e.g., be extracted from the imaginary part of the two-point (say one-loop at weak coupling) amplitude at fixed-loop momenta, $\textrm{Im}\,\mathcal{M}_{T^2}(\overline{\mathcal{O}_{z\bar{z}}},\mathcal{O}_{z\bar{z}};\mathbb{P})$, with two vertex operator insertions (related by Euclidean conjugation, more about which will be discussed in a sequel \cite{SklirosCopelandSaffin16bb}). 
In particular, for $D$ non-compact dimensions, the power associated to decay products of momentum $\mathbb{P}^{\mu}$, centred around some arbitrary spatial direction $\hat{\bf P}$, can be extracted from \cite{SklirosCopelandSaffin13}:\footnote{For this computation choosing the correct contour for the $\mathbb{P}^0$ integral is crucial and has been discussed very carefully and clearly for generic loop amplitudes in \cite{PiusSen16,Sen16b}.} 
\begin{equation}\label{eq:dP/dO}
\begin{aligned}
\frac{dP}{d\Omega_{S^{D-2}}} = \frac{1}{(2\pi)^{D}}\frac{2\pi\alpha'}{L}\,\int d\mathbb{P}^0 \mathbb{P}^0|{\bf P}|^{D-2}d|{\bf P}|\,\textrm{Im}\,\mathcal{M}_{T^2}(\overline{\mathcal{O}_{z\bar{z}}},\mathcal{O}_{z\bar{z}};\mathbb{P}),
\end{aligned}
\end{equation}
where it is seen that the loop momenta, $\mathbb{P}^{\mu}$, actually get identified with the momentum of the decay products, some of which end up in the detector, as alluded to above. Here $L$ is the length \cite{SklirosCopelandSaffin16bb} and $\frac{L}{2\pi\alpha'}$ the corresponding mass of the emitting string whose vertex operator, $\mathcal{O}_{z\bar{z}}$, is normalised by the leading singularity in the OPE:\label{page:OOnorm}\footnote{Newton's constant, $G_D$, is related to the gravitational coupling, $\kappa_D$, via $\kappa_D^2=8\pi G_D$. In $D=4$, $\kappa_4^{-1}=2.4\times 10^{18}{\rm GeV}$ is the reduced Planck mass. Note that $g_D=\kappa_D/(2\pi)$.} $\overline{\mathcal{O}_{z\bar{z}}}\mathcal{O}_{z'\bar{z}'}\simeq g_D^2/|z-z'|^4$. An `overline' represents taking the Euclidean adjoint, see \cite{Polchinski88,Polchinski_v1} and in particular \cite{SklirosCopelandSaffin16bb} for a refined discussion of this notion (in the presence of compact dimensions where there are some additional phases that are absent in \cite{Polchinski88,Polchinski_v1}), and we have decomposed the loop momentum integral as follows, $d^D\mathbb{P}^{\mu}=d\mathbb{P}^0d|{\bf P}|\,|{\bf P}|^{D-2}d\Omega_{S_{D-2}}$. Extracting the imaginary part will give rise (according to the Cutkosky rules \cite{PiusSen16})  to two delta functions, one of which places the emitted radiation onshell, and the second delta function quantises the spectrum of decay products  (in the case of massless radiation), leading to integer-valued energies, $\mathbb{P}^0=\omega_n$, of the form: 
$$
\qquad\qquad\omega_n=\frac{4\pi n}{L},\qquad {\rm with}\qquad n=1,2,\dots,
$$ 
and $n$ is summed over (subject to energy conservation), as expected, e.g., for {\it gravitational waves} from strings. This procedure was first carried out by the present authors in \cite{SklirosCopelandSaffin13}, where a brief summary of our results can be found, as well as the effective low energy theory that reproduces them. Clearly, if we want to extract information about the energy-dependence of the emitted radiation we do not want to perform the sum over $n$. If we were to integrate out the loop momentum completely the $\omega_n$ dependence of the power would be lost. We compute observables in explicitly in follow-up articles \cite{SklirosCopelandSaffin16dd,SklirosCopelandHindmarshSaffin16dd2}, but the purpose of the discussion in this paragraph is to show that it is sometimes desirable to {\it not} integrate out the loop momenta completely, and that this is of interest even in calculations of physical observables. Therefore, the resolution of the aforementioned ambiguity of Sen \cite{Sen16b} is not totally satisfactory (because it relies on the assumption that it is of interest to integrate out loop momenta completely).  Summarising, it would be desirable to adopt an approach that does not rely on the reverse engineering derivation of D'Hoker and Phong \cite{D'HokerPhong89} and that of Sen \cite{Sen16b} and show explicitly how to resolve this potential ambiguity observed by Sen in his study of two-point amplitudes of massive strings.

In this article we resolve this particular ambiguity completely\footnote{There remain field redefinition ambiguities familiar already from the string field theory context, see Sec.~4 in \cite{Sen16b}, and the authors are greatful to Ashoke Sen for extensive discussions on this.}, for completely generic vertex operator insertions (with arbitrary winding and KK charges), and for generic toroidally compactified backgrounds (with generic constant K\"ahler and complex structure target space moduli, background KK gauge fields and torsion), and to any finite order in the string loop expansion. Therefore, our derivation applies to all closed string amplitudes in target spacetimes $\mathbb{R}^{D-1,1}\times \mathbb{T}^{D_{\rm cr}-D}$.  We focus on the bosonic string for simplicity (or the bosonic sector of the superstring, the conclusions being independent of the chiral splitting statements given that up to fermion zero modes the fermionic contribution is already chirally factorised, as are the ghost contributions). 

Specifically, our approach will be to drop the `reverse engineering' approach of D'Hoker and Phong \cite{D'HokerPhong89} and Sen \cite{Sen16b} altogether, and to rather construct the fixed-loop momentum representations directly and explicitly, starting from a generic worldsheet path integral, leaving no room for ambiguities. This will be achieved by inserting momentum-conserving delta functions into the worldsheet path integral that explicitly determine the loop momentum contribution associated to $A_I$-cycle strings in the loops, as shown in the second equality in (\ref{eq:A(j)}) below. These loop momenta constitute an independent set and their presence are also a fundamental ingredient in obtaining a handle on the energies and momenta that contribute to loop corrections, thus for example bridging the gap between Wilson's approach to quantum field theory \cite{Wilson74,WilsonKogut74}, see e.g.~\cite{Polchinski84}, and an analogous approach in string theory (although we will not explore this connection further here), and Wilson's approach adapted to string field theory has recently been discussed by Sen in \cite{Sen16c}. Unless one wants to work directly with string field theory or in the old operator approach, instead of adopting the first-quantised covariant path integral formalism that we consider here, string amplitudes with fixed-loop momenta is the closest one can get to the corresponding field theory amplitudes without considering explicit pant decompositions and degenerations of the worldsheet. 

Fixed-loop momenta amplitudes are, arguably, one of the most natural approaches to (at least closed) perturbative string amplitude computations. First and foremost, it is considerably easier to write down amplitudes at fixed-loop momenta than it is to adopt the traditional approach \cite{Polchinski_v1} and write down the corresponding integrated loop momenta amplitudes directly -- a fact that is certainly not widely appreciated in the literature. This is largely due to the chiral splitting theorem of D'Hoker and Phong \cite{D'HokerPhong89}; recall that correlation functions for the worldsheet fields are carried out using the {\it chiral} propagators exhibited above (analogous relations for the fermionic sector or in terms of superfields can be found in \cite{D'HokerPhong89}), where also zero mode subtractions are absent (it is useful to compare with the non-chiral genus-$\mathfrak{g}$ propagator (\ref{eq: Green Function+regular*m})). Secondly, it is (apparently \cite{SklirosCopelandSaffin13,SklirosCopelandSaffin16dd,SklirosCopelandHindmarshSaffin16dd2}) considerably easier (from a technical point of view) to analytically compute explicit amplitudes in the chiral fixed-loop momenta formalism than it is to extract the corresponding integrated loop momenta expressions, as we briefly summarised in \cite{SklirosCopelandSaffin13}. The physical reason is that in the integrated loop momenta approach one is automatically resumming all momentum contributions inside loops (so that it is difficult to take a low energy limit as loop energies are already integrated out), and as a result  one ends up having to resort to saddle-point approximations or numerical methods in order to extract some physics or even order of magnitude estimates \cite{MitchellTurokWilkinsonJetzer89,WilkinsonTurokMitchell90,DabholkarMandalRamadevi98,IengoKalkkinen00,IengoRusso02,IengoRusso03,ChialvaIengoRusso03,ChialvaIengo04,ChialvaIengoRusso05,GutperleKrym06}, whereas in the fixed-loop momenta approach ({\it after} adopting a coherent vertex operator basis for external states \cite{HindmarshSkliros10,SklirosHindmarsh11,SklirosCopelandSaffin13}, see in particular \cite{SklirosCopelandSaffin16bb} for a recent analysis), things tend to resum into, e.g., Bessel functions, exponentials and related special functions \cite{SklirosCopelandSaffin13}. As the latter have been studied by mathematicians for centuries, with various of their properties examined in detail (such as asymptotic expansions, series and integral representations, etc.), this provides a useful and novel working handle on generic string amplitudes. 

Amplitudes with HES should be expected to reproduce various classical or effective field theory results in certain limits (with non-local stringy sources), and adopting the correct toolset is absolutely fundamental to exposing this simple structure, while also providing an explicit approach to computing various stringy and quantum corrections which may or may not be large compared to the effective description -- the effective approach and its link to string amplitudes with HES vertex operators will be presented in \cite{SklirosCopelandSaffin16ee}, where again the basic connection to the effective theory was presented in \cite{SklirosCopelandSaffin13}, building on an earlier conjecture by Dabholkar, Gibbons, Harvey and Ruiz \cite{DabholkarHarvey89,DabholkarGibbonsHarveyRuiz90}, see also \cite{Tseytlin90} for a very insightful complementary decription. The emphasis on the {\it naturalness} of adopting a coherent vertex operator basis in particular when discussing amplitudes with HES will be explained in sequels in much greater detail \cite{SklirosCopelandSaffin16bb,SklirosCopelandSaffin16cc,SklirosCopelandSaffin16dd,SklirosCopelandHindmarshSaffin16dd2}. The present contribution will not rely on any particular vertex operator basis, and also (with a bit a care and tweaking \cite{Sen15b}, see also \cite{Polchinski87,Polchinski88}) will apply to offshell as well as onshell string amplitudes.

Let us also re-emphasise that our initial focus will be on bosonic string theory in this series of papers, because as string amplitudes with coherent vertex operators is a novel and unexplored area of research it will be easiest to first focus on the bosonic string and understand that case well before moving on to the much more interesting but also more involved superstring framework. The bosonic string already contains most of the non-trivial features associated to HES and will provide the basic physics, with the additional complications associated to \cite{DHokerPhong,D'HokerPhong89,Witten12c,Witten13,Witten15,D'HokerPhong15,D'HokerPhong15b,Sen15b,SenWitten15} supermoduli space, gauge fixing, picture changing, etc., of the superstring providing a sharpening of the bosonic string results (by eliminating a tachyon, introducing supersymmetry to stabilise the vacuum and eliminate massless tadpoles, etc.), but it will not change much of the essential physics picture. 

Finally, for the more philosophically-minded readers, we will also discuss how wave/particle duality (or rather {\it wave/string duality}) manifests itself in string theory. We will discuss a simple example (the standard one-loop vacuum amplitude) and then generalise the argument to all string amplitudes at any string loop order. The resulting picture is rather simple, but we are not aware of it having been discussed before in the literature. The result is that fixed-loop momenta amplitudes can be thought of as corresponding to a {\it wave picture}, whereas the corresponding integrated-loop momenta amplitudes provide the corresponding {\it string picture}. There are four natural representations for string amplitudes in toroidally compactified spacetimes, corresponding to the fact that loop momenta in the compact or non-compact dimensions can be either integrated or fixed, so there are four possibilities. The intuitive statement, as we shall explain, is roughly that summing over all trajectories of a loop of string in (say) a compact target space, including the number of times a closed string can traverse the compact space is equivalent to summing over all frequencies of a standing wave in a ``box'', thus making wave/string duality manifest. This therefore provides a physical interpretation of the standard Poisson resummation in string partition functions (for the compact dimensions), and there is an analogous statement in the non-compact dimensions.

In {\bf Sec.~\ref{sec:O}} we provide a brief overview of our results (skipping almost all of the subtle and technical points). In {\bf Sec.~\ref{sec:SAGF}} we derive the generating function for generic string amplitudes in generic toroidal compactifications associated to arbitrary vertex operator insertions and at arbitrary string loop order; this is where the majority of the work lies.  The result here is extremely simple. In {\bf Sec.~\ref{sec:WSD}} we discuss the string theory manifestation of wave/particle duality of quantum mechanics, which is closely related to the presence or absence of fixed-loop momenta (in both compact and non-compact target spacetime dimensions). {\bf Sec.~\ref{sec:CF}} is a corollary of the preceding sections and completes the derivation of the D'Hoker and Phong chiral splitting theorem for generic HES vertex operator insertions (including KK and winding charges and general polarisation tensors and oscillators spanning all spacetime dimensions) and generic constant target space K\"ahler and complex structure moduli, KK gauge fields, as well as spacetime torsion.

\section{Overview}\label{sec:O}
In this section we provide a brief overview of the main results of the current document. The main objective underlying this series of papers is to provide a working handle on string amplitudes with HES vertex operator insertions.  
The first step in this direction, which is presented here, is thus to construct a {\it generating function}, $\mathcal{A}(j)$, for generic string amplitudes in generic toroidal compactifications:
\begin{equation}\label{eq:background topology}
\boxed{\mathcal{E}=\mathbb{R}^{D-1,1}\times \mathbb{T}^{D_{\rm cr}-D}}
\end{equation}
where $D$ denotes the number of non-compact dimensions (e.g.,~$D=4$) and $D_{\rm cr}$ the critical number of dimensions (e.g.,~$D_{\rm cr}=26$ or 10, in the bosonic string or superstring respectively). We will consider the exact (in the fundamental string length, $\ell_s\dfn \sqrt{\alpha'}$) string backgrounds where the spacetime metric, $G_{MN}$, antisymmetric tensor, $B_{MN}$, dilaton, $\Phi$, and tachyon, $U$, are general (bare) constants\footnote{Constants meaning that they are independent of worldsheet and target space embedding coordinates.}, subject only to the requirement that string perturbation theory is applicable, see (\ref{eq:generic GBPhi}). 
The first two of these contain \cite{NarainSarmadiWitten87} the K\"ahler, complex structure moduli and background Kaluza-Klein (KK) gauge fields associated to the compactification (\ref{eq:background topology}), as well as torsion, $B_{\mu\nu}$, all of which will be allowed to be turned on. In order to make contact with the NS sector of low energy supergravity, see e.g.~\cite{MaharanaSchwarz93}, it will sometimes be convenient to consider the parametrisation,\footnote{Detailed definitions appear in the main section and Appendices.}
\begin{equation}\label{eq:generic G_MN decomp}
G_{MN}=   \left(\begin{matrix} 
      g_{\mu\nu}+A_{\mu}^aG_{ab}A_{\nu}^b & G_{ab}A_{\mu}^a \\
      G_{ab}A_{\nu}^b & G_{ab} \\
   \end{matrix}\right),\quad 
G^{MN}=   \left(\begin{matrix} 
      \,\,(g_{\mu\nu})^{-1} & -(g_{\mu \rho})^{-1}A^{ b}_{\rho} \\
      -(g_{\nu \rho})^{-1}A^{a}_{\rho} &\,\, (G_{ab})^{-1}+A^a_{\mu}(g_{\mu\nu})^{-1}A^b_{\nu} \\
   \end{matrix}\right),
\end{equation}
where the $A_{\mu}^a$ are a subset of the aforementioned Kaluza-Klein gauge fields, the remaining ones being $B_{\mu a}$. We always raise and lower indices with $G_{MN}$, the inverse being defined by $G^{MN}G_{NL}=\delta^N_L$.

Using the fixed-loop momenta approach of D'Hoker and Phong \cite{D'HokerPhong89}, the first goal will be to show that generic correlation functions associated to asymptotic vertex operators with generic instanton contributions, KK and winding charges, and generic polarisation tensors can all be extracted from the following genus-$\mathfrak{g}$ contribution to the {\it generating function} in the aforementioned background:\footnote{The first equality here is in Euclidean worldsheet and target space signature whereas the second equality is in Lorentzian target spacetime signature and Euclidean worldsheet signature (which is always possible at generic points in the moduli space of the Riemann surface under consideration \cite{Witten13b,Sen16}).}
\begin{equation}\label{eq:Afull overview}
\begin{aligned}
\mathcal{A}(j)&\dfn \int \mathcal{D}(x,b,\tilde{b},c,\tilde{c})\!\!\prod_{j=1}^{\#_{\mathbb{C}}{\rm moduli}}\!\!\!\!\!|\langle \mu_j,b\rangle |^2\,\prod_{s=1}^{\#_{\mathbb{C}}{\rm CKVs}}\!\!\!|c(w_s)|^2\,\,e^{-I(x|j)-I_{\rm gh}}\\
&\,\,=i\bar{\delta}(j\ell_s)\,g_{\rm eff}^{2\mathfrak{g}-2}\,
\sumint\limits_{(\mathbb{Q},\bar{\mathbb{Q}})}\bigg|\mathcal{Z}_{\mathfrak{g}}\exp \bigg(\frac{\ell_s^2}{4}\int d^2z\int d^2z'\big(j+H\big)_MG^{MN} \big(j'+H'\big)_N\ln E(z,z')\bigg)\bigg|^{2\phantom{\Big(}}\!\!\!\!\!\!\!\!\!\!\!\!
\end{aligned}
\end{equation}
where, in a canonical intersection basis \cite{DHokerPhong} for the $2\mathfrak{g}$ homology cycles of the compact genus-$\mathfrak{g}$ Riemann surface, $\{A_1,B_1,\dots,A_{\mathfrak{g}},B_{\mathfrak{g}}\}$, the sum/integral appearing in the second equality is over loop momenta, $\mathbb{Q}_M^I,\bar{\mathbb{Q}}_M^I$, associated to $A_I$-cycle (with $I=1,\dots,\mathfrak{g}$) strings that span the {\it full} target spacetime $\mathcal{E}$, see (\ref{eq:sumintQQbar-Mink}) and (\ref{eq:PLPR}). $I_{\rm gh}$ is the usual $b,c$ ghost action (\ref{eq:GhostAction}), the $\mu_{j}$ are Beltrami differentials and specify a gauge slice in the space of worldsheet metrics \cite{DHokerPhong}, whereas, $I(x|j)$, encodes the  standard matter contribution with a source,
\begin{equation}\label{eq:action}
I(x|j) \dfn\frac{1}{2\pi\alpha'}\int_{\Sigma_{\mathfrak{g}}} \!\!d^2z\,\Big(\partial_zx^M\partial_{\bar{z}}x^N\big(G_{MN}+B_{MN}\big)+\alpha'R_{z\bar{z}}\Phi+g_{z\bar{z}}U\Big)-i\int_{\Sigma_{\mathfrak{g}}} \!\!d^2z\, j_Mx^M,
\end{equation}
where $j_M(z,\bar{z})$ is a generic source term, such that functional derivatives of $\mathcal{A}(j)$ with respect to it generate all (matter) correlation functions of interest (see below).  In going from the first to the second equality in (\ref{eq:Afull overview}) we have inserted loop-momentum conserving delta functions, see (\ref{eq:1=intPW}), expanded the embedding coordinate into a zero mode, instantons, and quantum fluctuations, 
$$
x^M=x^M_0+x^M_{\rm cl}+\tilde{x}^M,
$$ 
as discussed below (\ref{eq:xexp}), before finally integrating out $x_0,\tilde{x}$, and performing a Poisson resummation in the instanton sector. We also keep the constant tachyon background implicit throughout (this will play an explicit role in tadpole cancellation \cite{SklirosCopelandSaffin16cc}). The {\it effective coupling} appearing in (\ref{eq:Afull overview}) at fixed-loop momenta (in Lorentzian signature) is given by:
$$
g_{\rm eff}\dfn g_s\Big(\!-{\rm Det}G^{\mu\nu}{\rm Det}G_{ab}\Big)^{-\frac{1}{4}},
$$
whereas the delta function constraint in (\ref{eq:Afull overview}) enforces overall charge neutrality, see (\ref{eq:bardelta}). The quantity $\mathcal{Z}_{\mathfrak{g}}$ is determined entirely from the ghost contributions, see (\ref{eq:ghost}) and (\ref{eq:ghostb}). For example, at $\mathfrak{g}=1$, $\mathcal{Z}_1=\eta(\tau)^{-24}$, where $\eta(\tau)$ is the Dedekind eta function and $\tau$ the complex structure modulus of the torus \cite{Polchinski_v1}. The prime form \cite{DHokerPhong} is denoted by $E(z,w)$, and is the unique holomorphic function defined on a Riemann surface that has precisely one (simple) zero, which is at $z=w$. Finally, $H_M,\bar{H}_M$ are operators that encode the loop momentum (including instanton) contributions:
\begin{equation}\label{eq:HHbar overview}
H_M(z,\bar{z})\dfn \mathbb{Q}_{M}^I\oint_{B_I}dw\delta^2(w-z)\partial_z,
\qquad
\bar{H}_M(z,\bar{z})\dfn \bar{\mathbb{Q}}_{M}^I\oint_{B_I}d\bar{w}\delta^2(w-z)\partial_{\bar{z}},
\end{equation}
with implicit sums over repeated indices, 
whereas $\mathbb{Q}_M^I,\bar{\mathbb{Q}}^I_M$ are in turn related to the canonical momentum, $\mathbb{\Pi}_{I,M}$ and winding, $\mathbb{W}^M_I$, via (\ref{eq:QQbar hat}). When the indices $M$ span $\mathbb{T}^{D_{\rm cr}-D}$,
\begin{equation}\label{eq:PLPR overview}
\begin{aligned}
&\mathbb{Q}^{I}_a\equiv \frac{1}{\ell_s}\Big(M^{'I}_a+B_{ab}N^b_I+G_{ab}N^b_I\Big),\qquad \bar{\mathbb{Q}}^{I}_a\equiv \frac{1}{\ell_s}\Big(M^{'I}_a+B_{ab}N^b_I-G_{ab}N^b_I\Big),
\end{aligned}
\end{equation}
where $M^{'I}_a,N^a_I\in \mathbb{Z}$ are summed over.\footnote{In our conventions there is no notion of raising or lowering the indices $I,J,\dots$, whereas the location of spacetime indices, $M,N,\dots$, has a precise meaning, and we always raise and lower these with the full metric $G_{MN}$.}

When considering string amplitudes associated to HES vertex operator insertions it is extremely useful to have the result for a generic correlation function. 
Denoting expectation values by:
\begin{equation}\label{eq:<exp ijx> overview}
\mathcal{A}(j)\equiv \Big\langle \exp\Big(i\int d^2z j_M x^M(z,\bar{z})\Big) \Big\rangle,
\end{equation}
with $\mathcal{A}(j)$ defined in (\ref{eq:Afull overview}), we will show (using point-splitting) that generic correlation functions chirally factorise:\footnote{In writing down (\ref{<<V1...Vn>>fixedloop4chirallysplit overview}) we have taken $D_{\rm cr}=26$ (appropriate for the bosonic string), but in the main text the case $D_{\rm cr}\neq 26$ is also considered which is relevant for generalising this result to the superstring where $D_{\rm cr}=10$. (In all cases we ignore the Liouville factor \cite{D'HokerPhong89} that cancels in all critical bosonic and superstring theories.) }
\begin{equation}\label{<<V1...Vn>>fixedloop4chirallysplit overview}
\begin{aligned}
\Big\langle \mathcal{D}_1&x^{N_1}(z_1,\bar{z}_1)\dots\mathcal{D}_{\mathcal{I}}x^{N_{\mathcal{I}}}(z_{\mathcal{I}},\bar{z}_{\mathcal{I}}) \bar{\mathcal{D}}_1x^{\bar{N}_1}(w_1,\bar{w}_1)\dots\bar{\mathcal{D}}_{\bar{\mathcal{I}}}x^{\bar{N}_{\bar{\mathcal{I}}}}(w_{\mathcal{I}},\bar{w}_{\bar{\mathcal{I}}}) \exp\Big(i\int d^2z j\cdot x(z,\bar{z})\Big) \Big\rangle=\\
&=i\bar{\delta}(j\ell_s)\,g_{\rm eff}^{2\mathfrak{g}-2}\,
\sumint\limits_{(\mathbb{Q},\bar{\mathbb{Q}})}
\mathcal{Z}_{\mathfrak{g}} \Big\langle\mathcal{D}_1x_+^{N_1}(z_1)\dots\mathcal{D}_{\mathcal{I}}x_+^{N_{\mathcal{I}}}(z_{\mathcal{I}}) e^{i\int d^2z (j_L+H)\cdot x_+(z)}\Big\rangle_+\\
&\,\,\,\quad\qquad\qquad\qquad\times\bar{\mathcal{Z}}_{\mathfrak{g}} \Big\langle\bar{\mathcal{D}}_1x_-^{\bar{N}_1}(\bar{w}_1)\dots\bar{\mathcal{D}}_{\bar{\mathcal{I}}}x_-^{\bar{N}_{\bar{\mathcal{I}}}}(\bar{w}_{\bar{\mathcal{I}}})e^{i\int d^2z (\bar{j}_R+\bar{H})\cdot x_-(\bar{z})} \Big\rangle_-,
\end{aligned}
\end{equation}
generalising the classic result of D'Hoker and Phong \cite{D'HokerPhong89}, who showed that amplitudes with massless vertex operators chirally split in flat non-compact backgrounds (in target spacetimes $\mathbb{R}^{D_{\rm cr}-1,1}$ with $G_{MN}=\eta_{MN}$ and $B_{MN}=0$). In particular, we show by explicit calculation that chiral splitting holds for generic correlation functions (of generic vertex operators with arbitrary KK and winding charges and polarisation tensors) in constant backgrounds, $G_{MN}$, $B_{MN}$, $\Phi$ and $U$, in generic target spacetimes $\mathbb{R}^{D-1,1}\times \mathbb{T}^{D_{\rm cr}-D}$ with generic K\"ahler and complex structure moduli and background gauge fields and torsion. 
The $\{\mathcal{D}_j,\bar{\mathcal{D}}_j\}$ are arbitrary worldsheet derivative operators, which, together with the $j_{{\rm L}M}$, $j_{{\rm R}M}$ \mbox{(anti-)chiral} sources are (with an appropriate point-splitting procedure) read off from the specific vertex operator insertions of interest (in their chiral representation \cite{SklirosCopelandSaffin16bb}). 

We want to emphasise that the {\it left-hand side} of (\ref{<<V1...Vn>>fixedloop4chirallysplit overview}) contains insertions of the full path integral field, $x^M=x^M_0+x^M_{\rm cl}+\tilde{x}^M$, and its derivatives, including zero modes, instantons and quantum fluctuations (with Green function (\ref{eq: Green Function+regular*m})), whereas the chiral fields, $x_{\pm}^M$, of the {\it right-hand side} are {\it defined} by their correlation functions, according to the rule that Wick contractions are carried out using (anti-)chiral propagators, 
$
\langle x_+^M(z)x_+^N(w)\rangle_+ =-\frac{\alpha'}{2}G^{MN}\ln E(z,w)
$, 
$
\langle x_-^M(\bar{z})x_-^N(\bar{w})\rangle_- =-\frac{\alpha'}{2}G^{MN}\ln \bar{E}(\bar{z},\bar{w})
$, and do {\it not} contain zero modes or instantons. The latter have already been taken into account in writing down (\ref{<<V1...Vn>>fixedloop4chirallysplit overview}). Clearly, using the chiral representation on the right-hand side vastly simplifies computations. 
The result for the chiral half on the right-hand side of (\ref{<<V1...Vn>>fixedloop4chirallysplit overview}) is given by,
\begin{equation}\label{eq:<>_+ overview}
\begin{aligned}
\Big\langle& \mathcal{D}_1x_+^{N_1}(z_1)\dots\mathcal{D}_{\mathcal{I}}x_+^{N_{\mathcal{I}}}(z_{\mathcal{I}}) e^{i\int d^2z (j_{\rm L}+H)\cdot x_+(z)}\Big\rangle_+=\\
&
=\exp\bigg(\frac{\alpha'}{4}\int\! \!\!\int j_{{\rm L}M}G^{MN}j'_{{\rm L}N}\ln E(z,z')+i\frac{\pi\alpha'}{2}\mathbb{Q}_{I}^{M}G_{MN}\Omega_{IJ}\mathbb{Q}_{J}^{N}+i\pi\alpha' \mathbb{Q}_{I}^{M}\int\,\!j_{{\rm L}M}\int^z\!\!\omega_I\bigg)\\
&\quad\times\sum_{k=0}^{\lfloor\mathcal{I}/2\rfloor}\sum_{\pi\in S_{\mathcal{I}}/\sim}\prod_{l=1}^k\bigg{\{}-\frac{\alpha'}{2}G^{N_{\pi(2l-1)}N_{\pi(2l)}}(\mathcal{D}\mathcal{D}\ln E)_{\pi(2l-1)\pi(2l)})\bigg{\}}\\
&\quad\times\prod_{q=2k+1}^{\mathcal{I}}\left\{\pi\alpha' \mathbb{Q}_I^{N_{\pi(q)}}\mathcal{D}_{\pi(q)}\!\!\int^{z_{\pi(q)}}\!\!\!\!\omega_I-\frac{\alpha'}{2}i\int j_{\rm L}^{N_{\pi(q)}}(\mathcal{D}\ln E)_{\pi(q)}\right\}
\end{aligned}
\end{equation}
where the argument in the exponential equals $\frac{\ell_s^2}{4}\int d^2z\int d^2z'\big(j_{\rm L}+H\big)\cdot \big(j'_{\rm L}+H'\big)\ln E(z,z')$, and similarly for the anti-chiral half,
\begin{equation}\label{eq:<>_- overview}
\begin{aligned}
\Big\langle& \bar{\mathcal{D}}_1x_-^{\bar{N}_1}(\bar{w}_1)\dots\bar{\mathcal{D}}_{\bar{\mathcal{I}}}x_-^{\bar{N}_{\bar{\mathcal{I}}}}(\bar{w}_{\bar{\mathcal{I}}})e^{i\int d^2z (j_{\rm R}+\bar{H})\cdot x_-(\bar{z})} \Big\rangle_-=\\
&
=\exp\bigg(\frac{\alpha'}{4}\int\! \!\!\int j_{{\rm R}M}G^{MN}{\bar{j}}'_{{\rm R}N}\ln E(\bar{z},\bar{z}')-i\frac{\pi\alpha'}{2}\bar{\mathbb{Q}}_{I}^{M}G_{MN}\bar{\Omega}_{IJ}\bar{\mathbb{Q}}_{J}^{N}-i\pi\alpha' \bar{\mathbb{Q}}_{I}^{M}\int\,\! j_{{\rm R}M}\int^{\bar{z}}\!\!\bar{\omega}_I\bigg)\\
&\quad\times\sum_{k=0}^{\lfloor\bar{\mathcal{I}}/2\rfloor}\sum_{\pi\in S_{\bar{\mathcal{I}}}/\sim}\prod_{l=1}^k\bigg{\{}-\frac{\alpha'}{2}G^{\bar{N}_{\pi(2l-1)}\bar{N}_{\pi(2l)}}(\bar{\mathcal{D}}\bar{\mathcal{D}}\ln \bar{E})_{\pi(2l-1)\pi(2l)})\bigg{\}}\\
&\quad\times\prod_{q=2k+1}^{\bar{\mathcal{I}}}\left\{-\pi\alpha' \bar{\mathbb{Q}}_I^{\bar{N}_{\pi(q)}}\bar{\mathcal{D}}_{\pi(q)}\!\!\int^{\bar{w}_{\pi(q)}}\!\!\!\!\bar{\omega}_I-\frac{\alpha'}{2}i\int j^{\bar{N}_{\pi(q)}}_{\rm R}(\bar{\mathcal{D}}\ln \bar{E})_{\pi(q)}\right\}
\end{aligned}
\end{equation}
$S_{\mathcal{I}}$ is the {\it symmetric group} of degree $\mathcal{I}$ \cite{Hamermesh}, the group of all permutations of $\mathcal{I}$ elements, and the equivalence relation `$\sim$' is such that $\pi_i\sim\pi_j$ with $\pi_{i},\pi_j\in S_{\mathcal{I}}$ when they define the same element in (\ref{eq:<>_+ overview}), and similarly for (\ref{eq:<>_- overview}). 
In the case of coherent vertex operator insertions, as we will see in \cite{SklirosCopelandSaffin16bb,SklirosCopelandSaffin16cc,SklirosCopelandSaffin16dd}, the sum over permutations can be carried out explicitly, and the various quantities appearing can be rewritten in terms of exponentials and special functions, thus vastly simplifying amplitude computations compared to the traditional approach in the literature that adopts a momentum eigenstate basis for vertex operators.

One can think of the fixed-loop momenta representation of the generating function (\ref{eq:Afull overview}) as defining a {\it Hamiltonian formulation} of string theory, because the zero mode momenta in all spacetime dimensions are manifest. Integrating out the loop momenta leads to a {\it Lagrangian formulation}, which is the usual starting point for string amplitude computations in the path integral formalism. In addition to these two there are also two natural hybrid formulations (also called {\it Routhian formulations} by analogy to classical mechanics) whereby the loop momenta are manifest in the compact dimensions but integrated out in the non-compact dimensions and vice versa. All these cases are discussed explicitly in Sec.~\ref{sec:WSD}, where it is also argued that (by direct analogy to point-particle quantum mechanics) the Hamiltonian formulation may be regarded as a `{\it wave formulation of string theory}', whereas the Lagrangian formulation may correspondingly be thought of as a string formulation. The equivalence of all four formulations can thus be regarded as a stringy manifestation of `{\it wave/particle duality}' of quantum mechanics, and so by analogy we refer to it as `{\it wave/string duality}'. For instance, we will argue that (\ref{<<V1...Vn>>fixedloop4chirallysplit overview}) may be regarded as a string theory statement of wave/string duality, where the left-hand side is in a string picture whereas the right-hand side is the corresponding wave picture.  As one should expect (from our experience with point-particle quantum mechanics, such as the double-slit experiment), certain questions are more easily addressed in a wave rather than a string picture and vice versa. We provide flesh to this claim by explicit decay rate computations (in both pictures) whose details will be presented elsewhere \cite{SklirosCopelandSaffin16dd}. 

\section{Generating Function}\label{sec:SAGF}

The starting point is to obtain a simple expression for the generating function of interest, $\mathcal{A}(j)$, that is crucial in the discussion of string amplitudes, cross sections and decay rates, associated to generic HES vertex operator insertions. It is defined by (in Euclidean target space and worldsheet signature):
\begin{equation}\label{eq:Afull}
\mathcal{A}^{\rm Eucl}(j)\dfn \int \mathcal{D}(x,b,\tilde{b},c,\tilde{c})\!\!\prod_{j=1}^{\#_{\mathbb{C}}{\rm moduli}}\!\!\!\!\!|\langle \mu_j,b\rangle |^2\,\prod_{s=1}^{\#_{\mathbb{C}}{\rm CKVs}}\!\!\!|c(w_s)|^2\,\,e^{-I(x|j)-I_{\rm gh}},
\end{equation}
and we reserve the notation $\mathcal{A}(j)$ for the corresponding Lorentzian signature quantity, see below. The (complex) number of moduli and conformal Killing vectors (CKV) are:
\begin{equation}\label{eq:moduli, CKV}
\big(\#_{\mathbb{C}}{\rm moduli},\#_{\mathbb{C}}{\rm CKVs}\big)= 
\left\{ \begin{array}{ll}
(0,3)\qquad {\rm for}\quad \mathfrak{g=0}\phantom{\Big|}\\
(1,1)\qquad {\rm for}\quad \mathfrak{g=1}\phantom{\Big|}\\
(3\mathfrak{g}-3,0)\qquad {\rm for}\quad \mathfrak{g>1}\phantom{\Big|}\\
  \end{array} \right.
\end{equation}
The $b,c$ are the Grassmann-odd ghosts,
$(c,\tilde{c})=(c^z(dz)^{-1},c^{\bar{z}}(d\bar{z})^{-1})$ and $(b,\tilde{b})=(b_{zz}(dz)^{2},b_{\bar{z}\bar{z}}(d\bar{z})^{2})$, whereas the Beltrami differentials, $(\mu_j,\bar{\mu}_j)=(\mu_{\bar{z}}^{\phantom{z}z}(dz)^{-1}d\bar{z},\mu_{z}^{\phantom{z}{\bar{z}}}dz(d\bar{z})^{-1})_j$, provide a parametrisation of the space of metrics on the Riemann surface, $\Sigma_{\mathfrak{g}}$, and define a gauge slice.\footnote{Our complex tensor notation is explained in Appendix \ref{sec:RS}.} There are as many insertions of, $|\langle \mu_j,b\rangle|^2$, as there are moduli (equivalently $b$ zero modes), and the pairing, $\langle \mu_j,b\rangle$, is defined with respect to the natural inner product of the space and is independent of a metric, $\langle\mu,b\rangle=\int_{\Sigma}d^2z\,\mu_{\bar{z}}^{\phantom{z}z}b_{zz}$, see (\ref{eq: (V1,V2)}). Similarly, in our approach it will be convenient to have as many insertions of $c\tilde{c}$ as there are conformal Killing vectors (CKV) (equivalently $c$ zero modes) on the Riemann surface, i.e.~the minimal number of allowed $c\tilde{c}$-ghost zero insertions. More general ghost insertions are also of interest \cite{Polchinski87,Nelson89,BeckerBeckerRobbins12,Witten12c,Sen15b}, and it is straightforward to extend the results of this paper to include also these cases (although strict chiral splitting may be lost in these more general situations). In turn, every $x(z,\bar{z})$ represents an embedding of the worldsheet into spacetime, $x:\Sigma\rightarrow \mathbb{R}^{D-1,1}\times \mathbb{T}^{D_{\rm cr}-D}$.

In general, the (worldsheet) matter and ghost contributions factorise, 
$$
\mathcal{A}^{\rm Eucl}(j)=\mathcal{A}_{\rm gh}^{\rm Eucl}\mathcal{A}_x^{\rm Eucl}(j),
$$ 
so let us focus initially on the matter contribution,
\begin{equation}\label{eq:A(j)}
\begin{aligned}
\mathcal{A}_x^{\rm Eucl}(j)&=\int \mathcal{D}x\,e^{-I(x|j)}\\
&=\int \dslash^{D\mathfrak{g}}\mathbb{P}_I^{\mu}\int \dslash^{D\mathfrak{g}}\mathbb{W}_I^{\mu}\,\int \mathcal{D}x\,e^{-I(x|j)}\deltaslash^{D\mathfrak{g}}\big(\mathbb{P}_I^{\mu}-\hat{\mathbb{P}}_I^{\mu}\big)\deltaslash^{D\mathfrak{g}}\big(\mathbb{W}_I^{\mu}-\hat{\mathbb{W}}_I^{\mu}\big),
\end{aligned}
\end{equation}
with,
\begin{equation}\label{eq:action}
I(x|j) \dfn\frac{1}{2\pi\alpha'}\int_{\Sigma_{\mathfrak{g}}} d^2z\,\Big(\partial_zx^M\partial_{\bar{z}}x^N\big(G_{MN}+B_{MN}\big)+\alpha'R_{z\bar{z}}\Phi+g_{z\bar{z}}U\Big)-i\int_{\Sigma_{\mathfrak{g}}} d^2z\, j_Mx^M.
\end{equation}
The constant tachyon background term, $\frac{1}{2\pi\alpha'}\int d^2zg_{z\bar{z}}U$, will be kept implicit throughout, but it will play a role in tadpole cancellations as we will see in the context of coherent vertex operator 2-point amplitudes in \cite{SklirosCopelandSaffin16cc}. 
Notation-wise, it will be convenient to define $I_m\dfn I(x|0)$, so that the full (source-free) action reads $I=I_m+I_{\rm gh}$. 
We now define the various quantities appearing in (\ref{eq:A(j)}) and (\ref{eq:action}). 

\ssk
The quantity $j_M$ is a (possibly physical, either real or complex, possibly local) source, and as we also discuss below functional derivatives with respect to it (upon adopting an appropriate point-splitting procedure) generate the correlation functions and amplitudes of interest. The one condition it must satisfy will be: $\int d^2zj_M=0$, which is usually associated to charge and momentum conservation.

We consider the exact (in $\alpha'$) string background where the spacetime metric, $G_{MN}$, antisymmetric tensor, $B_{MN}$,  dilaton, $\Phi$, and tachyon, $U$, are generic\footnote{`Generic' meaning that, e.g., that the determinants of the block diagonal pieces, $G_{\mu\nu}$ and $G_{ab}$, are non-vanishing.} constants,
\begin{equation}\label{eq:generic GBPhi}
G_{MN}=   \left(\begin{matrix}       G_{\mu\nu} & G_{\mu b} \\
      G_{a \nu} & G_{ab} \\
   \end{matrix}\right),\qquad B_{MN}=\left(\begin{matrix} 
         B_{\mu\nu} & B_{\mu b} \\
      B_{a \nu} & B_{ab} \\
   \end{matrix}\right),\qquad {\rm and}\qquad \Phi,U={\rm const}.
\end{equation}
The first two of these parametrise \cite{NarainSarmadiWitten87} the K\"ahler and complex structure moduli of the target space torus, $\mathbb{T}^{D_{\rm cr}-D}$ (contained in $G_{ab}$ and $B_{ab}$), as well as KK gauge fields (contained in $G_{\mu a}$ and $B_{\mu a}$) and torsion (contained in $B_{\mu\nu}$). 
We work in Euclidean signature (to make sense of the path integral over $x^0$) and eventually analytically continue back to Lorentzian signature.\footnote{\label{foot:MinkEucl}
Wick-rotating to Lorentzian signature can be achieved by replacing $G_{MN}^{(\rm Eucl)}$ by $G_{MN}^{(\rm Lor)}$, 
such that $({\rm Det}G^{\mu\nu})_{\rm Eucl}=-({\rm Det}G^{\mu\nu})_{\rm Lor}$.  Note that $(j_Mx^M)_{\rm Eucl}\rightarrow (j_Mx^M)_{\rm Lor}$, where $G_{MN}^{(\rm Eucl)}$ and $G_{MN}^{(\rm Lor)}$ are used to raise and lower indices before and after this replacement respectively. Note however that one has to be extremely careful when trying to interpret the energy integrals of the loop momenta and this has been analysed in detail by Pius and Sen \cite{PiusSen16}; see also Witten \cite{Witten13b} for an alternative approach.} 
Modulo this comment, index contractions will henceforth be carried out using the spacetime metric, $A^MB_M=A^MB^NG_{MN}$, etc., so that we raise and lower indices with the full metric $G_{MN}$. We will state explicitly when we rotate to Lorentzian signature. 

\ssk
The coefficient of the constant dilaton, $\Phi$, in the action is a topological invariant, equal to the Euler character $\chi(\Sigma_{\mathfrak{g}})=2-2\mathfrak{g}$ of the Riemann surface; see (\ref{eq:chi}) and note that the Ricci tensor $R_{z\bar{z}}$ is related to the Ricci scalar $R_{(2)}$ in (\ref{eq:R}). It is convenient to also define the string coupling in the standard manner: 
\begin{equation}\label{eq:g_s=ephi}
g_{s}\dfn e^{\Phi},
\end{equation}
and so there is an overall factor $g_s^{-\chi(\Sigma_{\mathfrak{g}})}$ in the generating function, i.e.~$\mathcal{A}^{\rm Eucl}(j)\propto g_s^{-\chi(\Sigma_{\mathfrak{g}})}$. 

\ssk
In the second equality in (\ref{eq:A(j)}) we have inserted the unit operator:
\begin{equation}\label{eq:1=intPW}
\mathbb{1}=\int \dslash^{D\mathfrak{g}}\mathbb{P}_I^{\mu}\int \dslash^{D\mathfrak{g}}\mathbb{W}_I^{\nu}\,\deltaslash^{D\mathfrak{g}}\big(\mathbb{P}_I^{\mu}-\hat{\mathbb{P}}_I^{\mu}\big)\deltaslash^{D\mathfrak{g}}\big(\mathbb{W}_I^{\nu}-\hat{\mathbb{W}}_I^{\nu}\big),
\end{equation}
where $\hat{\mathbb{P}}_I^{\mu}$ is the standard momentum operator and $\hat{\mathbb{W}_I^{\mu}}$ the winding operator. For a generic homology cycle $\mathcal{C}$ of the compact genus-$\mathfrak{g}$ Riemann surface these are defined by:
\begin{equation}\label{eq:P}
\begin{aligned}
&\hat{\mathbb{P}}_{\mathcal{C}}^{M}\dfn \frac{1}{2\pi\alpha'}\oint_{\mathcal{C}} \!\!\big(\partial x^{M}-\bar{\partial}x^{M}\big),\qquad \hat{\mathbb{W}}_{\mathcal{C}}^{M}\dfn \frac{1}{2\pi\alpha'}\oint_{\mathcal{C}} \!\!\big(\partial x^{M}+\bar{\partial}x^{M}\big).
\end{aligned}
\end{equation}
The operator $\hat{\mathbb{W}}^M_{\mathcal{C}}$ \cite{Polchinski_v1} measures the winding of a string whose spacelike (worldsheet) dimension traverses a generic cycle $\mathcal{C}$ of the worldsheet. The eigenvalues $\mathbb{W}^M_{\mathcal{C}}$ will be non-vanishing when the spacetime embedding of this string (associated to the homology cycle $\mathcal{C}$ of interest) wraps topologically non-trivial cycles of the spacetime torus, $\mathbb{T}^{D_{\rm cr}-D}$. 
Let us also define the chiral and anti-chiral halves, $\hat{\mathbb{Q}}^M_I$, $\hat{\bar{\mathbb{Q}}}^M_I$, respectively, such that:
\begin{equation}\label{eq:PWQQ}
\hat{\mathbb{P}}^M_{\mathcal{C}}=\frac{1}{2}\big(\hat{\mathbb{Q}}^M_{\mathcal{C}}+\hat{\bar{\mathbb{Q}}}^M_{\mathcal{C}}\big),\qquad \hat{\mathbb{W}}^M_{\mathcal{C}}=\frac{1}{2}\big(\hat{\mathbb{Q}}^M_{\mathcal{C}}-\hat{\bar{\mathbb{Q}}}^M_{\mathcal{C}}\big),
\end{equation}
and so,
\begin{equation}\label{eq:QQ}
\hat{\mathbb{Q}}^M_{\mathcal{C}}\dfn \frac{1}{\pi\alpha'}\oint_{{\mathcal{C}}}\partial x^M,\qquad\hat{\bar{\mathbb{Q}}}^M_{{\mathcal{C}}}\dfn- \frac{1}{\pi\alpha'}\oint_{{\mathcal{C}}}\bar{\partial} x^M.
\end{equation}
Choosing a canonical intersection basis for the $2\mathfrak{g}$ homology cycles of the compact genus-$\mathfrak{g}$ Riemann surface \cite{DHokerPhong}, see Appendix \ref{sec:RS}, the operators appearing in the $2D\mathfrak{g}$ delta functions in (\ref{eq:A(j)}) or (\ref{eq:1=intPW}) correspond to the specific choice of contours $\mathcal{C}=A_I$, with $I=1,\dots,\mathfrak{g}$. For simplicity we write $\hat{\mathbb{P}}_I\equiv\hat{\mathbb{P}}_{A_I}$, and $\hat{\mathbb{W}}_I\equiv \hat{\mathbb{W}}_{A_I}$. In the corresponding eigenvalues we omit the `$\hat{\phantom{a}}$'. 

We want an expression for the amplitude at fixed-loop momenta, but in the presence of a $B_{MN}$ field, $\mathbb{P}^{M}$ is not the {\it physical} momentum ($\hat{\mathbb{P}}^{M}$ is not the charge associated to spacetime translations). In particular, the Noether current \cite{Polchinski_v1} associated to rigid spacetime translations, $x^M\rightarrow x^M+a^M$ of the theory (\ref{eq:action}), reads (with a stringy normalisation \cite{Polchinski_v1}),
$$
j_{z,M}=\frac{i}{\alpha'}\partial_zx^N(G_{MN}-B_{MN}),\qquad j_{\bar{z},M}=\frac{i}{\alpha'}\partial_{\bar{z}}x^N(G_{MN}+B_{MN}),
$$
and so the associated conserved charge flowing through an arbitrary closed contour, $\mathcal{C}$, of the Riemann surface instead reads,
\begin{equation}\label{eq:PiCM}
\begin{aligned}
\hat{\mathbb{\Pi}}_{\mathcal{C},M}&\dfn \frac{1}{2\pi i}\oint_{\mathcal{C}} \big(dzj_{z,M}-d\bar{z}j_{\bar{z},M}\big)\\
&= G_{MN}\hat{\mathbb{P}}^N_{\mathcal{C}}-B_{MN}\hat{\mathbb{W}}^N_{\mathcal{C}}.\\
\end{aligned}
\end{equation}
When $B_{MN}=0$ the quantity $\hat{\mathbb{P}}^M_{\mathcal{C}}$ indeed measures spacetime momentum, but in the presence of a $B_{MN}$ field  the notion of momentum is modified,  $\hat{\mathbb{P}}^M_{\mathcal{C}}$ being replaced by $\hat{\mathbb{\Pi}}_{\mathcal{C},M}$, the two being related as in (\ref{eq:PiCM}). This is much like the momentum of a particle of mass $m$, namely $m \dot{\bf r}$, is replaced by $m \dot{\bf r}+e{\bf A}$ in the presence of a U(1) charge, $e$, (corresponding to $\mathbb{W }$) and associated vector potential ${\bf A}$ (corresponding to $B_{MN}$). These statements hold for a generic closed contour ${\mathcal{C}}$, and holomorphicity allows one to continuously deform this across the various homology cycles of the Riemann surface, or it may be taken to encircle one or more punctures at which vertex operators are inserted. Momentum and winding conservation is of course closely related to this notion of holomorphicity \cite{Polchinski_v1}. As mentioned above, we identify the contour, $\mathcal{C}$, with the $A_I$-cycles in the above delta functions.

We are aiming for an expression for the generating function, $\mathcal{A}(j)$, at physical fixed-loop momenta, and on account of the above discussion we should think of $\mathbb{\Pi}_{I\mu}$ as the {\it physical momentum} (i.e.~the momentum dumped into a detector) and so insert one more delta function constraint into the amplitude (\ref{eq:A(j)}):
\begin{equation}\label{eq:1=int Pi}
\begin{aligned}
1&=\int d^{D\mathfrak{g}}\mathbb{\Pi}_{I\mu}\delta^{D\mathfrak{g}}\big(\mathbb{\Pi}_{I\mu}-G_{\mu N}\mathbb{P}_{I}^{N}+B_{\mu N}\mathbb{W}^{N}_I\big),
\end{aligned}
\end{equation}
before finally integrating out $\mathbb{P}^{\mu}_I$ and $\mathbb{W}^{\mu}_I$, {\it after} having evaluated the path integral over embeddings in (\ref{eq:A(j)}) at fixed loop momenta. One thing to note is that when $B_{\mu a}=G_{\mu a}=0$, then $\mathbb{\Pi}_{\mu}=G_{\mu\nu}\mathbb{P}^{\nu}$ (there will also be an independent delta function constraint $\delta^{D\mathfrak{g}}(\mathbb{W}_I^{\mu})$ as we discuss momentarily), so when this is the case it is natural to define $\mathbb{P}_{\mu}\dfn \mathbb{\Pi}_{\mu}$. In the current document however all components of background fields will be kept generic. 

Of course, strings cannot wrap around a non-compact dimension, and so $\mathbb{W}^{\mu}_I$ should vanish identically. An important consistency check therefore will be to show that in fact:
$$
\langle\deltaslash^{D\mathfrak{g}}\big(\mathbb{W}_I^{\mu}-\hat{\mathbb{W}}_I^{\mu}\big)\dots\rangle_x\equiv\langle\deltaslash^{D\mathfrak{g}}\big(\mathbb{W}_I^{\mu}\big)\dots\rangle_x.
$$ 
On the other hand, winding in the compact dimensions will generically be non-trivial (see below), and so $\mathbb{\Pi}_{I\mu}$ will also receive contributions from $\mathbb{W}^{a}_I$ when $B_{\mu a}\neq0$, as can be seen from (\ref{eq:PiCM}). 

Before embarking on the evaluation of the path integral it will be important to make two final remarks. 
Even though the quantity $\mathbb{\Pi}_M$ is the physical momentum, the quantities that appear most naturally in loop amplitudes will actually be $\mathbb{Q}^M,\bar{\mathbb{Q}}^M$, and these are also the quantities that enter the mass formulas and vertex operators. It will be useful for later reference to have at hand an expression for the latter in terms of $\hat{\mathbb{\Pi}}_M$ and $\hat{\mathbb{W}}^M$,
\begin{equation}\label{eq:QQbar hat}
\begin{aligned}
\hat{\mathbb{Q}}^M_{\mathcal{C}}&=G^{MN}\hat{\mathbb{\Pi}}_{\mathcal{C},N}+G^{MN}B_{NK}\hat{\mathbb{W}}^K_{\mathcal{C}}+\hat{\mathbb{W}}^M_{\mathcal{C}}\\
\hat{\bar{\mathbb{Q}}}^M_{\mathcal{C}}&=G^{MN}\hat{\mathbb{\Pi}}_{\mathcal{C},N}+G^{MN}B_{NK}\hat{\mathbb{W}}^K_{\mathcal{C}}-\hat{\mathbb{W}}^M_{\mathcal{C}},
\end{aligned}
\end{equation}
and these follow from the above expressions by trivial rearrangement. That chiral and anti-chiral vertex operator momenta are actually constructed out of eigenvalues of $\hat{\mathbb{Q}}^M_{\mathcal{C}}$, $\hat{\bar{\mathbb{Q}}}^M_{\mathcal{C}}$ is clear from the definitions in (\ref{eq:QQ}); for example, $\hat{\mathbb{Q}}^M_{\mathcal{C}}e^{i\mathbb{k}\cdot x_+(z)+i\bar{\mathbb{k}}\cdot x_-(\bar{z})}=\mathbb{k}^Me^{i\mathbb{k}\cdot x_+(z)+i\bar{\mathbb{k}}\cdot x_-(\bar{z})}$, etc., where the (anti-)chiral fields $x_+(z)$, ($x_-(\bar{z})$) are related to the full path integral field $x(z,\bar{z})$ by a very subtle and indirect (yet remarkable) relation that we derive below, see (\ref{<<V1...Vn>>fixedloop4chirallysplit}), and as is well-known it is {\it not} correct to identify $x(z,\bar{z})$ with $x_+(z)+x_-(\bar{z})$ in general, although this may sometimes be justified.\footnote{Suffice it to say here that (as mentioned in the Overview section) $x(z,\bar{z})$ contains zero modes, instanton contributions and quantum fluctuations, $x(z,\bar{z})=x_0+x_{\rm cl}(z)+x_{\rm cl}(\bar{z})+\tilde{x}(z,\bar{z})$, whereas the chiral fields $x_+(z)$, $x_-(\bar{z})$ are {\it defined} (for any genus $\mathfrak{g}=0,1,\dots$) by their correlation functions, $\langle x_+^M(z)x_+^N(w)\rangle_+ =-\frac{\alpha'}{2}G^{MN}\ln E(z,w)$ and $\langle x_-^M(\bar{z})x_-^N(\bar{w})\rangle_- =-\frac{\alpha'}{2}G^{MN}\ln \bar{E}(\bar{z},\bar{w})$ (and $\langle x_+^M(z)x_-^N(\bar{w})\rangle=0$) with $E(z,w)$ Fay's prime form, and do not contain zero modes or instanton contributions. This observation is closely related to the observation of D'Hoker and Phong \cite{D'HokerPhong89} that fixing the loop momenta in {\it all} spacetime directions leads to chirally factorised amplitudes. The map between the two (anti-)chiral and full path integral fields is given in (\ref{<<V1...Vn>>fixedloop4chirallysplit}) for generic constant backgrounds, $G_{MN}$, $B_{MN}$ and $\Phi$.} We then choose the contour $\mathcal{C}$ to encircle the vertex operator (with any other features or insertions outside the contour), use holomorphicity to shrink the contour, in which case only the leading singular piece $\langle x_+^M(z)x_+^N(w)\rangle_+\big|_{z\rightarrow w} =-\frac{\alpha'}{2}G^{MN}\ln (z-w)$ contributes, and similarly for the anti-chiral sector of vertex operators.

Because it is $\hat{\mathbb{\Pi}}_M$ that generates spacetime translations, the usual argument concerning single-valuedness of the wavefunction \cite{Polchinski_v1} implies that eigenvalues of $\hat{\mathbb{\Pi}}_M$ must be discrete. To make this statement sharp, note that we absorb all K\"ahler and complex structure moduli into the background fields, $G_{MN}$, $B_{MN}$. This allows us to compactify the $x^a$ on a ($D_{\rm cr}-D$)-dimensional hypercube, such that for any $a$ spanning $\mathbb{T}^{D_{\rm cr}-D}$, we make the identification:
$$x^a\sim x^a+2\pi \ell_s,$$ with $\ell_s=\sqrt{\alpha'}$ the string length. (In this approach the actual compactification radius is determined by the moduli in $G_{ab}$, $B_{ab}$, and it is not $\ell_s$ as one might naively conclude.) Then, under a {\it lattice} translation $x^a\rightarrow x^a+2\pi\ell_s$, the equation $e^{i(2\pi \ell_s)\hat{\mathbb{\Pi}}_a}=\mathbb{1}$ (for every $a$ spanning $\mathbb{T}^{D_{\rm cr}-D}$) should hold as an operator statement in the string Hilbert space, so that its eigenvalues must be quantised in units of $1/\ell_s$: 
\begin{equation}\label{eq:PiMl}
\mathbb{\Pi}_a= M_a/\ell_s,\qquad{\rm with} \qquad M_a\in \mathbb{Z},\qquad (a \,\,\,\,{\rm spans}\,\,\,\, \mathbb{T}^{D_{\rm cr}-D}).
\end{equation} 
The position of the indices is important; recall that we generically raise and lower spacetime indices with $G_{MN}$, and we do not assume $G_{\mu a}=0$ (or $B_{\mu a}=0$) in this document.  

\ssk
We are now ready to evaluate the matter generating function (\ref{eq:A(j)}) in target spacetimes of the form $\mathbb{R}^{D-1,1}\times \mathbb{T}^{D_{\rm cr}-D}$ for generic constant backgrounds (\ref{eq:generic GBPhi}). 
We expand around classical instanton solutions, $x_{\rm cl}^M$, defined to solve the classical equation of motion of  $I(x|j)$, see (\ref{eq:ddbarx_cl=j}),
\begin{equation}\label{eq:xexp}
x^M=x^M_0+x^M_{\rm cl}+\tilde{x}^M,
\end{equation}
where we denote  quantum fluctuations by $\tilde{x}^M$ and have also extracted out a constant zero mode $x^M_0$. Before inserting this into the action (\ref{eq:action}), and then into the path integral (\ref{eq:A(j)}), let us determine the classical instanton solution. There are various subtleties (as well as new features) that are not discussed in the standard literature, so we will be fairly explicit. 

\ssk
Note primarily that $x_{\rm cl}$ encodes the information that closed cycles on the worldsheet may wrap around non-trivial cycles of the torus $\mathbb{T}^{D_{\rm cr}-D}$. As discussed above, all K\"ahler and complex structure moduli will be absorbed into $G_{MN},B_{MN}$, and so we are free to normalise the $x^a_{\rm cl}$ such that $x^a_{\rm cl}\sim x^a_{\rm cl}+2\pi \ell_s$, for all $a$ spanning $\mathbb{T}^{D_{\rm cr}-D}$. The quantity $x_{\rm cl}^M$ by definition satisfies the classical equations of motion\footnote{Note that the constant $B_{MN}$ does not contribute to the classical equations of motion, and neither does it enter the worldsheet energy-momentum tensor, and so the Virasoro constraints are as in the non-compact theory with the replacement $\eta_{MN}\rightarrow G_{MN}$.} of (\ref{eq:action}), 
\begin{equation}\label{eq:ddbarx_cl=j}
\partial_z\partial_{\bar{z}}x_{\rm cl}^M\equiv -\pi i\alpha'G^{MN}j_N,
\end{equation}
and is transverse to the constant zero mode, $x_0^M$. First consider the case $j_M=0$, the case of interest always being $\int d^2zj_M=0$, that ensures overall charge neutrality (this is enforced upon us by the zero mode integrals\footnote{It is sometimes of interest to relax momentum conservation either at an intermediate stage in a calculation (see \cite{Minahan87} and recently emphasised in \cite{BergBuchbergerSchlotterer16}, as a means of regularisation while preserving onshell conditions), or relax momentum conservation all together (which is of interest for perturbation theory on non-trivial backgrounds), so we try to state explicitly throughout where momentum conservation is assumed so as to allow for appropriate generalisations.}). The solution that describes the soliton contribution of interest can be expanded in a complete basis, $\omega_I$, $\bar{\omega}_I$, as follows:
\begin{equation}\label{eq:xcl sol}
x_{\rm cl}^M = \left\{ \begin{array}{ll}
 \gamma_I^M\displaystyle\int^z_{\wp}\omega_I+\bar{\gamma}_I^M\int^{\bar{z}}_{\bar{\wp}}\bar{\omega}_I&\quad \textrm{if $M$ spans $\mathbb{T}^{D_{\rm cr}-D}$}\phantom{\Big]}\\
0 & \quad\textrm{if $M$ spans $\mathbb{R}^D$},\phantom{\Big]}
  \end{array} \right.
\end{equation}
where $\wp,\bar{\wp}\in \Sigma_{\mathfrak{g}}$ denote an arbitrary reference point on which amplitudes do not depend (see below), the $\omega_I=\omega_I(z)dz$, (with $I=1,\dots,\mathfrak{g}$ and an implicit sum over repeated indices) denote a basis for the $\mathfrak{g}$ abelian holomorphic differentials associated to a compact genus-$\mathfrak{g}$ Riemann surface, normalised by their $A_I$-cycles, $\oint_{A_I}\omega_J=\delta_{IJ}$, and similarly for $\bar{\omega}_I=\bar{\omega}_I(\bar{z})d\bar{z}$, namely $\oint_{A_I}\bar{\omega}_J=\delta_{IJ}$. The existence of the $\omega_I,\bar{\omega}_I$ is guaranteed by the Atiyah-Singer-Riemann-Roch index theorem, see (\ref{eq:Atiyah-Singer}) and the discussion following (\ref{eq:intersectionbasis}). Working with a canonical intersection basis (\ref{eq:intersectionbasis}) we denote the corresponding period matrix by $\Omega_{IJ}$, defined by $\oint_{B_I}\omega_J=\Omega_{IJ}$, $\oint_{B_I}\bar{\omega}_J=\bar{\Omega}_{IJ}$. The quantities $\gamma_I^M,\bar{\gamma}_I^M$ in (\ref{eq:xcl sol}) read:
\begin{equation}\label{eq:gammas}
\begin{aligned}
&\gamma_I^M=-i\pi ({\rm Im}\Omega)^{-1}_{IJ}(M_J^M-\bar{\Omega}_{JL}N_L^M)\ell_s,\\
&\bar{\gamma}_I^M=i\pi ({\rm Im}\Omega)^{-1}_{IJ}(M_J^M-\Omega_{JL}N_L^M)\ell_s,
\end{aligned}
\end{equation}
where $\{N_I^M,M_I^M\}\in \mathbb{Z}$ (see Appendix \ref{sec:RS} for a more explicit overview of conventions). 

\ssk
In order to arrive at (\ref{eq:xcl sol}) and (\ref{eq:gammas}), note that in toroidal compactifications as we go around an $A_I$- or $B_I$-cycle of the worldsheet, the spacetime embedding should return to itself up to an integer multiple of $2\pi \ell_s$,
\begin{equation}\label{eq:periodicity x AB cycles}
\oint_{A_I}dx_{\rm cl}^a=2\pi N_I^a\ell_s,\qquad \oint_{B_I}dx_{\rm cl}^a=2\pi M_I^a\ell_s,
\end{equation}
where we write $d=dz\partial_z+d\bar{z}\partial_{\bar{z}}$ for total differentials in the $(z,\bar{z})$ coordinate system, with $\bar{z}^*=z$. We solve these constraints by expanding in a  complete basis, 
$
\partial x_{\rm cl}^a= \sum_I\gamma_I^a\omega_I$, $\bar{\partial}x_{\rm cl}^a=\sum_I\bar{\gamma}_I^a\bar{\omega}_I
$, 
and then the $\gamma_I^a,\bar{\gamma}_I^a$ are determined immediately from (\ref{eq:periodicity x AB cycles}), (\ref{eq:omegaA}) and (\ref{eq:omegaB}), leading to (\ref{eq:gammas}).  
As a consistency check, notice that under $A_I$-cycle translations\footnote{\label{foot:z+A}We are being a little bit sloppy here. The coordinate $z$ should really be thought of as the image ${\bf z}(\wp)$ of a point $\wp\in \Sigma_{\mathfrak{g}}$ under the Jacobi (or Abel) map,  
$\mathbb{I}:\wp\rightarrow {\bf z}(\wp)=\big(\int_{\wp_0}^{\wp}\omega_1,\dots,\int_{\wp_0}^{\wp}\omega_{\mathfrak{g}}\big)$,  
with $\wp_0$ some (universal) reference point on which physical observables do not depend. 
In particular, by transport $z$ around a cycle $A_I$ we mean ${\bf z}(\wp)\rightarrow{\bf z}(\wp+A_I)$, and similarly for the $B$-cycles. 
The vector ${\bf z}$ is an element of the complex torus $J(\Sigma_{\mathfrak{g}})\equiv  \mathbb{C}^h/(\mathbb{Z}^h+\Omega\mathbb{Z}^h)$.
 } 
$z\rightarrow z+A_I$, (\ref{eq:xcl sol}) implies that $x_{\rm cl}^a\rightarrow x_{\rm cl}^a+2\pi N_I^a\ell_s$ (with $x_{\rm cl}^{\mu}$ invariant),  where we have used (\ref{eq:omegaA}), and similarly for translations around $B$-cycles, under $z\rightarrow z+B_I$ on account of (\ref{eq:omegaB}) we have $x_{\rm cl}^a\rightarrow x_{\rm cl}^a+2\pi M_I^a\ell_s$. Given the identification $x^a\sim x^a+2\pi \ell_s$, the embedding of the worldsheet into spacetime is single-valued under $z\rightarrow z+A_I$ and $z\rightarrow z+B_I$. 

Now let us turn on a general source term, $j_M(z,\bar{z})$, subject to $\int d^2zj_M(z,\bar{z})=0$, and consider the set of solutions to (\ref{eq:ddbarx_cl=j}). Making use of the defining equation for the Green function transverse to zero modes, see Appendices \ref{sec:GF} and \ref{sec:RS}, (we have factored out the zero mode $x_0^M$ as displayed in (\ref{eq:xexp})),
$$
\partial_z\partial_{\bar{z}}G(z,w)=-\pi \alpha'\delta^2(z-w)+\frac{\pi\alpha'g_{z\bar{z}}}{\int_{\Sigma_{\mathfrak{g}}}d^2z\sqrt{g}},
$$
the soliton solution of interest that solves the full equation of motion (\ref{eq:ddbarx_cl=j}) can now be seen to take the form:
\begin{equation}\label{eq:xcl sol2}
x_{\rm cl}^M(z,\bar{z}) = \left\{ \begin{array}{ll}
 \gamma_I^M\displaystyle\int^z_{\wp}\omega_I+\bar{\gamma}_I^M\int^{\bar{z}}_{\bar{\wp}}\bar{\omega}_I+i \int d^2wG^{MN}j_N(w,\bar{w})G(w,z)&\quad \textrm{if $M$ spans $\mathbb{T}^{D_{\rm cr}-D}$}\phantom{\Big]}\\
i \int d^2wG^{MN}j_N(w,\bar{w})G(w,z) & \quad\textrm{if $M$ spans $\mathbb{R}^D$},\phantom{\Big]}
  \end{array} \right.
\end{equation}
with $\gamma_I^M,\bar{\gamma}_I^M$ as displayed above.\footnote{In principle we could add to $x_{\rm cl}^M(z,\bar{z})$ arbitrary well-behaved functions $f(a_I\int^z\omega_I)+\bar{f}(\bar{a}_I\int^{\bar{z}}\bar{\omega}_I)$ subject to (\ref{eq:periodicity x AB cycles}), but the aforementioned choice (which will be referred to as the `basic' one below) will be sufficient for our purposes.} This satisfies all the monodromy requirements, given that (in addition to the above observations concerning the $j=0$ piece) the Green function is by construction periodic under translations $z\rightarrow z+A_I$ and $z\rightarrow z+B_I$ (see Appendix \ref{sec:GF}). 

For a given source term, $j_M(z,\bar{z})$, the set of topologically distinct classical solutions is still classified by the set of integers in $\gamma_I^M,\bar{\gamma}_I^M$, (i.e.~the topological winding numbers associated to $A_I$ and $B_I$ cycles wrapping $\mathbb{T}^{D_{\rm cr}-D}$) as in (\ref{eq:xcl sol}). Secondly, the theory is Gaussian and so on account of the decomposition (\ref{eq:xexp}) we are free to absorb the $j$-dependent terms in (\ref{eq:xcl sol2}) into a redefinition of the fluctuations, $\tilde{x}^M(z,\bar{z})$, without affecting the background around which we are expanding. (We will give this last comment more flesh at the end of this section, where we will derive the effect of this shift in the final answer for the generating function.)\footnote{This would not generically be the case in the context of non-linear sigma model background field perturbation theory (in $\alpha'$), where such shifts can take us to a new vacuum that is physically distinct from the previous one (within perturbation theory).} This amounts to the simultaneous shifts $x_{\rm cl}\rightarrow y_{\rm cl}$ and $\tilde{x}\rightarrow y$, with:
 \begin{subequations}\label{eq:yycl}
\begin{align}
&y_{\rm cl}^M(z,\bar{z}) \dfn x_{\rm cl}^M(z,\bar{z})-i\int d^2wG^{MN}j_N(w,\bar{w})G(z,w)\\
&y^M(z,\bar{z})\dfn \tilde{x}^M(z,\bar{z})+i\int d^2wG^{MN}j_N(w,\bar{w})G(z,w),\label{eq:x->y shift}
\end{align}
\end{subequations}
where note that $y_{\rm cl}+y=x_{\rm cl}+\tilde{x}$. In particular, $y_{\rm cl}^M(z,\bar{z})$ is identified with (\ref{eq:xcl sol}) and $y^M(z,\bar{z})$ is the new quantum field. Given that such a shift (\ref{eq:x->y shift}), being field-independent, will certainly leave the path integral measure invariant, the quantity (\ref{eq:xcl sol}), equivalently $y_{\rm cl}^M$, can be taken to be the complete set of the basic classical solutions. 

We next substitute:
\begin{equation}\label{eq:x=x0+ycl+y}
x^M=x_0^M+y_{\rm cl}^M+y^M
\end{equation}
into the full action (\ref{eq:action}), {\it without} dropping any boundary terms, so that on account of (\ref{eq:yycl}) and (\ref{eq:xcl sol2}) we can recast the result into the form:
\begin{equation}\label{eq:I(x0+ycl+y)}
\begin{aligned}
I(x_0+y_{\rm cl}+y|j)&=\frac{1}{2\pi\alpha'}\int_{\Sigma_{\mathfrak{g}}} d^2z\,\partial_zy^M\partial_{\bar{z}}y^N\big(G_{MN}+B_{MN}\big)-i\int_{\Sigma_{\mathfrak{g}}} d^2z\, j_My^M-i\int_{\Sigma_{\mathfrak{g}}} d^2z\, j_Mx_0^M\\
&\qquad +\gamma_I^a\Big(\frac{1}{\pi\alpha'}{\rm Im}\,\Omega_{IJ}\big(G_{ab}+B_{ab}\big)\Big)\bar{\gamma}_J^b-i\Phi_a^I\gamma_I^a-i\bar{\Phi}^I_a\bar{\gamma}_I^a+\chi \Phi,
\end{aligned}
\end{equation}
where we have defined (note that $dz\wedge dz=0$):
\begin{equation}\label{eq:PhibarPhi}
\begin{aligned}
&\Phi_a^I\dfn \int_{\Sigma_{\mathfrak{g}}} d^2z\,j_a\int^z\!\omega_I-\frac{1}{2\pi\alpha'}\int_{\Sigma_{\mathfrak{g}}} \omega_I\wedge dy^N(G_{Na}-B_{Na})\\
&\bar{\Phi}_a^I\dfn \int_{\Sigma_{\mathfrak{g}}} d^2z\,j_a\int^{\bar{z}}\!\bar{\omega}_I+\frac{1}{2\pi\alpha'}\int_{\Sigma_{\mathfrak{g}}} \bar{\omega}_I\wedge dy^N(G_{Na}+B_{Na}).
\end{aligned}
\end{equation}

Let us now return to the full path integral over matter fields (\ref{eq:A(j)}). It is necessary to integrate out the zero modes, $x_0^M$, first, because this leads to constraints that will be enforced when integrating out $y^M$ and $y_{\rm cl}^M$.\footnote{Not integrating out the zero modes, $x_0^M$, at this stage of the calculation is certainly interesting, as it is relevant for string scattering in curved backgrounds (in a background field expansion sense) for strings whose spatial extent is smaller than any background curvature scale, and we hope to return to this point in the future.} We expand 
$x_0^M+y^M(z,\bar{z})$ in a complete orthonormal basis $\{\phi_{\alpha}\}$ as follows, $\sum_{\alpha\in \mathbb{Z}}A_{\alpha}^M\,\phi_{\alpha}(z,\bar{z})$, with canonical normalisation $\int d^2z\sqrt{g}\phi_{\alpha}(z,\bar{z})\phi_{\beta}(z,\bar{z})=\delta_{\alpha\beta}$, with $\phi_{\alpha}(z,\bar{z})$ an eigenfunction of the Laplacian, $\Delta_{(0)}\phi_{\alpha}=\omega_{\alpha}^2\phi_{\alpha}$, and $\omega_0^2\dfn 0$ defining the constant zero mode, $x_0=A_0\phi_0$. The natural measure is then, $\mathcal{D}x=``d^{D_{\rm cr}}x_0\mathcal{D}y\sum_{y_{\rm cl}}{\textrm{''}}=\prod_{\alpha\in\mathbb{Z}}\big(d^{D_{\rm cr}}A_{\alpha}\sqrt{{\rm Det} G_{MN}}\big)\sum_{y_{\rm cl}}$, and we factor out the zero modes, $d^{D_{\rm cr}}x_0=d^{D_{\rm cr}}A_0\sqrt{{\rm Det}G_{MN}}$, with $\mathcal{D}y=\prod_{\alpha\neq0}\big(d^{D_{\rm cr}}A_{\alpha}\sqrt{{\rm Det} G_{MN}}\big)$ a remaining fluctuation contribution and a sum over topologically distinct classical instanton contributions, $\sum_{y_{\rm cl}}$. The zero mode integral then factorises into a piece associated to $\mathbb{R}^D$ and one associated to $\mathbb{T}^{D_{\rm cr}-D}$:
\begin{equation}\label{eq:zeromodes}
\begin{aligned}
\Psi_{\rm Eucl}^0 &\dfn \int d^{D_{\rm cr}}A_0\sqrt{{\rm Det} G_{MN}}\,e^{i\int j\cdot A_0\,\phi_0}\,g_s^{-\chi}\\
&\,=(2\pi)^D\delta^D(\smallint j_{\mu})\,(2\pi \ell_s)^{D_{\rm cr}-D}\delta^{D_{\rm cr}-D}_{(\smallint j_a),0}\,\sqrt{{\rm Det} G_{MN}}\Big(\int d^2z\sqrt{g}\Big)^{D_{\rm cr}/2}\,g_s^{-\chi},
\end{aligned}
\end{equation}
and for convenience we have also included the dilaton contribution in the definition of $\Psi_{\rm Eucl}^0$. The $d$-dimensional {\it Kronecker} delta is denoted by $\delta^d_{(\cdot),0}$, and $\delta^d(\cdot )$ is a $d$-dimensional Dirac delta function (whose argument has indices ``downstairs''\footnote{When $G_{\mu a}=0$, we can raise indices in the Dirac delta function using the following rule: $\delta^D(C_{\mu})=\frac{1}{{\det}\,G_{\mu\nu}}\delta^D(C^{\mu})$, with $C_{\mu}=G_{\mu\nu}C^{\nu}$, assuming ${\det }G_{\mu\nu}$ is positive definite. Note also that $d^DC_{\mu}={\det} \,G_{\mu\nu}d^DC^{\mu}$, by which we mean $d^DC^{\mu}=dC^0\wedge dC^1\wedge\dots$. We do not assume $G_{\mu a}=0$ in this section however.}), that arise from $\mathbb{T}^{D_{\rm cr}-D}$ and $\mathbb{R}^D$ respectively. The identifications $x^a\sim x^a+2\pi \ell_s$ lead to the factor $(2\pi \ell_s)^{D_{\rm cr}-D}$. Rotating back to Lorentzian signature target spacetime amounts to replacing 
$(G_{MN})^{\rm Eucl}$ by $(G_{MN})^{\rm Mink}$ (and hence $({\rm Det}\,G_{MN})^{\rm Eucl}$ by $({\rm Det}\,G_{MN})^{\rm Mink}=-(-{\rm Det}\,G_{MN})^{\rm Mink}$), so that the right-hand side of (\ref{eq:zeromodes}) in Lorentzian target space signature reads:
\begin{equation}\label{eq:zeromodesL}
\begin{aligned}
\Psi_{\rm Mink}^0\dfn  i(2\pi)^D\delta^D(\smallint j_{\mu})\,(2\pi \ell_s)^{D_{\rm cr}-D}\delta^{D_{\rm cr}-D}_{(\smallint j_a),0}\,\sqrt{-{\rm Det} G_{MN}}\Big(\int d^2z\sqrt{g}\Big)^{D_{\rm cr}/2}\,g_s^{-\chi},
\end{aligned}
\end{equation}
the branch of the square root being convention-dependent (our choice is in agreement with Polchinski \cite{Polchinski_v1}). 
Note that the source (Dirac and Kronecker) delta functions enforce charge neutrality for the asymptotic states, $\int d^2zj_M(z,\bar{z}) =0$ for all $M$ spanning $\mathbb{R}^{D-1,1}\times \mathbb{T}^{D_{\rm cr}-D}$.

\begin{figure}
\begin{center}
\vspace{-2cm}
\includegraphics[angle=-90,origin=c,width=0.7\textwidth]{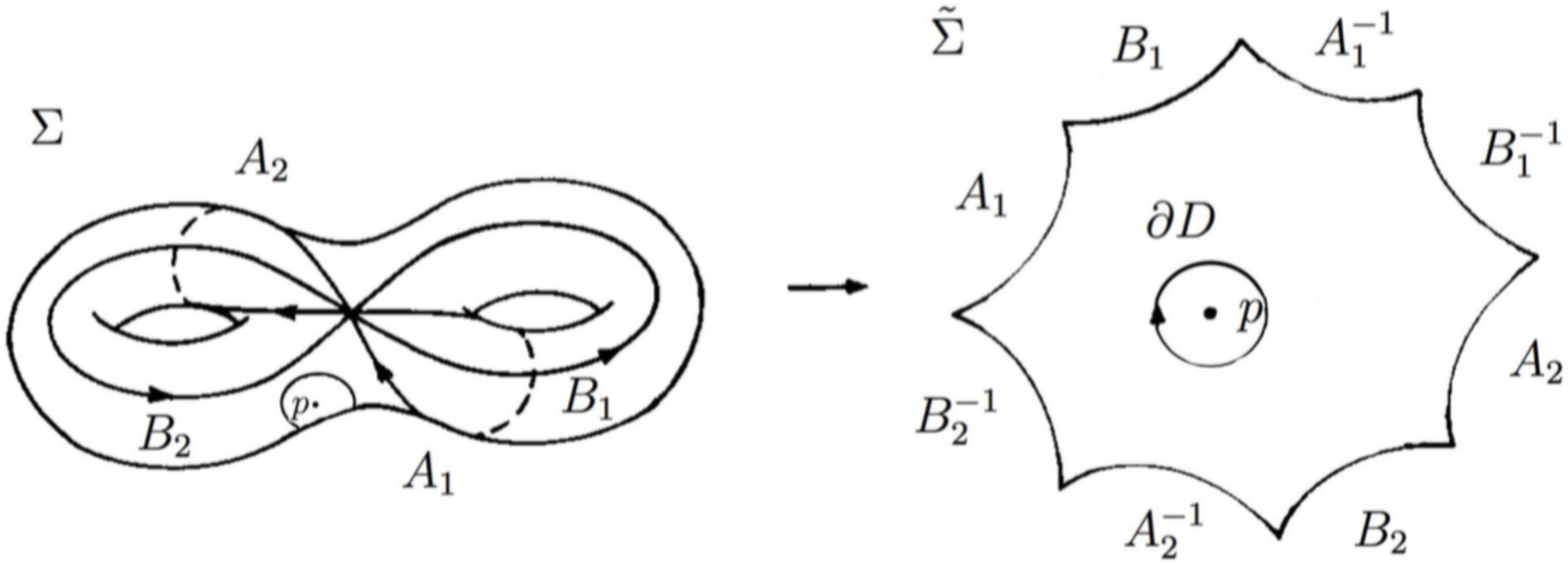}
\vspace{-3.5cm}
\caption{Pictorial representation of a genus-$2$ Riemann surface, $\Sigma_{2}$, (on the left) and the corresponding cut surface, $\tilde{\Sigma}_2$, (on the right), obtained from the former by smoothly (isotopically in $\tilde{\Sigma}_2$) dragging all cycles associated to the canonical intersection basis of the homology group so that they meet at a point (an 8-point vertex, see the first image), and subsequently ``deleting'' these homology lines from the surface.  This leads to the cut Riemann surface, $\tilde{\Sigma}_2$, which has a boundary, $\partial\tilde{\Sigma}_2$, on which functions are single-valued. The point $p$ indicates a point at which the integrand (\ref{eq:intermediate integral}) is singular and the disc $D$ of infinitesimal radius $\epsilon$ is defined to be centred at $p$ (denoted by local coordinates $w,\bar{w}$ in the text) with boundary $\partial D$. A similar picture is to be understood for any genus $\mathfrak{g}\geq1$ surface where the resulting cut surface is a $4\mathfrak{g}$ polygon.}\label{Fig:CutRiemannSurface}
\end{center}
\end{figure}

Having determined the zero mode contribution, let us turn our attention to the $y$-dependent pieces, starting from the $y$-dependent integrals in (\ref{eq:PhibarPhi}). 
It is sometimes convenient to write the holomorphic one forms, $\omega_I$, $\bar{\omega}_I$, in terms of the Abel map $g_I,\bar{g}_I$:
$$
\omega_I=dg_I,\qquad \bar{\omega}_I=d\bar{g}_I,\qquad{\rm with}\qquad g_I(z)=\int_{\wp}^z\omega_I,\qquad \bar{g}_I(\bar{z})=\int_{\bar{\wp}}^{\bar{z}}\bar{\omega}_I.
$$
On the cut Riemann surface \cite{Mumford_v12}, see Fig.~\ref{Fig:CutRiemannSurface}, the $g_I,\bar{g}_I$ are single-valued and $g_I({\wp})=\bar{g}_I({\bar \wp})=0$ for an arbitrary point $\wp,\bar{\wp}$ on the surface. 
The above definition makes manifest the fact that the $j$-independent integrals in (\ref{eq:PhibarPhi}) are integrals of exact forms,
\begin{equation}\label{eq:FFbar}
F_I\dfn \int_{\Sigma_{\mathfrak{g}}} \omega_I\wedge dy=\int_{\Sigma_{\mathfrak{g}}} d\big(g_I\wedge dy\big),\qquad \bar{F}_I\dfn \int _{\Sigma_{\mathfrak{g}}}\bar{\omega}_I\wedge dy=\int_{\Sigma_{\mathfrak{g}}} d\big(\bar{g}_I\wedge dy\big).
\end{equation}
We might therefore be tempted to drop these integrals, given that the integration domain is a compact Riemann surface, but the integrand has non-trivial monodromies around $A_I$ and $B_I$ cycles, and there is also the possibility of $dy$ contributing poles that may lead to a non-vanishing result, see e.g.~\cite{Mumford_v12} (p.~150) and also Appendix A of \cite{LugoRusso89}. 
Given that $y$ is a quantum field, what we need to check is whether $F_I$, $\bar{F}_I$ contribute to correlation functions. That is, if we can show that (when $\int d^2zj_M(z,\bar{z})=0$):
\begin{equation}\label{eq:<FF>,<FFbar>,<jyF>}
\begin{aligned}
& \Big\langle F_{I}^M\,\bar{F}_{J}^N\Big\rangle=\Big\langle F_{I}^M\,F_{J}^N\Big\rangle=0,\qquad \Big\langle \int j_M y^M\,F_{I}^N\Big\rangle=0,
\end{aligned}
\end{equation}
and similar expressions with $\bar{F}_I$ replacing $F_I$, then (given the theory is free) the following equality holds (correlators being with respect to the $y$ path integral),
\begin{equation}\label{eq:<ejyeFeFbar>}
\Big\langle e^{i\int j_M y^M}e^{C_{IN}F_I^N}e^{\bar{C}_{IN}\bar{F}_I^N}\Big\rangle=\Big\langle e^{i\int j_M y^M}\Big\rangle,\qquad {\rm when}\qquad \int d^2zj_M(z,\bar{z})=0,
\end{equation}
for any set of constants\footnote{The case of interest above being: $ C_{IN}=-\frac{i}{2\pi\alpha'}(G_{Na}-B_{Na})\gamma_I^a$ and $\bar{C}_{IN}=\frac{i}{2\pi\alpha'}(G_{Na}+B_{Na})\bar{\gamma}_I^a$.} $C_{IN}$, $\bar{C}_{IN}$, and we can effectively set $F_I=\bar{F}_I=0$ from the outset. That (\ref{eq:<FF>,<FFbar>,<jyF>}) holds is indeed the case, but because the reasoning is somewhat subtle we will be explicit.  
Considering first $\langle F_I \bar{F}_J\rangle$ and $\langle F_I F_J\rangle$, we make use of the explicit expression for the propagator, $\langle y^M(z,\bar{z})y^N(w,\bar{w})\rangle =G^{MN}G(z,w)$, see (\ref{eq: Green Function+regular*m}), the Riemann bilinear identity, $\int_{\Sigma_{\mathfrak{g}}}\omega_I\wedge \bar{\omega}_J=-2i({\rm Im}\Omega)_{IJ}$, and the monodromy properties of the prime form \cite{Fay,Mumford_v12,D'HokerPhong89}, see also Appendix \ref{sec:RS}:
$$
\oint_{B_I}dz\partial_z\partial_w\ln E(z,w)=2\pi i\,\omega_I(w),\qquad \oint_{A_I}dz\partial_z\partial_w\ln E(z,w)=0.
$$ 
It is straightforward to then show that the following contractions of $F_I,\bar{F}_I$ vanish identically:
$$
\Big\langle \bar{F}_I \bar{F}_J\Big\rangle=\Big\langle F_I F_J\Big\rangle =\Big\langle F_I \bar{F}_J\Big\rangle =0.
$$
The remaining correlators we need to compute in order to establish (\ref{eq:<ejyeFeFbar>}) are $\langle \int j_My^M\, \bar{F}_J^N\rangle$ and $\langle  \int j_My^M\, F_I^N\rangle$. Given that $\int j_M=0$, all we need to check is that the correlator $\langle y(w,\bar{w})F_I\rangle$ is $w,\bar{w}$-independent (and similarly for the anti-chiral sector) -- that is, the $w$ or $\bar{w}$ derivative of this correlator must vanish. Therefore, carrying out the contraction using the full propagator, it suffices to check that the integrals:
\begin{equation}\label{eq:intermediate integral}
\begin{aligned}
\partial_{\bar{w}}\int_{\Sigma_{\mathfrak{g}}}d\big(g_I\wedge dG\big)&=\partial_{\bar{w}}\int_{\partial\Sigma_{\mathfrak{g}}}\big(g_I\wedge dG\big)\\
&=\int_{\partial\Sigma_{\mathfrak{g}}}d\bar{z}g_I(z)\partial_{\bar{z}}\partial_{\bar{w}}G(z,w)
\end{aligned}
\end{equation}
vanish, and similar expressions with $\bar{g}_I$ replacing $g_I$, and also $\partial_w$ replacing $\partial_{\bar{w}}$; four integrals in total, but two are related by complex conjugation. Here $dG\equiv dG(z,w)$, $d=dz\partial_z+d\bar{z}\partial_{\bar{z}}$, and $G(z,w)$ is the full propagator on the genus-$\mathfrak{g}$ surface. 
For these boundary integrals one may consider the polygon representation of the cut surface, $\tilde{\Sigma}_{\mathfrak{g}}$, see Fig.~\ref{Fig:CutRiemannSurface}, cut out small discs ($D$) of infinitesimal radius $|\epsilon|$ centred around the pole that comes from $dG(z,w)$ for $z\rightarrow w$, and write the integral over the full cut surface as an integral over $\tilde{\Sigma}_{\mathfrak{g}}-D$ plus an integral over $D$. Both of these can be written as boundary integrals using Stoke's theorem with $\partial\tilde{\Sigma}_{\mathfrak{g}}=\sum_{I=1}^{\mathfrak{g}}(A_I+B_I+A_I^{-1}+B_I^{-1})$, and a careful consideration of each of the terms that arise (along the lines of Mumford \cite{Mumford_v12} or Lugo and Russo \cite{LugoRusso89}, although in the present context the integrand contains both meromorphic {\it and} anti-meromorphic quantities), after various cancellations, leads to the vanishing of the correlator in question,
$$
\Big\langle \int j_M y^M\,F_{I}^N\Big\rangle=0,\qquad {\rm when}\qquad \int d^2zj_M(z,\bar{z})=0.
$$
Therefore, repeating the argument for $\bar{F}_I^N$ as well the naive assumption that we can set $F_I=\bar{F}_I=0$ is, effectively, correct. Note that on $\tilde{\Sigma}_{\mathfrak{g}}-D$, $\partial_z\partial_{\bar{w}}\ln|E(z,w)|^2=0$, whereas on $D$, $\partial_z\partial_{\bar{w}}\ln|E(z,w)|^2=-2\pi\delta^2(z-w)$. We also use the fact that $g_I$ is on $A_J^{-1}$ the same as $g_I$ on $A_J$ plus $\oint_{B_J}\omega_I=\Omega_{IJ}$, and that $g_I$ is on $B_J^{-1}$ the same as $g_I$ on $B_J$ plus $\oint_{A_J^{-1}}\omega_I=-\oint_{A_J}\omega_I=-\delta_{IJ}$, see Fig.~\ref{Fig:CutRiemannSurface}, and similarly for the remaining terms in the integrand.

Having established (\ref{eq:<ejyeFeFbar>}), from (\ref{eq:I(x0+ycl+y)}), (\ref{eq:PhibarPhi}) and (\ref{eq:zeromodes}) we learn that the decomposition $x=x_0+y_{\rm cl}+y$ factorises the action into three distinct pieces, and substituting these into (\ref{eq:A(j)}), leads to:
\begin{equation}\label{eq:A(j)2}
\begin{aligned}
\mathcal{A}_x(j)&=\Psi_{\rm Eucl}^0\Psi_{\rm Eucl}^{\rm cl}\Psi_{\rm Eucl}^{\rm q}\\
&=\Psi_{\rm Eucl}^0\bigg(\sum_{y_{\rm cl}}e^{-I(y_{\rm cl}|j)}\bigg)\bigg(\int \mathcal{D}y\,e^{-I(y|j)}\bigg)\\
&=\Psi_{\rm Eucl}^0\bigg(\sum_{y_{\rm cl}}e^{-I(y_{\rm cl}|j)}\bigg)\,\bigg(\int \dslash^{D\mathfrak{g}}\mathbb{P}_I^{\mu}\int \dslash^{D\mathfrak{g}}\mathbb{W}_I^{\mu}\int \mathcal{D}y\,e^{-I(y|j)}\deltaslash^{D\mathfrak{g}}\big(\mathbb{P}_I^{\mu}-\hat{\mathbb{P}}_I^{\mu}\big)\deltaslash^{D\mathfrak{g}}\big(\mathbb{W}_I^{\mu}-\hat{\mathbb{W}}_I^{\mu}\big)\bigg),
\end{aligned}
\end{equation}
where $\Psi_{\rm Eucl}^{\rm cl}$ and $\Psi_{\rm Eucl}^{\rm q}$ are to be identified with the second and third parentheses respectively in the second or third lines, and we have defined:
\begin{equation}\label{eq:IclIy}
\begin{aligned}
I(y_{\rm cl}|j)&=\gamma_I^a\Big(\frac{1}{\pi\alpha'}{\rm Im}\,\Omega_{IJ}\big(G_{ab}+B_{ab}\big)\Big)\bar{\gamma}_J^b-i\Phi_a^I\gamma_I^a-i\bar{\Phi}^I_a\bar{\gamma}_I^a\\
I(y|j)&=\frac{1}{2\pi\alpha'}\int_{\Sigma_{\mathfrak{g}}} d^2z\,\partial_zy^M\partial_{\bar{z}}y^N\big(G_{MN}+B_{MN}\big)-i\int_{\Sigma_{\mathfrak{g}}} d^2z\, j_My^M,
\end{aligned}
\end{equation}
with the understanding that, according to the above discussion, the $\Phi_a^I$, $\bar{\Phi}_a^I$ in (\ref{eq:PhibarPhi}) reduce to (generalising the definition to generic spacetime components for later convenience):
\begin{equation}\label{eq:PhiPhibar}
\Phi_M^I\dfn \int_{\Sigma_{\mathfrak{g}}} d^2z\,j_M\int^z\!\omega_I\qquad {\rm and} \qquad\bar{\Phi}_M^I\dfn \int_{\Sigma_{\mathfrak{g}}} d^2z\,j_M\int^{\bar{z}}\!\bar{\omega}_I.
\end{equation}

We next evaluate each of the two remaining factors, $\Psi_{\rm Eucl}^{\rm cl}$ and $\Psi_{\rm Eucl}^{\rm q}$, below in the generic case where {\it all} components of the source $j_M$ are potentially non-vanishing (or possibly physical). This latter point will ensure that we can extract amplitudes from $\mathcal{A}_x(j)$ by functional differentiation with respect to the source $j$ associated to S-matrix elements whose asymptotic states have generic Neveu-Schwarz (NS) charges: we allow for vertex operators whose polarisation tensors and momenta potentially span the full space $\mathbb{R}^{D-1,1}\times \mathbb{T}^{D_{\rm cr}-D}$ or any subspace of interest, with generic KK and winding charges.

\ssk
Let us now consider the classical instanton contributions, 
\begin{equation}\label{eq:Psi^cl original}
\Psi_{\rm Eucl}^{\rm cl}\dfn \sum_{\{N^a_I,M^a_I\}\in \mathbb{Z}}\exp \bigg\{-\gamma_I^a\Big(\frac{1}{\pi\alpha'}{\rm Im}\,\Omega_{IJ}\big(G_{ab}+B_{ab}\big)\Big)\bar{\gamma}_J^b+i\Phi_a^I\gamma_I^a+i\bar{\Phi}^I_a\bar{\gamma}_I^a\bigg\},
\end{equation}
with $\gamma_I^a$, $\bar{\gamma}_I^a$ given explicitly in (\ref{eq:gammas}). To make the loop momenta in the compact dimensions manifest it is desirable to perform a Poisson resummation on the integers $\{M_I^a\}$ appearing in $\gamma_I^a$, $\bar{\gamma}_I^a$, the particular identity of interest being (when the matrix $A_{IJ}^{ab}$ is not necessarily diagonal in `$ab$' or `$IJ$'),
$$
\sum_{\{M_I^{a}\}\in\mathbb{Z}}e^{-\pi M_I^{\alpha}A^{IJ}_{ab}M_J^{b}+2\pi B_{a}^IM_I^{a}} = \sum_{\{{M'}^I_{a}\}\in\mathbb{Z}} {\det}^{-1/2}\big({\rm Det} A^{IJ}_{ab}\big)\,e^{-\pi({M'}^I_a-iB^I_a)(A^{-1})_{IJ}^{ab}({M'}^J_b-iB^J_b)},
$$
with `${\rm Det}$' with respect to the `$ab$' indices and `$\det$' with respect to `$IJ$'. (This can be thought of as a first step towards a Hamiltonian formulation of string theory as opposed to a Lagrangian formulation, because this Poissson resummation makes the compact loop momenta manifest, see below.) The invertible ($\mathfrak{g}\times \mathfrak{g}$, real and symmetric) matrices of interest are $A_{ab}^{IJ}=G_{ab}({\rm Im}\Omega)^{-1}_{IJ}$, and the complex vectors $B_{a}^I=(2{\rm Im}\Omega)^{-1}_{IJ}\big[\Omega_{JK}N^b_K(G_{ba}-B_{ba})+\bar{\Omega}_{JK}N^b_K(G_{ba}+B_{ba})+\ell_s\big(\Phi_a^J-\bar{\Phi}_a^J\big)\big]$. Note that 
\begin{equation}\label{eq:Gab-1}
(G_{ab})^{-1}=G^{ab}-G^{a\mu}(G^{\mu\nu})^{-1}G^{\nu b}, 
\end{equation}
which follows from the defining relations $G^{MN}G_{NL}\equiv \delta^M_L$ and $(G_{ab})^{-1}G_{bc}\equiv \delta^{a}_c$. After a certain amount of algebra we learn that:
\begin{equation}\label{eq:InstSum}
\begin{aligned}
&\Psi_{\rm Eucl}^{\rm cl}=\sum_{\{N^a_I,M^{'I}_a\}\in \mathbb{Z}}
({\rm Det}\,G_{ab})^{-\frac{\mathfrak{g}}{2}}\,
\big({\det}\,{\rm Im}\Omega_{IJ}\big)^{\frac{D_{\rm cr}-D}{2}}\\
&\qquad\qquad\times \exp \bigg(i\frac{\pi\alpha'}{2}\mathbb{Q}^{I}_a(G_{ab})^{-1}\Omega_{IJ}\mathbb{Q}_b^J+i\pi\alpha' \mathbb{Q}_a^I(G_{ab})^{-1}\Phi_b^I\bigg)\\
&\qquad\qquad\times\exp\bigg(-i\frac{\pi\alpha'}{2}\bar{\mathbb{Q}}^{I}_a(G_{ab})^{-1}\bar{\Omega}_{IJ}\bar{\mathbb{Q}}_b^J-i\pi\alpha' \bar{\mathbb{Q}}_a^I(G_{ab})^{-1}\bar{\Phi}_b^I\bigg)\\
&\qquad\qquad\times \exp\bigg(i\,\frac{\pi \alpha'}{2}\big(\Phi_a^I-\bar{\Phi}_a^I\big)(G_{ab})^{-1}\big(\Omega_{IJ}-\bar{\Omega}_{IJ}\big)^{-1}\big(\Phi_b^J-\bar{\Phi}_b^J\big)\bigg).
\end{aligned}
\end{equation}
The (non-chirally split) exponent in the last factor in (\ref{eq:InstSum}) is closely related to the zero mode of the multi-loop propagator, see below.\footnote{We often use the convention that under complex conjugation one is to take (in addition to $z\rightarrow \bar{z}$, $\omega_I\rightarrow \bar{\omega}_I$, $\Omega_{IJ}\rightarrow \bar{\Omega}_{IJ}$) $\mathbb{Q}\rightarrow \bar{\mathbb{Q}}$ and $j_a\rightarrow j_a$ ({\it independently} of whether $j_M$ or $\mathbb{Q}$ is real or complex), as this allows us to rewrite the two exponents in the second line of (\ref{eq:InstSum}) as $\big|\exp(\dots)\big|^2$.} 
We have defined the quantities:
\begin{equation}\label{eq:PLPR}
\begin{aligned}
&\mathbb{Q}^{I}_a\dfn \frac{1}{\ell_s}\Big(M^{'I}_a+B_{ab}N^b_I+G_{ab}N^b_I\Big),\qquad \bar{\mathbb{Q}}^{I}_a\dfn \frac{1}{\ell_s}\Big(M^{'I}_a+B_{ab}N^b_I-G_{ab}N^b_I\Big),
\end{aligned}
\end{equation}
where the indices $a,b$ span $\mathbb{T}^{D_{\rm cr}-D}$. Given our discussion leading to (\ref{eq:QQbar hat}), these are interpreted as (anti-)chiral momenta, related to momentum and winding operators as in (\ref{eq:PWQQ}). The correspondence with (\ref{eq:PiMl}) is clear, implying that the first terms in (\ref{eq:PLPR}), namely $M^{'I}_a/\ell_s$, should be identified with eigenvalues of $\hat{\mathbb{\Pi}}_{a,A_I}$, whereas from (\ref{eq:PLPR}), (\ref{eq:periodicity x AB cycles}) and (\ref{eq:P}) we see that $N^a_I/\ell_s$ is precisely the expected $A_I$-cycle eigenvalue of the winding operator, $\hat{W}^a_I$, as always paying careful attention to the location of spacetime indices. (The Poisson resummation maps integers $M^b_I$ to a new dual set of integers ${M'_b}^I$.) 

\ssk
We next consider the fluctuations in (\ref{eq:A(j)2}) (still working in Euclidean target space and worldsheet signature),
\begin{equation}\label{eq:inttildex}
\begin{aligned}
\Psi_{\rm Eucl}^{q}\dfn \int \dslash \mathbb{P}^{\mu}_I \int \dslash \mathbb{W}^{\mu}_I\int \mathcal{D}y\,e^{-I(y|j)}\deltaslash^{D\mathfrak{g}}\big(\mathbb{P}_{I}^{\mu}-\hat{\mathbb{P}}_{I}^{\mu}\big)\deltaslash^{D\mathfrak{g}}\big(\mathbb{W}_I^{\mu}-\hat{\mathbb{W}}_I^{\mu}\big),
\end{aligned}
\end{equation} 
To evaluate (\ref{eq:inttildex}) for arbitrary sources we break the calculation down into 4 steps: 
\begin{itemize}
\item[{\bf (i)}] introduce integral representations for each of the $2D\mathfrak{g}$ delta functions, $\deltaslash(a)=\int d\lambda\,e^{i\lambda a}$; 
\item[{\bf (ii)}] integrate out $y^M$ in resulting expression; 
\item[{\bf (iii)}] evaluate resulting $A_I$-cycle contour integrals, see (\ref{eq:P});
\item[{\bf (iv)}] integrate out the $\lambda$, in resulting expression.
\end{itemize}

{\bf Step (i)}, introducing an integral representation for the delta functions, leads to:
\begin{equation}\label{eq:step(i)}
\begin{aligned}
\int \mathcal{D}y\,e^{-I(y|j)}\deltaslash^{D\mathfrak{g}}\big(\mathbb{P}_{I}^{\mu}-\hat{\mathbb{P}}_{I}^{\mu}\big)\deltaslash^{D\mathfrak{g}}\big(\mathbb{W}_I^{\mu}-\hat{\mathbb{W}}_I^{\mu}\big)=\int d^{D\mathfrak{g}}\lambda_{I\mu} e^{i\lambda_{I\mu} \mathbb{P}_{I}^{\mu}} \int d^{D\mathfrak{g}}\bar{\lambda}_{I\mu}\,e^{i\bar{\lambda}_{I\mu}\mathbb{W}_{I}^{\mu}}\int \mathcal{D}y\,e^{-I(y|B+j)}
\end{aligned}
\end{equation}
where we made use of (\ref{eq:P}) in order to define the quantity $B_M=(B_{\mu},B_a)$ with:
\begin{equation}\label{eq:Bsource}
\begin{aligned}
B_{\mu}(z,\bar{z})&\dfn -\frac{1}{2\pi\alpha'}\big(\lambda_{I\mu}+\bar{\lambda}_{I\mu}\big)\oint_{A_I}du\delta^2(u-z)\partial_z+ \frac{1}{2\pi\alpha'}\big(\lambda_{I\mu}-\bar{\lambda}_{I\mu}\big)\oint_{A_I}d\bar{u}\delta^2(u-z)\partial_{\bar{z}},
\end{aligned}
\end{equation}
and $B_a(z,\bar{z})\dfn0$ when $\mu$ spans $\mathbb{R}^D$ and $a$ spans $\mathbb{T}^{D_{\rm cr}-D}$ respectively, and there is an implicit sum over repeated indices $I$, with $I=1,\dots,\mathfrak{g}$. (Note that $B_a$ vanishes because the delta functions are associated to the {\it non-compact} dimensions; the corresponding momentum and winding in the compact dimensions has already been fixed by the Poisson resummation.)

There are two slightly subtle points that go into deriving the equality (\ref{eq:step(i)}). The first concerns an apparent interchange of the orders of integration: for some generic homology cycle, $\mathcal{C}$, we have been somewhat cavalier in going from the first to the second equality in:
\begin{equation}
\begin{aligned}
\oint_{\mathcal{C}}du\partial_ux&=\oint_{\mathcal{C}}du\Big(\int d^2z\delta^2(z-u)\partial_ux\Big)\\
&=\int d^2z\Big(\oint_{\mathcal{C}}du\delta^2(z-u)\partial_ux\Big),
\end{aligned}
\end{equation}
without discussing the issue of absolute convergence. 
However, the key word in the above statement is the word `apparent', because in evaluating these integrals (after integrating out $x$) it will always be understood that we {\it first} carry out the area integrals and subsequently the contour integrals. The second subtlety is potentially more serious, namely a real interchange in the order of integration: $\int \mathcal{D}y \int d^{Dg}\lambda(\dots)=\int d^{Dg}\lambda\int \mathcal{D}y (\dots)$. This interchange is potentially subtle (given that, e.g., we have not addressed the issue of absolute convergence of the $y$ integral), but will nevertheless proceed in this manner and rest assured on the fact that our result for $\mathcal{A}_x(j)$ will be consistent (in certain limiting cases) with the result of D'Hoker and Phong \cite{D'HokerPhong89} who proceed without introducing such delta function insertions, and so this procedure is not expected to introduce any spurious terms.

\ssk
{\bf Step (ii)}, integrating out $y^M$ in (\ref{eq:step(i)}), is carried out by writing $Q_M\dfn  B_M+j_M$ and making use of the mode expansion depicted above (\ref{eq:zeromodes}). Zeta function regularisation (to show that $\prod_{\alpha>0}c=\frac{1}{\sqrt{c}}$) is used to rewrite the measure as follows, $\mathcal{D}y={\rm Det}^{-1/2}G_{MN}\big(\prod_{\alpha\neq0}d^{D_{\rm cr}}A_{\alpha}\big)$, and a standard calculation \cite{Polchinski_v1} then leads to:
\begin{equation}\label{eq:xtilde}
\begin{aligned}
\int\mathcal{D}y\,e^{-I(y|Q)}= \,\big(4\pi^2\alpha'{\det}'\Delta_{(0)}\big)^{-D_{\rm cr}/2}e^{-\frac{1}{2}\int d^2z\int d^2z' Q_M(z,\bar{z}) G^{MN}Q_N(z',\bar{z}')G(z,z')}.
\end{aligned}
\end{equation}
Note that the determinant, ${\rm Det}^{-1/2}G_{MN}$, crucially, has cancelled out of (\ref{eq:xtilde}). 
Here $\Delta_{(0)}=-\frac{1}{\sqrt{g}}\partial_{\alpha}\sqrt{g}g^{\alpha\beta}\partial_{\beta}$ is the standard Laplacian (which in the $z,\bar{z}$ coordinates reduces to $-2g^{z\bar{z}}\partial_z\partial_{\bar{z}}$), the prime on the determinant indicates that $\Delta_{(0)}$ acts in the space orthogonal to zero modes. The (multi-loop) propagator, $G(z,w)\dfn \sum_{\alpha\neq0}\frac{2\pi\alpha'}{\omega_{\alpha}^2}\phi_{\alpha}(z,\bar{z})\phi_{\alpha}(w,\bar{w})$, and satisfies $\Delta_{(0)} G(z,w)=\frac{2\pi\alpha'}{\sqrt{g}}\delta^2(z-w)-\frac{2\pi\alpha'}{\int_{\Sigma_{\mathfrak{g}}}d^2z\sqrt{g}}$, with the completeness relation $\frac{1}{\sqrt{g}}\delta^2(z-w)=\sum_{\alpha\in \mathbb{Z}}\phi_{\alpha}(z,\bar{z})\phi_{\alpha}(w,\bar{w})$; see also Appendix \ref{sec:GF}. For example, for compact genus-$\mathfrak{g}$ Riemann surfaces (the case of interest in the current document) we can take \cite{VerlindeVerlinde87},
\begin{equation}\label{eq: Green Function+regular*m}
G(z,z') = -\frac{\alpha'}{2}\ln \left|E(z,z')\right|^2 + \pi\alpha'\, {\rm Im}\int\limits_{z}^{z'}\omega_I\left({\rm Im}\Omega\right)_{IJ}^{-1}{\rm Im}\int\limits_{z}^{z'}\omega_J.
\end{equation}
The zero mode piece appearing here is the main obstruction to chiral splitting in amplitudes. Fixing the loop momenta \cite{D'HokerPhong89} removes these zero mode contributions from correlation functions, thus significantly facilitating amplitude computations as we shall see, especially in the context of highly excited strings.

\ssk
{\bf Step (iii)}, the evaluation of the $A_I$-cycle contour integrals is carried out by noting that in the exponent in (\ref{eq:xtilde}) there arise the terms: 
$
e^{-\frac{1}{2}\int \int Q_MG^{MN}Q'_NG}
$ 
which when expanded (on account of $B_a=0$) read 
$e^{-\frac{1}{2}\int\int B_{\mu}G^{\mu\nu}B'_{\nu}G}e^{-\int\int (j_{\mu}G^{\mu\nu}+j_aG^{a\nu})B'_{\nu}G}e^{-\frac{1}{2}\int \int j_MG^{MN}j'_NG},
$ 
and from the definition of $B_{\mu}$ therefore, we must interpret the quantities $\oint_{A_I} dz\partial_zG(z,z')$,  and $\oint_{A_I}d\bar{z}\partial_{\bar{z}}G(z,z')$, and various related combinations -- these are the contour integrals referred to in item {\bf (iii)} above. 
According to (\ref{eq:quasiperiodprimeform}), the prime form is periodic around the $A_I$-cycles, 
$$
\oint_{A_I}dz\,\partial_z\ln E(z,w)=0,
$$
and therefore the sole contribution will come from  zero modes, see (\ref{eq: Green Function+regular*}). (If we had instead fixed the $B$-cycle momenta, or a linear combination of $A$- and $B$-cycle momenta, the non-zero mode components would also contribute, and we would have to add additional pieces to the propagator.\footnote{DPS thanks Eric D'Hoker for a discussion on this point.}) 
Analytically continuing $z,\bar{z}$ to independent variables, a short calculation\footnote{Here we write ${\rm Im}\int_z^w\omega_I=\frac{1}{2i}(\int_z^w\omega_I-\int_{\bar{z}}^{\bar{w}}\bar{\omega}_I)$ and then express $\int_z^{\wp+A_I}\omega_J$ as $\int_z^{\wp}\omega_J+\oint_{A_I}\omega_J$ for some reference point $\wp\in \Sigma_{\mathfrak{g}}$, and similarly for the antiholomorphic sector.} using the defining relations for the holomorphic differentials, $\oint_{A_I}\omega_J = \delta_{IJ}$ and $\oint_{A_I}\bar{\omega}_J = \delta_{IJ}$, yields,
\begin{equation}\label{eq:oint_AI dG1}
\begin{aligned}
&\frac{1}{\pi\alpha'}\oint_{A_I}dz\partial_zG(z,z')=-\frac{1}{4}({\rm Im}\Omega)_{II}^{-1}+i({\rm Im}\Omega)_{IJ}^{-1}{\rm Im}\int_{\wp}^{z'}\omega_J\\
&\frac{1}{\pi\alpha'}\oint_{A_I}d\bar{z}\partial_{\bar{z}}G(z,z')=-\frac{1}{4}({\rm Im}\Omega)_{II}^{-1}-i({\rm Im}\Omega)_{IJ}^{-1}{\rm Im}\int_{\wp}^{z'}\omega_J,
\end{aligned}
\end{equation}
with $\wp\in\Sigma_{\mathfrak{g}}$ a reference point on which amplitudes do not depend. This step is somewhat naive (due to the continuation of $z,\bar{z}$ to independent variables), but it gives the correct answer. One of the relevant integrals for the second (for $M={\mu}$) and third (for $M=a$) terms in the above exponential is then,
\begin{equation}\label{eq:int jBG compact1}
\begin{aligned}
-\int d^2z&\int d^2z' \Big(j_{\mu}(z,\bar{z})G^{\mu\nu}+j_a(z,\bar{z})G^{a\nu}\Big)B_{\nu}(z',\bar{z}')G(z,z')\\
&\qquad =i\lambda_{I\mu}({\rm Im}\Omega)_{IJ}^{-1}\int d^2z\Big(G^{\mu\nu}j_{\nu}(z,\bar{z})+G^{\mu a}j_a(z,\bar{z})\Big){\rm Im}\int_p^{z}\omega_J+\dots,
\end{aligned}
\end{equation}
where the dots denote the contribution coming from the terms $-\frac{1}{4}({\rm Im}\Omega)_{II}$ in (\ref{eq:oint_AI dG1}), which do not contribute because $\int d^2zj_{M}(z,\bar{z})=0$, see (\ref{eq:zeromodes}). Notice we are not assuming the cross terms, $G_{\mu a}$, vanish. Furthermore, the difference in sign in the chiral and anti-chiral halves in the second terms of the right-hand sides on (\ref{eq:oint_AI dG1}) is crucial: it implies that the $\bar{\lambda}$-dependent terms in (\ref{eq:Bsource}) precisely cancel out (on account of the constraint $\int d^2zj_M(z,\bar{z})=0$). 

Finally, there is an integral that is quadratic in $B_{\mu}(z,\bar{z})$. This is equivalent to (\ref{eq:int jBG compact1}) (up to a factor of two) but with $B_M(z,\bar{z})$ replacing $j_M(z,\bar{z})$,
\begin{equation}
\begin{aligned}
-\frac{1}{2}\int d^2z\int d^2z'  &B_{\mu}(z,\bar{z})G^{\mu\nu} B_{\nu}(z',\bar{z}')G(z,z')=-\frac{1}{2}\lambda_{I\mu}\Big(\frac{G^{\mu\nu}}{2\pi\alpha'}\big({\rm Im}\Omega\big)_{IJ}^{-1}\Big)\lambda_{J\nu}.
\end{aligned}
\end{equation}
Also here the $\bar{\lambda}$ terms cancel out and do not appear on the right-hand side, here because of the equality $\oint_{A_I}\omega_J=\oint_{A_I}\bar{\omega}_J$.  Gathering the above results, making explicit use of the propagator (\ref{eq: Green Function+regular*m}) and writing the result in terms of $\Phi_a^I,\bar{\Phi}_a^I$ as defined in (\ref{eq:PhiPhibar}), we learn that the exponent in (\ref{eq:xtilde}), namely $e^{-\frac{1}{2}\int d^2z\int d^2z' Q_M(z,\bar{z}) G^{MN}Q_N(z',\bar{z}')G(z,z')}$, is precisely equal to:
\begin{equation}
\begin{aligned}
&e^{\frac{\alpha'}{4}\int d^2z\int d^2z' j_M(z,\bar{z}) G^{MN}j_N(z',\bar{z}')\ln|E(z,z')|^2}e^{-i\frac{\pi \alpha'}{2}(\Phi_M^I-\bar{\Phi}_M^I)G^{MN}(\Omega_{IJ}-\bar{\Omega}_{IJ})^{-1}(\Phi_N^J-\bar{\Phi}_N^J)}\\
&\qquad\times e^{-\frac{1}{2}\lambda_{I\mu}\big(\frac{G^{\mu\nu}}{2\pi\alpha'}({\rm Im}\Omega)_{IJ}^{-1}\big)\lambda_{J\nu}}e^{i\lambda_{I\mu}\big[(\Omega_{IJ}-\bar{\Omega}_{IJ})^{-1}G^{\mu M}(\Phi_M^J-\bar{\Phi}_M^J)\big]}
\end{aligned}
\end{equation}
Substituting this into (\ref{eq:xtilde}), which is in turn substituted back into (\ref{eq:step(i)}), we obtain the following expression for the delta function expectation values in the presence of a source:
\begin{equation}\label{eq:step(iii)}
\begin{aligned}
\int &\mathcal{D}y\,e^{-I(y|j)}\deltaslash^{D\mathfrak{g}}\big(\mathbb{P}_{I}^{\mu}-\hat{\mathbb{P}}_{I}^{\mu}\big)\deltaslash^{D\mathfrak{g}}\big(\mathbb{W}_I^{\mu}-\hat{\mathbb{W}}_I^{\mu}\big)=\\
&=\big(4\pi^2\alpha'{\det}'\Delta_{(0)}\big)^{-D_{\rm cr}/2}\,e^{\frac{\alpha'}{4}\int \!\!\int j_MG^{MN}j'_N\ln |E(z,z')|^2}e^{-i\frac{\pi \alpha'}{2}(\Phi_M^I-\bar{\Phi}_M^I)G^{MN}(\Omega_{IJ}-\bar{\Omega}_{IJ})^{-1}(\Phi_N^J-\bar{\Phi}_N^J)}\\
&\quad\times \int d^{D\mathfrak{g}}\bar{\lambda}_{I\mu}\,e^{i\bar{\lambda}_{I\mu}\mathbb{W}_{I}^{\mu}}\int d^{D\mathfrak{g}}\lambda_{I\mu} \,e^{-\frac{1}{2}\lambda_{I\mu}\big(\frac{G^{\mu\nu}}{2\pi\alpha'}({\rm Im}\Omega)_{IJ}^{-1}\big)\lambda_{J\nu}}e^{i\lambda_{I\mu}\big[\mathbb{P}^{\mu}_I+(\Omega_{IJ}-\bar{\Omega}_{IJ})^{-1}G^{\mu M}(\Phi_M^J-\bar{\Phi}_M^J)\big]}.
\end{aligned}
\end{equation}

\ssk
{\bf Step (iv)} of the computation is to carry out the remaining integrations over the $\lambda_{I\mu}$, $\bar{\lambda}_{I\mu}$ in (\ref{eq:step(iii)}).\footnote{Defining $\mathcal{A}_{IJ}^{\mu\nu}\dfn\frac{G^{\mu\nu}}{2\pi\alpha'}\big({\rm Im}\Omega\big)_{IJ}^{-1}$ and $G_{I}^{\mu} \dfn\mathbb{P}^{\mu}_I+(\Omega_{IJ}-\bar{\Omega}_{IJ})^{-1}G^{\mu M}(\Phi_M^J-\bar{\Phi}_M^J)$, the following integral is required:
\begin{equation}\label{eq:V1...Vnfixedloop}
\begin{aligned}
\Big(\prod_{I,\mu}\int_{-\infty}^{\infty}&d\lambda_{I\mu}\Big)\exp\Big\{-\frac{1}{2}\lambda_{I\mu} \mathcal{A}_{IJ}^{\mu\nu}\lambda_{I\nu}+iG_{I}^{\mu} \lambda_{I\mu}\Big\}={\det}^{-1/2}\big({\rm Det}\tfrac{\mathcal{A}_{IJ}^{\mu\nu}}{2\pi}\big)\exp\bigg\{-\frac{1}{2}G_{I}^{\mu}(\mathcal{A}^{-1})_{\mu\nu}^{IJ}G_{J}^{\nu}\bigg\}.
\end{aligned}
\end{equation}
} After some trivial rearrangement,
\begin{equation}\label{eq:step(iv)}
\begin{aligned}
\int \mathcal{D}y\,&e^{-I(y|j)}\deltaslash^{D\mathfrak{g}}\big(\mathbb{P}_{I}^{\mu}-\hat{\mathbb{P}}_{I}^{\mu}\big)\deltaslash^{D\mathfrak{g}}\big(\mathbb{W}_I^{\mu}-\hat{\mathbb{W}}_I^{\mu}\big)=\\
&=(2\pi \ell_s)^{D\mathfrak{g}-D_{\rm cr}}\big({\det}'\Delta_{(0)}\big)^{-D_{\rm cr}/2}{\rm Det}^{-\mathfrak{g}/2}G^{\mu\nu}\,{\det}^{D/2}({\rm Im}\Omega_{IJ}) \deltaslash^{D\mathfrak{g}}\big(\mathbb{W}_I^{\mu}\big)\\
&\quad\times e^{\frac{\alpha'}{4}\int \!\!\int j_MG^{MN}j'_N\ln E(z,z')}e^{i\frac{\pi\alpha'}{2}\mathbb{P}_{I}^{\mu}(G^{\mu\nu})^{-1}\Omega_{IJ}\mathbb{P}_{J}^{\nu}+i\pi\alpha' \mathbb{P}_{I}^{\mu}(G^{\mu\nu})^{-1}G^{\nu N}\Phi^I_{N}}\\
&\quad\times e^{\frac{\alpha'}{4}\int \!\!\int j_MG^{MN}j'_N\ln \bar{E}(\bar{z},\bar{z}')}e^{-i\frac{\pi\alpha'}{2}\mathbb{P}_{I}^{\mu}(G^{\mu\nu})^{-1}\bar{\Omega}_{IJ}\mathbb{P}_{J}^{\nu}-i\pi\alpha' \mathbb{P}_{I}^{\mu}(G^{\mu\nu})^{-1}G^{\nu N}\bar{\Phi}^I_{N}}\\
&\quad\times e^{-i\frac{\pi \alpha'}{2}(\Phi_M^I-\bar{\Phi}_M^I)\big[G^{MN}-G^{M\mu}(G^{\mu\nu})^{-1}G^{\nu N}\big](\Omega_{IJ}-\bar{\Omega}_{IJ})^{-1}(\Phi_N^J-\bar{\Phi}_N^J)}
\end{aligned}
\end{equation}
where we note that the $\bar{\lambda}$ integrals lead to the delta function constraint $\deltaslash^{D\mathfrak{g}}\big(\mathbb{W}_I^{\mu}\big)$, which on account of (\ref{eq:PWQQ}) allows us to identify $\mathbb{P}_{I}^{\mu}$ with $\mathbb{Q}_{I}^{\mu}$ or $\bar{\mathbb{Q}}_{I}^{\mu}$, given that in the absence of winding all these are equivalent. 

The second and third lines of the RHS in (\ref{eq:step(iv)}) are chirally split, whereas the last line is not. (The non-chirally split terms in the first line will cancel when ghost contributions are included.) The term in the last line will ultimately cancel a similar quantity that arose from the instanton contribution, $\Psi_{\rm Eucl}^{\rm cl}$, see the last line in (\ref{eq:InstSum}), but to make this cancellation manifest let us consider the quantity:
$$
G^{MN}-G^{M\mu}(G^{\mu\nu})^{-1}G^{\nu N}
$$
in (\ref{eq:step(iv)}). When $M$ and/or $N$ span $\mathbb{R}^D$ this quantity vanishes, given that by definition $(G^{\mu\nu})^{-1}G^{\nu \rho}\equiv \delta^{\rho}_{\mu}$. Therefore, only if {\it both} $M$ and $N$ span $\mathbb{T}^{D_{\rm cr}-D}$ will the non-chirally split term in the last line of (\ref{eq:step(iv)}) contribute. That is,
\begin{equation}\label{eq:non-chirallysplitpiece}
\begin{aligned}
&e^{-i\frac{\pi \alpha'}{2}(\Phi_M^I-\bar{\Phi}_M^I)\big[G^{MN}-G^{M\mu}(G^{\mu\nu})^{-1}G^{\nu N}\big](\Omega_{IJ}-\bar{\Omega}_{IJ})^{-1}(\Phi_N^J-\bar{\Phi}_N^J)}\\
&\qquad=e^{-i\frac{\pi \alpha'}{2}(\Phi_a^I-\bar{\Phi}_a^I)\big[G^{ab}-G^{a\mu}(G^{\mu\nu})^{-1}G^{\nu b}\big](\Omega_{IJ}-\bar{\Omega}_{IJ})^{-1}(\Phi_b^J-\bar{\Phi}_b^J)}.
\end{aligned}
\end{equation}
But according to (\ref{eq:Gab-1}) the quantity in the brackets, $G^{ab}-G^{a\mu}(G^{\mu\nu})^{-1}G^{\nu b}$, is precisely $(G_{ab})^{-1}$, and so the non-chirally split factor (\ref{eq:non-chirallysplitpiece}) is also equal to:
$$
e^{-i\frac{\pi \alpha'}{2}(\Phi_a^I-\bar{\Phi}_a^I)(G_{ab})^{-1}(\Omega_{IJ}-\bar{\Omega}_{IJ})^{-1}(\Phi_b^J-\bar{\Phi}_b^J)}.
$$
This exponent is (up to a crucial minus sign) identical to that in the last line of (\ref{eq:InstSum}), implying that in the product $\Psi_{\rm Eucl}^{\rm cl}\Psi_{\rm Eucl}^{\rm q}$ the non-chirally split exponential will cancel out of the full generating function, $\mathcal{A}_x^{\rm Eucl}(j)=\Psi_{\rm Eucl}^{0}\Psi_{\rm Eucl}^{\rm cl}\Psi_{\rm Eucl}^{\rm q}$. This generalises a similar observation by D'Hoker and Phong \cite{D'HokerPhong89} (in the context of a $\mathbb{R}^{D_{\rm cr}}$ target spacetime with $G_{MN}=\delta_{MN}$ and $B_{MN}=0$ string backgrounds) to completely generic constant string backgrounds $G_{MN},B_{MN}$ and $\Phi$ in $\mathbb{R}^D\times \mathbb{T}^{D_{\rm cr}-D}$. 

\ssk
The full result for the quantum fluctuations reads, on account of (\ref{eq:step(iv)}) and (\ref{eq:inttildex}) and the above discussion,
\begin{equation}\label{eq:Psi^q_Eucl}
\begin{aligned}
&\Psi_{\rm Eucl}^{\rm q}
=(2\pi \ell_s)^{D\mathfrak{g}-D_{\rm cr}}\big({\det}'\Delta_{(0)}\big)^{-D_{\rm cr}/2}{\rm Det}^{-\mathfrak{g}/2}G^{\mu\nu}\,{\det}^{D/2}({\rm Im}\Omega_{IJ}) \int \dslash^{D\mathfrak{g}} \mathbb{W}^{\mu_I}\deltaslash^{D\mathfrak{g}}\big(\mathbb{W}_I^{\mu}\big)\int \dslash^{D\mathfrak{g}} \mathbb{P}^{\mu_I}\\
&\,\,\times \exp\bigg(\frac{\alpha'}{4}\int \!\!\int j_MG^{MN}j'_N\ln E(z,z')+i\frac{\pi\alpha'}{2}\mathbb{P}_{I}^{\mu}(G^{\mu\nu})^{-1}\Omega_{IJ}\mathbb{P}_{J}^{\nu}+i\pi\alpha' \mathbb{P}_{I}^{\mu}(G^{\mu\nu})^{-1}G^{\nu N}\Phi^I_{N}\bigg)\\
&\,\,\times \exp\bigg(\frac{\alpha'}{4}\int \!\!\int j_MG^{MN}j'_N\ln \bar{E}(\bar{z},\bar{z}')-i\frac{\pi\alpha'}{2}\mathbb{P}_{I}^{\mu}(G^{\mu\nu})^{-1}\bar{\Omega}_{IJ}\mathbb{P}_{J}^{\nu}-i\pi\alpha' \mathbb{P}_{I}^{\mu}(G^{\mu\nu})^{-1}G^{\nu N}\bar{\Phi}^I_{N}\bigg)\\
&\quad\qquad\qquad\qquad\times \exp\bigg(-i\frac{\pi \alpha'}{2}(\Phi_a^I-\bar{\Phi}_a^I)(G_{ab})^{-1}(\Omega_{IJ}-\bar{\Omega}_{IJ})^{-1}(\Phi_b^J-\bar{\Phi}_b^J)\bigg)
\end{aligned}
\end{equation}

Let us now gather all the results for the various terms appearing in (\ref{eq:A(j)2}), starting from the chirally split exponentials in $\Psi_{\rm Eucl}^{\rm q}$ in (\ref{eq:Psi^q_Eucl}) and $\Psi_{\rm Eucl}^{\rm cl}$ in (\ref{eq:InstSum}). It is straightforward to show (using $G^{MN}G_{NL}=\delta^M_L$, $(G_{ab})^{-1}G_{bc}\equiv \delta^a_c$ and $(G^{\mu\nu})^{-1}G^{\nu\sigma}\equiv \delta_{\mu}^{\sigma}$, always raising and lowering indices with $G_{MN}$, and taking into account the delta function constraint, $\delta^{D\mathfrak{g}}(\mathbb{W}^{\mu}_I)$, in $\Psi_{\rm Eucl}^{\rm q}$ which enforces $\mathbb{P}^{\mu}=\mathbb{Q}^{\mu}=\bar{\mathbb{Q}}^{\mu}$) that: 
\begin{equation}\label{eq:QGQ QGPhi}
\begin{aligned}
\mathbb{Q}_M^IG^{MN}\Omega_{IJ}\mathbb{Q}_N^J&=\mathbb{Q}_a^I(G_{ab})^{-1}\Omega_{IJ}\mathbb{Q}_b^J+\mathbb{P}^{\mu}_I(G^{\mu\nu})^{-1}\Omega_{IJ}\mathbb{P}^{\nu}_J\\
\mathbb{Q}_M^IG^{MN}\Phi_N^I&=\mathbb{Q}_a^I(G_{ab})^{-1}\Phi_b^I+\mathbb{P}^{\mu}_I(G^{\mu\nu})^{-1}G^{\nu N}\Phi_N^I,
\end{aligned}
\end{equation}
with similar relations for the anti-chiral sector with the replacements $(\mathbb{Q},\Omega,\Phi)\rightarrow (\bar{\mathbb{Q}},\bar{\Omega},\bar{\Phi})$. Taking (\ref{eq:QGQ QGPhi}) into account, the full result for the matter contribution to the generating function, $\mathcal{A}_x^{\rm Eucl}(j)=\Psi_{\rm Eucl}^0\Psi_{\rm Eucl}^{\rm cl}\Psi_{\rm Eucl}^{\rm q}$, from (\ref{eq:zeromodes}), (\ref{eq:InstSum}) and (\ref{eq:Psi^q_Eucl}) reads:
\begin{equation}\label{eq:A(j)3}
\begin{aligned}
&\mathcal{A}_x^{\rm Eucl}(j)=\delta^D(\ell_s\smallint j^{\mu})\,\delta^{D_{\rm cr}-D}_{(\ell_s\smallint j_a),0}\,\,\bigg(g_s\big({\rm Det}G^{\mu\nu}{\rm Det}G_{ab}\big)^{-\frac{1}{4}}\bigg)^{2\mathfrak{g}-2}\bigg(\frac{{\det}'\Delta_{(0)}}{{\det}({\rm Im}\Omega_{IJ})\int_{\Sigma_{\mathfrak{g}}}d^2z\sqrt{g}}\bigg)^{-\frac{D_{\rm cr}}{2}}\\
&\,\,\,\times \sumint\limits_{(\mathbb{Q},\bar{\mathbb{Q}})}\bigg|\exp\bigg(\frac{\alpha'}{4}\int\! \!\!\int j_MG^{MN}j'_N\ln E(z,z')+i\frac{\pi\alpha'}{2}\mathbb{Q}_{I}^{M}G_{MN}\Omega_{IJ}\mathbb{Q}_{J}^{N}+i\pi\alpha' \mathbb{Q}_{I}^{M}\int\,\!j_M\int^z\!\!\omega_I\bigg)\bigg|^2\\
\end{aligned}
\end{equation}
where we used (\ref{eq:PhiPhibar}) and took into account that $\delta^D(\ell_s\smallint j_{\mu})\frac{1}{{\rm Det}G^{\mu\nu}}=\delta^D(\ell_s\smallint j^{\mu})$ (given that although we raise indices with the full metric, $G^{MN}$, we also have $\int j_a=0$ implying that effectively $G^{\mu M}\int j_M=G^{\mu \nu}\int j_{\nu}$). 
The (dimensionless) sum/integral over $(\mathbb{Q},\bar{\mathbb{Q}})$ should be understood as an integral over non-compact momenta, $\mathbb{P}_{I}^{\mu}$ (with $\mathbb{Q}_{I}^{\mu}=\bar{\mathbb{Q}}_{I}^{\mu}$), and a sum over compact momenta, $\mathbb{Q}_{Ia},\bar{\mathbb{Q}}_{Ia}$, defined in (\ref{eq:PLPR}),
\begin{equation}\label{eq:sumint QQbar}
\sumint\limits_{(\mathbb{Q},\bar{\mathbb{Q}})}\dfn \ell_s^{D\,\mathfrak{g}}\int d^{D\mathfrak{g}}\mathbb{Q}_{I}^{\mu}\!\!\!\!\!\sum_{\{N^{a}_I,M^{'I}_{a}\}\in \mathbb{Z}}=\ell_s^{D\,\mathfrak{g}}\int d^{D\mathfrak{g}}\mathbb{P}_{I}^{\mu}d^{D\mathfrak{g}}\mathbb{W}^{\mu}_I\delta^{D\mathfrak{g}}(\mathbb{W}^{\mu}_I)\!\!\!\!\!\sum_{\{N^{a}_I,M^{'I}_{a}\}\in \mathbb{Z}}
\end{equation}

Clearly, from (\ref{eq:A(j)3}) we see that the natural expansion parameter at fixed-loop momenta is:
\begin{equation}\label{eq:geff_Eucl}
g_{\rm eff}\dfn g_s\Big({\rm Det}G^{\mu\nu}{\rm Det}G_{ab}\Big)^{-\frac{1}{4}}.
\end{equation}
The $G_{ab}$ and $g_s$ dependence of $g_{\rm eff}$ is as expected, since this combination has precisely the form required in order for $g_{\rm eff}$ to be invariant under T-duality, more about which later. The $G^{\mu\nu}$ dependence is novel and deserves further elaboration; we elaborate on this below. 
Note that $g_{\rm eff}$ is also precisely the dimensionless version of the coupling $g_D$ that appears in vertex operators, with $g_D=g_{\rm eff}\ell_s^{\frac{D}{2}-1}$, the metric dependence being dictated by the fact that vertex operators are composed of (possibly linear superpositions of) {\it momentum} eigenstates with fixed KK and winding charges and momenta. 

\ssk 
It is natural and convenient when considering functional derivatives of $\mathcal{A}_x^{\rm Eucl}(j)$ to complete the square in the exponent in (\ref{eq:A(j)3}), and so we reach the main expression for the matter contribution to the (dimensionless) generating function of generic fixed-loop momenta amplitudes in target spacetimes $\mathbb{R}^D\times \mathbb{T}^{D_{\rm cr}-D}$:
\begin{equation}\label{eq:A(j)4}
\begin{aligned}
\mathcal{A}_x^{\rm Eucl}(j)
&=\delta^D(\ell_s\smallint j^{\mu})\delta_{(\ell_s\smallint j_a),0}^{D_{\rm cr}-D}\,\,
g_{\rm eff}^{2\mathfrak{g}-2}
\bigg(\frac{{\det}'\Delta_{(0)}}{{\det}({\rm Im}\Omega_{IJ})\int_{\Sigma_{\mathfrak{g}}}d^2z\sqrt{g}}\bigg)^{-D_{\rm cr}/2}\\
&\quad\times\sumint\limits_{(\mathbb{Q},\bar{\mathbb{Q}})}\bigg|\exp \bigg(\frac{\ell_s^2}{4}\int d^2z\int d^2z'\big(j_M+H_M\big)G^{MN}\big(j'_N+H'_N\big)\ln E(z,z')\bigg)\bigg|^2,
\end{aligned}
\end{equation}
and we have defined:
\begin{equation}\label{eq:HHbar}
H_M(z,\bar{z})\dfn \mathbb{Q}_{M}^I\oint_{B_I}dw\delta^2(w-z)\partial_z,
\qquad
\bar{H}_M(z,\bar{z})\dfn \bar{\mathbb{Q}}_{M}^I\oint_{B_I}d\bar{w}\delta^2(w-z)\partial_{\bar{z}},
\end{equation}
with an implicit sum over $I=1,\dots,\mathfrak{g}$, and the spacetime indices $M,N$ span the full target space $\mathbb{R}^{D}\times \mathbb{T}^{D_{\rm cr}-D}$. This is the analogue of the classic textbook formula of Polchinski (equation (6.2.6) in \cite{Polchinski_v1}), generalised here to target spacetimes $\mathbb{R}^{D}\times \mathbb{T}^{D_{\rm cr}-D}$ with generic K\"ahler and complex structure moduli (generic constant background fields $G_{MN}$, $B_{MN}$ and $\Phi$), fixed-loop momenta, and genus-$\mathfrak{g}>0$ worldsheets. By convention, complex conjugation in (\ref{eq:A(j)4}) takes $(\mathbb{Q}^I_{M},\Omega_{IJ},j_{M},z)$ to $(\bar{\mathbb{Q}}^I_{M},\bar{\Omega}_{IJ},j_{M},\bar{z})$ ({\it independently} of whether the $\mathbb{Q}_{M}^I$ and $j_M$ are real or complex), and in the compact and non-compact dimensions we have, respectively, (\ref{eq:PLPR}) and $\mathbb{P}^{\mu}_I=\mathbb{Q}_I^{\mu}=\bar{\mathbb{Q}}^{\mu}_I$.\footnote{\label{foot:ext source}The result (\ref{eq:A(j)4}) is consistent with the chiral splitting theorem of D'Hoker and Phong \cite{D'HokerPhong89} and reproduces the tachyon $n$-point amplitude of \cite{DijkgraafVerlindeVerlinde88} in $D_{\rm cr}=26$  when:
$
j(z,\bar{z}) = \sum_{i}\big(k_{{\rm L},i}\int^{z_i}_{\wp}du\delta^2(u-z)\partial_z+k_{{\rm R},i}\int^{\bar{z}_i}_{\bar{\wp}}d\bar{u}\delta^2(u-z)\partial_{\bar{z}}\big),
$
for vertex operators with total momentum $k_{i}=\frac{1}{2}(k_{\rm L}+k_{\rm R})_i$ and winding $w_i=\frac{1}{2}(k_{\rm L}-k_{\rm R})_i$, and $({\wp},\bar{\wp})$ a universal generic point on the worldsheet on which amplitudes do not depend due to momentum conservation, $\sum_ik_i=0$ that in turn arises from the delta-function constraint. When the external states have zero winding, $w_i=0$, with $w=\frac{1}{2}(k_{\rm L}-k_{\rm R})$, the source reduces to 
$j(z,\bar{z})=\sum_{i}\delta^2(z_{i}-z)k_{i}.$\label{foot:j}
}

To see that (\ref{eq:A(j)4}) indeed follows from (\ref{eq:A(j)3}), the quasi-periodicity property of the prime form around $B_I$ cycles is useful:
\begin{equation}\label{eq:BIntPrimeForm}
\oint_{B_I}dw\partial_w\ln E(w,z) = 2\pi i\int^z_{\wp}\omega_I+\dots
\end{equation}
where the `\dots' denote terms that drop out of the amplitude due to the constraint $\int j_M=0$. Similarly, amplitudes do not depend on the lower limit, $\wp$, of the integral on the right-hand side. Note also that the amplitude (\ref{eq:A(j)4}) is symmetric under $H\rightarrow -H$, and that the various factors of $2\pi$ in (\ref{eq:zeromodesL}) and (\ref{eq:step(iv)}) have cancelled out of the final result. It is apparently natural to write the result (for the {\it fixed-loop} momenta generating function) in terms of the string length:
$$
\ell_s\dfn \sqrt{\alpha'}
$$ 
in (\ref{eq:A(j)4}). 

\ssk
To complete the story we now include the ghost contribution, $\mathcal{A}_{\rm gh}$, to extract the {\it full} generating function, $\mathcal{A}_{\rm Eucl}(j)=\mathcal{A}_{\rm gh}\mathcal{A}_x^{\rm Eucl}(j)$. At this point the ghost insertions can be completely general, but we restrict here to the minimal number of ghost insertions that lead to a non-vanishing result,
\begin{equation}\label{eq:ghost}
\begin{aligned}
\mathcal{A}_{\rm gh}&=\int \mathcal{D}(b,\tilde{b},c,\tilde{c})\!\!\prod_{j=1}^{\#_{\mathbb{C}}{\rm moduli}}\!\!\!\!\!|\langle \mu_j,b\rangle|^2\,\prod_{s=1}^{\#_{\mathbb{C}}{\rm CKVs}}\!\!\!|c(w_s)|^2\,\,e^{-I_{\rm gh}},
\end{aligned}
\end{equation}
with,
\begin{equation}\label{eq:GhostAction}
I_{\rm gh}= \frac{1}{2\pi} \int_{\Sigma_{\mathfrak{g}}}d^2z\sqrt{g} \big(b\nabla^z_{(-1)}c+\tilde{b}\nabla^{\bar{z}}_{(1)}\tilde{c}\big).
\end{equation}
$\mathcal{A}_{\rm gh}$ has been well-studied for arbitrary genus \cite{VerlindeVerlinde87,DHokerPhong} and we have nothing new to add here, so we will be brief. Suffice it to say that the operator-product expansions (OPE's) (\ref{eq:OPE'sn}) imply $\mathcal{A}_{\rm gh}$ has various obvious zeros and poles due to the explicit $b,c$ insertions. In addition, viewed as a function of, say, $w_1$, it has $\mathfrak{g}$ additional zeros that are determined uniquely by the Jacobi inversion theorem \cite{VerlindeVerlinde87}, while the related Riemann vanishing theorem \cite{DHokerPhong} further ensures that $\mathcal{A}_{\rm gh}$ can be expressed entirely in terms of Riemann theta functions and related quantities, allowing it to be evaluated explicitly \cite{BelavinKnizhnik86,VerlindeVerlinde87,DHokerPhong}; see also \cite{DHoker} for a pedagogical account. What we will make use of here is the following generic result (when $\mathfrak{g}>0$):\footnote{We are neglecting the contribution of the Liouville factor that will always cancel in the final answer for the full amplitude in the critical dimension.}
\begin{equation}\label{eq:ghostb}
\begin{aligned}
\mathcal{A}_{\rm gh}
&= \bigg(\frac{{\det}'\Delta_{(0)}}{{\det}\,{\rm Im}\Omega_{IJ}\int_{\Sigma_{\mathfrak{g}}}d^2z\sqrt{g}}\bigg)^{13}|\mathcal{Z}_{\mathfrak{g}}|^2.
\end{aligned}
\end{equation}
As above, a prime on determinants always signifies that it is to be computed in the space transverse to zero modes of the associated operator. 
That $\mathcal{A}_{\rm gh}$ chirally factorises up to the term in the parenthesis (and a Liouville factor that we are suppressing) is well understood and holds for arbitrary genus $\mathfrak{g}$. The quantity $\mathcal{Z}_{\mathfrak{g}}$ is in turn a certain combination of modular functions. For example, at genus $\mathfrak{g}=1$,  
it may be written in terms of the Dedekind eta function, 
$$
\qquad\qquad\qquad\qquad\mathcal{Z}_1=\eta(\tau)^{-24}\qquad\qquad (\mathfrak{g}=1),
$$ 
with $\tau=\tau_1+i\tau_2$ the complex structure modulus of the torus \cite{Polchinski_v1}. The quantity in the parenthesis can also be expressed in terms of modular functions \cite{VerlindeVerlinde87,DHokerPhong}, but in fact the above form will be more useful in what follows, because it precisely cancels a similar factor from the matter sector in the critical dimension at fixed internal loop momenta, as we will elaborate on explicitly next.

Let us now collect all the pieces and write down an expression for the full ($\mathfrak{g}>0$) generating function (\ref{eq:Afull}) for closed string scattering in target spacetimes $\mathbb{R}^{D}\times \mathbb{T}^{D_{\rm cr}-D}$ with generic K\"ahler and complex structure moduli, background KK gauge fields and torsion, see (\ref{eq:generic GBPhi}), on account of the ghost  (\ref{eq:ghostb}) and matter contribution (\ref{eq:A(j)4}),
\begin{equation}\label{eq:A_Eucl(j) non-crit}
\begin{aligned}
\mathcal{A}_{\rm Eucl}(j)
&=\delta^D(\ell_s\smallint j^{\mu})\,\delta^{D_{\rm cr}-D}_{(\ell_s\smallint j_a),0}
\,\bigg(g_s\,\big({\rm Det}G^{\mu\nu}{\rm Det}G_{ab}\big)^{-\frac{1}{4}}\bigg)^{2\mathfrak{g}-2}
\bigg(\frac{{\det}'\Delta_{(0)}}{{\det}({\rm Im}\Omega_{IJ})\int_{\Sigma_{\mathfrak{g}}}d^2z\sqrt{g}}\bigg)^{(26-D_{\rm cr})/2}\\
&\!\!\!\!\!\!\!\!\!\!\!\!\times\ell_s^{D\mathfrak{g}}\int d^{D\mathfrak{g}}\mathbb{Q}_{I}^{\mu}\!\!\!\!\!\sum_{\{N^{a}_I,M^{'I}_{a}\}\in \mathbb{Z}}\!\!\!\!\!\bigg|\mathcal{Z}_{\mathfrak{g}}\exp \bigg(\frac{\ell_s^2}{4}\int d^2z\int d^2z'\big(j_M+H_M\big)G^{MN}\big(j'_N+H'_N\big)\ln E(z,z')\bigg)\bigg|^2
\end{aligned}
\end{equation}
Notice that the loop momenta contribution, $H_M$, $\bar{H}_M$, is nothing but an operator shift in $j_M$. 

In the critical dimension of bosonic string theory, $D_{\rm cr}=26$, the non-chirally split terms cancel out completely, and (unless we include a fermionic sector to extend this result to the superstring) this is precisely where this computation is valid. In non-critical bosonic string theory, where $D_{\rm cr}\neq 26$, there is an additional Liouville factor that contributes to restore Weyl invariance. In what follows we focus on $D_{\rm cr}=26$, but we emphasise that the above expression for $\mathcal{A}_{\rm Eucl}(j)$ holds true also for the superstring when $D_{\rm cr}=10$ and the sources $j_M$ are shifted by worldsheet fermions \cite{D'HokerPhong89}. The superstring will be discussed in detail elsewhere.

Wick-rotating $\mathcal{A}_{\rm Eucl}(j)$ to Lorentzian target space signature\footnote{As discussed above, Wick rotating back to Lorentzian target spacetime signature can be achieved by replacing 
 $G_{\mu\nu}^{\rm Eucl}\rightarrow G_{\mu\nu}^{\rm Mink}$ and $\sqrt{{\rm Det}G^{\mu\nu}_{\rm Eucl}}\rightarrow i\sqrt{-{\rm Det}G^{\mu\nu}_{\rm Mink}}$ (the branch of the square root being convention-dependent). This is equivalent to starting from a Euclidean signature generating function and then interpreting all spacetime contractions as being with respect to a Lorentzian signature metric, $G_{\mu\nu}^{\rm Mink}$, while rotating the {\it coupling} $(g_{\rm eff}^2)^{\rm Eucl}\rightarrow -i (g_{\rm eff}^2)^{\rm Mink}$, leaving other quantities unchanged, with $(g_{\rm eff})^{\rm Mink}$ positive definite as defined in (\ref{eq:geff}). This approach leads to an overall factor of $i^{1-\mathfrak{g}}$, with the $i$ displayed explicitly in (\ref{eq:A(j)full}) and the $i^{-\mathfrak{g}}$ absorbed into (\ref{eq:sumintQQbar-Mink}). Upon rotating to Lorentzian signature, the contours of energy loop integrals are to be interpreted as in \cite{PiusSen16,Sen16,Sen16b}.}, and denoting the resulting object by $\mathcal{A}(j)$, we can then very concisely write the full result for the generating function as follows:
\begin{empheq}[box=\widefbox]{align}\label{eq:A(j)full}
\mathcal{A}(j)
&=i\bar{\delta}(j\ell_s)
\,g_{\rm eff}^{2\mathfrak{g}-2}\,
\sumint\limits_{(\mathbb{Q},\bar{\mathbb{Q}})}\bigg|\mathcal{Z}_{\mathfrak{g}}\exp \bigg(\frac{\ell_s^2}{4}\int d^2z\int d^2z'\big(j+H\big)\cdot \big(j'+H'\big)\ln E(z,z')\bigg)\bigg|^{2\phantom{\Big(}}\!\!\!\!\!\!\!\!\!\!\!\!
\end{empheq}
The (dimensionless) sum/integral over $(\mathbb{Q},\bar{\mathbb{Q}})$ should be understood as an integral over the non-compact momenta, $\mathbb{P}_{I}^{\mu}$ (with $\mathbb{Q}_{I}^{\mu}=\bar{\mathbb{Q}}_{I}^{\mu}$), and a sum over the compact momenta, $\mathbb{Q}_{Ia},\bar{\mathbb{Q}}_{Ia}$, which lie on the genus-$\mathfrak{g}$ torus lattice, $\Gamma^{\mathfrak{g}}_{D_{\rm cr}-D,D_{\rm cr}-D}$, and are labelled by integers $\{{M'}^I_{a},N_I^{a}\}\in \mathbb{Z}$, where $a$ spans $\mathbb{T}^{D_{\rm cr}-D}$ and $I=1,\dots,\mathfrak{g}$,
\begin{equation}\label{eq:sumintQQbar-Mink}
\sumint\limits_{(\mathbb{Q},\bar{\mathbb{Q}})}\dfn \ell_s^{D\,\mathfrak{g}}i^{-\mathfrak{g}}\int d^{D\mathfrak{g}}\mathbb{Q}_{I}^{\mu}\!\!\!\!\!\sum_{\{N^{a}_I,M^{'I}_{a}\}\in \mathbb{Z}}=\ell_s^{D\,\mathfrak{g}}i^{-\mathfrak{g}}\int d^{D\mathfrak{g}}\mathbb{P}_{I}^{\mu}d^{D\mathfrak{g}}\mathbb{W}^{I\mu}\delta^{D\mathfrak{g}}(\mathbb{W}^{I\mu})\!\!\!\!\!\sum_{\{N^{a}_I,M^{'I}_{a}\}\in \mathbb{Z}},
\end{equation}
and (for transparency of exposition) have also defined the following dimensionless combination of Dirac and Kronecker delta functions:
\begin{equation}\label{eq:bardelta}
\boxed{\bar{\delta}(j\ell_s)\dfn \delta^D(\smallint j^{\mu}\ell_s)\delta_{(\smallint j_a\ell_s),0}^{D_{\rm cr}-D}}
\end{equation}
with the property $\bar{\delta}(j\ell_s)=\ell_s^{-D}\bar{\delta}(j)$. It is customary for S-matrix calculations, see (\ref{eq:SfiM}),  to work in terms of the dimensionful delta function $\bar{\deltaslash}(j)=(2\pi)^D\bar{\delta}(j)$, i.e.,
\begin{equation}\label{eq:bardelta-Smatrix}
\bar{\deltaslash}(j)\dfn (2\pi)^D\delta^D(\smallint j^{\mu})\delta_{(\smallint j_a\ell_s),0}^{D_{\rm cr}-D}.
\end{equation}
It should also be understood that all implicit spacetime index contractions in (\ref{eq:A(j)full}) are carried out with the full metric $G_{MN}$, which now has Lorentzian signature, and that the T-duality {\it invariant} $g_{\rm eff}$ is defined in terms of this:
\begin{equation}\label{eq:geff}
\boxed{g_{\rm eff}\dfn g_s\Big(\!-{\rm Det}G^{\mu\nu}{\rm Det}G_{ab}\Big)^{-\frac{1}{4}}}
\end{equation}
The above holds for arbitrary constant backgrounds, $G_{MN}$, $B_{MN}$, but a standard example is the following (torsion-free) background:
\begin{equation}\label{eq:diagonal GB..}
G_{MN}=   \left(\begin{matrix} 
      \eta_{\mu\nu} & 0 \\
      0 & G_{ab} \\
   \end{matrix}\right), \qquad  
   G_{ab}=   \left(\begin{matrix} 
         (R^{1}/\ell_s)^2 &  & 0\\
          & \ddots & \\
         0 &  & (R^{D_{\rm cr}-D}/\ell_s)^2\\
      \end{matrix}\right),\qquad {\rm and }\qquad B_{MN}=0,
\end{equation}
where now $g_{\rm eff}=g_s\,({\rm Det}G_{ab})^{-\frac{1}{4}}=g_s\prod_a(\ell_s/R^a)^{\frac{1}{2}}$, and T-dualising along one compact dimension of radius $R^1$ \cite{Polchinski_v1}:
\begin{equation}\label{eq:Tdual}
\frac{R^1}{\ell_s}\rightarrow \frac{\ell_s}{R^1},\qquad \Phi\rightarrow \Phi+\ln \Big(\frac{\ell_s}{R^1}\Big),
\end{equation}
leaves $g_{\rm eff}$ invariant; recall from (\ref{eq:g_s=ephi}) that $g_s=e^{\Phi}$. A full discussion of T-duality for generic correlation functions is outside the scope of the current document. There are numerous discussions of T-duality from a worldsheet perspective; a recent exposition that is particularly transparent and relevant for constant backgrounds $G_{MN}, B_{MN}$, and $\Phi$ is \cite{BakasLust15}. 

\ssk
The full generating function is explicitly dimensionless, as it should be. Vertex operators should of course also then be dimensionless in order to lead to a dimensionless S matrix whose modulus-square yields a probability. This is indeed the case when kinematic factors for each of the vertex operators are included, $1/\sqrt{2\mathbb{k}^0 V_{D-1}}$, (with $V_{D-1}=(2\pi)^{D-1}\delta^{D-1}(0)$ the formal volume of non-compact space that always cancels out of observables (precisely as in standard field theory \cite{LandauLifshitzRQF}) and $\mathbb{k}^0$ the expectation value of the energy of the vertex operator) whose mass dimension precisely cancels that of the string coupling $g_D\equiv g_{\rm eff} \ell_s^{\frac{D}{2}-1}$. (A nice way of tracing back the origin of these factors was presented in \cite{SklirosHindmarsh11}.) As in field theory, these kinematic factors will not appear in the (Lorentz-invariant) invariant amplitudes, $\mathcal{M}_{\rm fi}(1,2,\dots)$.

We would like to end this subsection by briefly returning to the discussion associated to the shift of quantum fluctuations (\ref{eq:yycl}) that subtracts the source dependent piece from the classical solitons (\ref{eq:xcl sol2}). We stated there that (being a field redefinition) such a shift will not affect amplitudes or the generating function. The manner in which this invariance manifests itself is quite interesting, so we will discuss it briefly. Suppose that instead of computing quantum fluctuations around the soliton solution $y_{\rm cl}$ we computed quantum fluctuations around the original soliton solution,
$$
x_{\rm cl}^M(z,\bar{z})=y_{\rm cl}^M(z,\bar{z}) +i\int d^2wG^{MN}j_N(w,\bar{w})G(z,w).
$$
By explicit computation one can show that the effect of the new source-dependent shift in the classical soliton sector, $i\int d^2wG^{MN}j_N(w,\bar{w})G(z,w)$, is to {\it undo} the chiral splitting in the sense that the exponential,
$$
\bigg(\frac{\alpha'}{4}\int\! \!\!\int j_MG^{MN}j'_N\ln E(z,z')+i\frac{\pi\alpha'}{2}\mathbb{Q}_{I}^{M}G_{MN}\Omega_{IJ}\mathbb{Q}_{J}^{N}+i\pi\alpha' \mathbb{Q}_{I}^{M}\int\,\!j_M\int^z\!\!\omega_I\bigg)+{\rm c.c.},
$$
in (\ref{eq:A(j)3}) or (\ref{eq:A(j)full}) would get replaced by:
\begin{equation*}
\begin{aligned}
&-\frac{1}{2}\int\! \!\!\int j_MG^{MN}j'_N\Big(\! -\frac{\alpha'}{2}\ln \left|E(z,z')\right|^2 + \pi\alpha'\, {\rm Im}\int\limits_{z}^{z'}\omega_I\left({\rm Im}\Omega\right)_{IJ}^{-1}{\rm Im}\int\limits_{z}^{z'}\omega_J\Big)\\
&\qquad\qquad\qquad+i\frac{\pi\alpha'}{2}\mathbb{Q}_{I}^{M}G_{MN}\Omega_{IJ}\mathbb{Q}_{J}^{N}-i\frac{\pi\alpha'}{2}\bar{\mathbb{Q}}_{I}^{M}G_{MN}\bar{\Omega}_{IJ}\bar{\mathbb{Q}}_{J}^{N},
\end{aligned}
\end{equation*}
where the term in the parenthesis in the latter expression is precisely the propagator for the full non-chirally split quantum fluctuations (\ref{eq: Green Function+regular*m}). A further integral over the non-compact loop momenta produces the standard \cite{Polchinski_v1} non-chirally split generating function. Given that integrating out the loop momenta does of course leave the generating function invariant, it follows that the field redefinition (\ref{eq:yycl}) also leaves the amplitudes invariant. That is, doing perturbation theory around either of the soliton solutions (\ref{eq:xcl sol2}) or (\ref{eq:yycl}) leads to identical results, thus justifying our original claim.

\section{Wave/String Duality}\label{sec:WSD}
In this subsection we discuss the sense in which wave/particle duality of point-particle quantum mechanics arises in string theory. The analogous relation in string theory will be referred to as {\it wave/string duality}, because in string perturbation theory the notion of particle is replaced by the notion of string. In passing we will also elaborate on some aspects of target space effective actions.

The generating function (\ref{eq:A(j)full}) is given in the fixed-loop momenta representation, in both compact and non-compact sectors, associated to $\mathbb{T}^{D_{\rm cr}-C}$ and $\mathbb{R}^{D-1,1}$ respectively, for generic constant K\"ahler moduli, complex structure moduli, background KK gauge fields and torsion. It is also useful to display the analogous expressions for integrated loop momenta. There are four natural possibilities: 
$$
(\textrm{loop momenta in }\mathbb{R}^{D-1,1},\textrm{loop momenta in }\mathbb{T}^{D_{\rm cr}-D})={\bf (F,F)},\,\, {\bf (F,I)}, \,\,{\bf (I,F)} \,\,{\rm and}\,\, {\bf (I,I)},
$$
where\footnote{`Integrated' here means that the associated loop momenta have been integrated out, whereas `fixed' means that the loop momenta appear explicitly in the integrand/summand.} {\bf I}$=$`integrated' and {\bf F}$=$`fixed'. The associated generating functions are all equal as they are related by Fourier transforms (for the non-compact sector) and Poisson resummations (for the compact sector),
\begin{equation}\label{eq:A(j)wave=A(j)string}
\mathcal{A}(j)_{\bf \textrm{(F,F)}}=\mathcal{A}(j)_{\bf \textrm{(F,I)}}=\mathcal{A}(j)_{\bf \textrm{(I,F)}}=\mathcal{A}(j)_{\bf \textrm{(I,I)}}.
\end{equation}
We will now argue that the displayed equalities (\ref{eq:A(j)wave=A(j)string}) may be regarded as a stringy manifestation of {\it wave/string duality}, generalising the well-known wave/particle duality of quantum mechanics; a principle that applies to {\it all} scattering amplitudes in string theory, to all orders in perturbation theory. In this language the correspondences are:
$$
\textrm{\bf `F'}\,\, \equalhat \,\,\textrm{\bf wave picture},\qquad  {\rm and}\qquad \textrm{\bf `I'} \,\,\equalhat\,\, \textrm{\bf string picture}. 
$$
In the above discussion $\mathcal{A}(j)$ corresponds to a {\bf (wave,\,wave)} formulation of the generating function, $\mathcal{A}(j)=\mathcal{A}(j)_{\bf \textrm{(F,F)}}$, but considering also the other three pictures provides some additional insight. 

The {\bf (string,\,string)} generating function, $\mathcal{A}(j)_{\bf \textrm{(I,I)}}$, is the formulation that naturally arises out of the Lagrangian formulation of string theory, which is the usual starting point for string calculations in the path integral language \cite{Polchinski_v1}. Here one (generically) sums over all string trajectories for some fixed set of boundary conditions and asymptotic states, so it is natural to associate this with a {\it string picture} (analogous to a particle picture in the Feynman formulation of quantum mechanics where one sums over all trajectories of one or more particles given a set of boundary conditions). A good example that provides some further insight arises from considering the one-loop partition function in the Lagrangian formulation. Let us in particular focus on the compact dimensions, there being analogous statements in the non-compact dimensions. This contains the instanton action associated to classical trajectories (\ref{eq:xcl sol}). Setting $N_I^a=0$,  one can compute the associated momentum of an $A_I$-cycle string using (\ref{eq:P}), and hence notice that the winding number $M^a_I$ that one is summing over, see (\ref{eq:gammas}) and (\ref{eq:periodicity x AB cycles}), may be interpreted as the number of times an $A_I$-cycle closed string traverses a compact dimension of size $2\pi R$ in worldsheet time interval ${\rm Im}\,\Omega_{11}$ (more precisely in the analytically continued worldsheet real time interval $-i{\rm Im}\,\Omega_{11}$, recall the worldsheet theory is in Euclidean signature). 

\ssk
Let us now think about the corresponding interpretation in the {\bf (wave,\,wave)} picture where the relevant quantity is $\mathcal{A}(j)_{\bf \textrm{(F,F)}}$. This is closely related to a Hamiltonian formulation of string theory, given that all (independent) loop momenta in this formulation are explicit. Considering again the one-loop partition function referred to above, $\mathcal{A}(j)_{\bf \textrm{(F,F)}}$ is obtained from $\mathcal{A}(j)_{\bf \textrm{(I,I)}}$ by performing a Poisson resummation in the compact dimensions with a momentum-conserving delta function insertion in the non-compact dimensions. The Poisson resummation maps the aforementioned integer $M^a_I$ to a new integer ${M'}_a^I$ (for every $I,a$) whose interpretation is now the {\it mode number} associated to a wave in a periodic box of dimension $2\pi R$. So summing over the number of times, $M^a_I$, a loop of string travels around a compact dimension of size $2\pi R$ can be equivalently written as a sum over mode numbers, ${M'}_a^I$, of a standing wave in a box of size $2\pi R$, hence making wave/string duality completely manifest. T-duality and modular invariance provide alternative geometrical pictures (all of which are physically equivalent).

\ssk
The remaining two (hybrid) cases, $\mathcal{A}(j)_{\bf \textrm{(F,I)}}$ and $\mathcal{A}(j)_{\bf \textrm{(I,F)}}$ may be thought of as Routhians (analogous to the Routhians of classical mechanics) given that they correspond to a Hamiltonian formulation in the non-compact and compact dimensions respectively, with a Lagrangian formulation in the remaining dimensions. 

In examining further these four pictures let us primarily zoom in on the $G^{\mu\nu}$ dependence in the effective coupling (\ref{eq:geff_Eucl}). The quantity ${\rm Det}\,G^{\mu\nu}$ is present in $g_{\rm eff}$ because the associated generating function (\ref{eq:A(j)3}), being in the ({\bf F},{\bf F}) picture, has fixed non-compact loop momenta. To see that this enters in precisely the expected manner it proves useful to consider the usual dimensional reduction on $\mathbb{R}^D\times \mathbb{T}^{D_{\rm cr}-D}$ of the NS-NS sector of low energy supergravity, see e.g.~\cite{MaharanaSchwarz93,Kiritsis}. The relevant metric decomposition that leads to a natural expression for the dimensionally reduced target space effective action is:
\begin{equation}\label{eq:generic G_MN decomp}
G_{MN}=   \left(\begin{matrix} 
      g_{\mu\nu}+A_{\mu}^aG_{ab}A_{\nu}^b & G_{ab}A_{\mu}^a \\
      G_{ab}A_{\nu}^b & G_{ab} \\
   \end{matrix}\right),\quad 
G^{MN}=   \left(\begin{matrix}
      \,\,(g_{\mu\nu})^{-1} & -(g_{\mu \rho})^{-1}A^{ b}_{\rho} \\
      -(g_{\nu \rho})^{-1}A^{a}_{\rho} &\,\, (G_{ab})^{-1}+A^a_{\mu}(g_{\mu\nu})^{-1}A^b_{\nu} \\
   \end{matrix}\right),
\end{equation}
where the $A_{\mu}^a$ are Kaluza-Klein gauge fields. It is useful to compare with (\ref{eq:generic GBPhi}). The usefulness of this  parametrisation is that:
$$
({\rm Det}\,G^{\mu\nu})^{-1}={\rm Det}\,g_{\mu\nu},\qquad {\rm Det}\,G_{MN}={\rm Det}\,g_{\mu\nu}\,{\rm Det}\,G_{ab}.
$$
We are free to define what we mean by $g^{\mu\nu}$, and it is consistent \cite{MaharanaSchwarz93} to simply define $g^{\mu\nu}\dfn (g_{\mu\nu})^{-1}$. (Note however that generically $G^{ab}\neq (G_{ab})^{-1}$, because $G^{ab}$ is the $ab$ component of $G^{MN}$ that is completely fixed by the defining relation $G^{MN}G_{NL}=\delta^M_L$.) We can then rewrite the following term in (\ref{eq:A_Eucl(j) non-crit}) in terms of $g_{\mu\nu}$,
\begin{equation}\label{eq:zero modes G->g}
\begin{aligned}
\delta^D(\ell_s\smallint j^{\mu})&\Big(g_s\big({\rm Det}G^{\mu\nu}{\rm Det}G_{ab}\big)^{-\frac{1}{4}}\Big)^{2\mathfrak{g}-2}\int d^{D\mathfrak{g}}\mathbb{P}_{I}^{\mu}\,(\dots) =\\
&=\delta^D(\ell_s\smallint j_{\mu})\sqrt{{\rm Det}\,g_{\mu\nu}}\prod_{I=1}^{\mathfrak{g}}\Big(\int d^{D}\mathbb{P}_{I}^{\mu}\sqrt{{\rm Det}\,g_{\mu\nu}}\Big)\,e^{(2\mathfrak{g}-2)\Phi_D}\,(\dots),
\end{aligned}
\end{equation}
where following standard practice we have defined a dimensionally-reduced dilaton:
\begin{equation}\label{eq:Phi_D dfn}
\boxed{\Phi_D\dfn \Phi-\tfrac{1}{4}\ln{\rm Det}\,G_{ab}}
\end{equation}
Fourier transforming the depicted delta function in (\ref{eq:zero modes G->g}) and the integrand of the $\mathfrak{g}$ $A_I$-cycle loop momentum integrals leads to factors:
\begin{equation}\label{eq:dxsqrtg ints string}
\prod_{I=0}^{\mathfrak{g}}\Big(\int d^Dx_I\sqrt{{\rm Det}\,g_{\mu\nu}}\Big)\,e^{(2\mathfrak{g}-2)\Phi_D},
\end{equation}
which is clearly a collection of natural position space measures with a dilaton dependence that is precisely that expected from low energy supergravity \cite{MaharanaSchwarz93}, whose tree-level in $g_s$ contribution contains the universal factor:
$$
\int d^Dx_I\sqrt{{\rm Det}\,g_{\mu\nu}}\,e^{-2\Phi_D}(\dots).
$$

For the reader that is trying to make contact with the quantum effective action of quantum field theory \cite{EllisMavromatosSkliros15} note that the momenta of strings propagating through the various {\it isotopically} (in $\Sigma_{\mathfrak{g}}$) distinct cycles of the underlying genus-$\mathfrak{g}$ Riemann surface with  $n$ external vertex operators (equalling \cite{Sen15b} $3\mathfrak{g}-3+2n$ in number) are completely fixed by momentum conservation once the $A_I$-cycle loop momenta and the $n$ vertex operator momenta are fixed (as we have done above). When the remaining internal momenta become manifest there will be additional Fourier transforms leading to additional $x_I$ integrals. For example, consider the case of 3-point interaction vertices, associated to degeneration limits of the underlying Riemann surface so that a decomposition into pant diagrams \cite{Sen15b} becomes natural. Then, the number of {\it topologically} distinct Feynman diagrams for fixed loop order, $\mathfrak{g}$, and fixed number of external vertex operators, $n$, in the presence 3-point interaction vertices (the number of internal vertices) equals the number of distinct pant decompositions of the Riemann surface. The total number of pants,  $2\mathfrak{g}-2+n$, in either one of the complete set of pant decompositions in turn equals the number of vertices in the corresponding low energy field theory, and hence also equals the number of position space integrals (as one expects from a perturbative expansion of the field theory path integral \cite{EllisMavromatosSkliros15}). Hence there will be a universal factor in this particular degeneration:
\begin{equation}\label{eq:dxsqrtg ints total}
\prod_{\alpha=1}^{2\mathfrak{g}-2+n}\Big(\int d^Dx_{\alpha}\sqrt{{\rm Det}\,g_{\mu\nu}}\Big)\,e^{(2\mathfrak{g}-2)\Phi_D},
\end{equation}
$\mathfrak{g}+1$ integrals of which are manifest from the derived explicit factor (\ref{eq:dxsqrtg ints string}) above, while the remaining $\mathfrak{g}+n-3$ integrals are not manifest in the above decomposition because we have {\it only} fixed the independent (in particular $A_I$-cycle) loop momenta (and implicitly the vertex operator momenta). This is a well-known peculiarity of string theory \cite{DHokerPhong95}, in that it is not so natural to exhibit all intermediate propagators in a string theory amplitude until we reach a field theory limit. The remaining internal momenta are nevertheless all fixed by momentum conservation and so can be made explicit by introducing momentum conserving delta functions {\it for a given} pant decomposition. From standard field theory and Feynman diagram topology considerations Fourier transforming the resulting momentum integrands must lead precisely to an overall factor (\ref{eq:dxsqrtg ints total}), thus making the quantum effective action and corresponding field theory limit manifest. Clearly, the $\mathfrak{g}=0$ terms all have one overall factor $\int d^Dx_I\sqrt{{\rm Det}\,g_{\mu\nu}}e^{-2\Phi_D}$, as one expects for the classical contribution to the effective action, with higher loop orders introducing additional integrals. The overall measure (\ref{eq:dxsqrtg ints string}) is the closest one can get to obtaining an expression resembling the standard renormalised quantum effective action, $\Gamma(\phi)$, of quantum field theory (the Legendre transform, $-\Gamma(\phi)+\int J\phi=W(J)$, of the renormalised generating function of correlation functions, $W(J)$, with $\phi$ the renormalised fields under consideration) without considering explicit pant decompositions or degenerations. A much more complete discussion on some of these aspects can be found in \cite{Sen15b}. 

\ssk
From these considerations it is clear that the ${\rm Det}\,G^{\mu\nu}$ dependence in (\ref{eq:A_Eucl(j) non-crit}) is completely natural and  necessary in order to make contact with quantum field theory considerations. Given that the ${\rm Det}\,G^{\mu\nu}$ dependence is (according to the above discussion) associated to the explicit presence of internal non-compact loop momenta, $\mathbb{P}^{\mu}_I$, we can remove it by integrating them out. Returning to the original parametrisation of the metric (\ref{eq:generic GBPhi}), on account of (\ref{eq:sumint QQbar}), (\ref{eq:QGQ QGPhi}), $\delta^D(\smallint j^{\mu})=\delta^D(\smallint j_{\mu})\frac{1}{{\rm Det}G^{\mu\nu}}$ (recall that $\int j_a=0$ and that in Euclidean space $G^{\mu\nu}$ is positive definite), and a slight variation of the Gaussian integral (\ref{eq:V1...Vnfixedloop}) we obtain the {\bf (string,\,wave)} (or {\bf (I,\,F)}) representation,\footnote{The following relations (and analogous expressions obtained by interchanges $(\mu,\nu)\leftrightarrow (a,b)$) are useful:
\begin{equation}\label{eq:detGrelations}
\begin{aligned}
&({\rm Det}G^{\mu\nu})^{-1} = {\rm Det}\,G_{MN}\,({\rm Det}G_{ab})^{-1},\qquad {\rm Det}\,G_{MN}={\rm Det}(G_{\mu\nu}-G_{\mu a}G_{ab}^{-1}G_{b\nu })\,{\rm Det}\,G_{ab},\\
&\qquad\qquad (G^{ab})^{-1}=G_{ab}-G_{a\mu}G_{\mu\nu}^{-1}G_{\nu b},\qquad 
(G_{ab})^{-1}=G^{ab}-G^{a\mu}(G^{\mu\nu})^{-1}G^{\nu b}.
\end{aligned}
\end{equation}
}
\begin{equation}\label{eq:A(j)3 int loop mom}
\begin{aligned}
&\mathcal{A}_x^{\rm Eucl}(j)_{\rm (I,F)}=(2\pi)^D\delta^D(\smallint j_{\mu})\delta^{D_{\rm cr}-D}_{(\ell_s\smallint j_a),0}\,\frac{1}{\sqrt{{\rm Det}\,G^{\mu\nu}}}\,\Big(g_s\big({\rm Det}G_{ab}\big)^{-\frac{1}{4}}\Big)^{2\mathfrak{g}-2}\\
&\,\quad \times\bigg(\frac{{\det}'\Delta_{(0)}}{{\det}({\rm Im}\Omega_{IJ})\int_{\Sigma_{\mathfrak{g}}}d^2z\sqrt{g}}\bigg)^{-\frac{D_{\rm cr}}{2}}\!\!\! (4\pi\alpha'{\det}\,{\rm Im}\Omega_{IJ})^{-D/2}\exp\bigg(-\frac{1}{2}\int\! \!\!\int j_MG^{MN}j'_NG(z,z')\bigg)\\
&\,\,\,\qquad \qquad \times\!\!\!\!\!\sum_{\{N^{a}_I,M^{'I}_{a}\}\in \mathbb{Z}}\Big|e^{i\frac{\pi\alpha'}{2}\mathbb{Q}_a^I(G_{ab})^{-1}\Omega_{IJ}\mathbb{Q}_b^J+i\pi\alpha'\mathbb{Q}_a^I(G_{ab})^{-1}\Phi_b^I}\Big|^2e^{i\frac{\pi\alpha'}{2}(\Phi-\bar{\Phi})_a^I(G_{ab})^{-1}(\Omega-\bar{\Omega})_{IJ}^{-1}(\Phi-\bar{\Phi})_b^J}\\
\end{aligned}
\end{equation}
where $G(z,z')$ denotes the full Green function (\ref{eq: Green Function+regular*m}). We have presented the result for the matter contribution for clarity, the full generating function, $\mathcal{A}_{\rm Eucl}(j)_{\rm (I,F)}$, being obtained by multiplying the right-hand side of (\ref{eq:A(j)3 int loop mom}) by the ghost contribution (\ref{eq:ghostb}).  (In reconstructing the worldsheet Green function we have made use of the constraint $\int d^2zj_M=0$.) It is seen that the effective coupling in this representation is $g_s\big({\rm Det}G_{ab}\big)^{-\frac{1}{4}}$, as one expects from Kaluza-Klein reduction of low energy supergravity \cite{MaharanaSchwarz93,Kiritsis} on $\mathbb{R}^{D}\times \mathbb{T}^{D_{\rm cr}-D}$.  

\ssk
The corresponding generating function in the {\bf (wave,\,string)} (or {\bf (F,\,I)}) picture is similarly obtained from (\ref{eq:zeromodes}), (\ref{eq:Psi^q_Eucl}) and (\ref{eq:Psi^cl original}),
\begin{equation}
\begin{aligned}
\mathcal{A}_x(j&)_{\rm (F,I)}=\delta^D(\ell_s\smallint j^{\mu})\,\delta^{D_{\rm cr}-D}_{(\smallint j_a),0}\,V_C\,\Big(g_s\big({\rm Det}G^{\mu\nu}\big)^{-\frac{1}{4}}\Big)^{2\mathfrak{g}-2}\\
&\times\,\Big(\frac{{\det}'\Delta_{(0)}}{{\det}\,{\rm Im}\Omega_{IJ}\int d^2z\sqrt{g}}\Big)^{-D_{\rm cr}/2}(4\pi^2\alpha'{\det}\,{\rm Im}\Omega_{IJ})^{-(D_{\rm cr}-D)/2} \\
&\times  \sum_{\{N^a_I,M^a_I\}\in \mathbb{Z}}\exp \bigg\{-\gamma_I^a\Big(\frac{1}{\pi\alpha'}{\rm Im}\,\Omega_{IJ}\big(G_{ab}+B_{ab}\big)\Big)\bar{\gamma}_J^b+i\Phi_a^I\gamma_I^a+i\bar{\Phi}^I_a\bar{\gamma}_I^a\bigg\}\\
&\,\,\times  \ell_s^{D\mathfrak{g}}\int d^{D\mathfrak{g}} \mathbb{P}^{\mu_I} \Big|e^{\frac{\alpha'}{4}\int \!\!\int j_MG^{MN}j'_N\ln E(z,z')+i\frac{\pi\alpha'}{2}\mathbb{P}_{I}^{\mu}(G^{\mu\nu})^{-1}\Omega_{IJ}\mathbb{P}_{J}^{\nu}+i\pi\alpha' \mathbb{P}_{I}^{\mu}(G^{\mu\nu})^{-1}G^{\nu N}\Phi^I_{N}}\Big|^2\\
&\qquad \qquad \times \exp\Big(\!-i\frac{\pi \alpha'}{2}(\Phi_a^I-\bar{\Phi}_a^I)(G_{ab})^{-1}(\Omega_{IJ}-\bar{\Omega}_{IJ})^{-1}(\Phi_b^J-\bar{\Phi}_b^J)\Big)
\end{aligned}
\end{equation}
where we have defined the compactification volume:
$$
V_C\dfn (2\pi\ell_s)^{D_{\rm cr}-D}\sqrt{{\rm Det}\,G_{ab}},
$$
whereas for the {\bf (string,\,string)} picture generating function (i.e.~{\bf (I,\,I)}) we may consider (\ref{eq:A(j)3 int loop mom}), and perform a (or more precisely undo the) Poisson resummation in the compact dimensions. This will also remove the $G_{ab}$ determinant from the effective coupling, $g_s\big({\rm Det}G_{ab}\big)^{-\frac{1}{4}}$. 
Making further use of the relations for determinants (\ref{eq:detGrelations}) in the footnote  
and taking (\ref{eq:InstSum}) into account we obtain the {\bf (string,\,string)} (or {\bf (I,\,I)}) picture representation,
\begin{equation}\label{eq:A(j)3 int loop mom 2}
\begin{aligned}
\mathcal{A}_x^{\rm Eucl}&(j)_{\rm (I,I)}=(2\pi)^D\delta^D(\smallint j_{\mu})\,\delta^{D_{\rm cr}-D}_{(\ell_s\smallint j_a),0}\,\sqrt{{\rm Det}\,G_{MN}}(2\pi\ell_s)^{D_{\rm cr}-D}\,g_s^{2\mathfrak{g}-2}\\
&\times\sum_{\{N^a_I,M^a_I\}\in \mathbb{Z}}\exp \bigg\{-\gamma_I^a\Big(\frac{1}{\pi\alpha'}{\rm Im}\,\Omega_{IJ}\big(G_{ab}+B_{ab}\big)\Big)\bar{\gamma}_J^b+i\Phi_a^I\gamma_I^a+i\bar{\Phi}^I_a\bar{\gamma}_I^a\bigg\}\\
&\times\bigg(\frac{4\pi^2\alpha'{\det}'\Delta_{(0)}}{\int_{\Sigma_{\mathfrak{g}}}d^2z\sqrt{g}}\bigg)^{-\frac{D_{\rm cr}}{2}}\!\!\!\!\exp\bigg(\!-\frac{1}{2}\int d^2z\int d^2z' j_M(z,\bar{z}) G^{MN}j_N(z',\bar{z}')G(z,z')\bigg),
\end{aligned}
\end{equation}
This expression (\ref{eq:A(j)3 int loop mom 2}) is in precise agreement with more standard expressions \cite{Polchinski_v1,D'HokerPhong89} when the target space metric associated to $\mathbb{R}^{D}\times \mathbb{T}^{D_{\rm cr}-D}$ is parametrised as in (\ref{eq:generic G_MN decomp}). Note that when this is the case, $\sqrt{{\rm Det}\,G_{MN}}(2\pi\ell_s)^{D_{\rm cr}-D}=\sqrt{{\rm Det}\,g_{\mu\nu}}\,V_{C}$. 
Using zeta function regularisation to place the explicit factor $4\pi^2\alpha'$ in (\ref{eq:A(j)3 int loop mom 2}) inside the determinant, analytically continuing to Lorentzian signature, where $\sqrt{{\rm Det}\,g_{\mu\nu}}\rightarrow i\sqrt{-{\rm Det}\,g_{\mu\nu}}$, and when $D=D_{\rm cr}$ we clearly reproduce Polchinski's expression for the generating function \cite{Polchinski_v1} (equation (6.2.6) there), thus providing a non-trivial check of the normalisation and of the various factors present.

\ssk
It is important to mention that when $\mathfrak{g}=1$ the above expressions for the generating function assume there is at least one vertex operator insertion. Let us briefly discuss the vacuum amplitude which (although standard) is the one example where this is not the case.

In the absence of vertex operators and for $\mathfrak{g}>1$ integrating over moduli gives the vacuum amplitude, usually denoted by $Z_{\mathfrak{g}}$,
\begin{equation}
\begin{aligned}
Z_{\mathfrak{g}}&=\int_{\mathscr{F}_{\mathfrak{g}}} d{\bf M}_{\mathfrak{g}}\mathcal{A}(0),\qquad (\mathfrak{g}>1)\\
&=-i\int d^Dx_0\sqrt{-{\rm Det}\,g_{\mu\nu}}\,\Lambda_D^{\mathfrak{g}},
\end{aligned}
\end{equation}
because in this case there are no CKV's; see (\ref{eq:intdM}) for the definition of $d{\bf M}_{\mathfrak{g}}$. 
The genus $\mathfrak{g}$ cosmological constant, $\Lambda_D^{\mathfrak{g}}$, is defined by:
\begin{equation}
\begin{aligned}
\Lambda_D^{\mathfrak{g}}&\dfn -e^{-\chi(\Sigma_{\mathfrak{g}})\Phi}\,V_C\!\int_{\mathscr{F}_{\mathfrak{g}}} \!\!d{\bf M}_{\mathfrak{g}}\!\!\,\,\,\big(4\pi^2\alpha'{\det}\,{\rm Im}\Omega_{IJ}\big)^{-13}|\mathcal{Z}_{\mathfrak{g}}|^2\Psi^{\rm cl}|_{j=0}\\
&\,\,=-e^{-\chi(\Sigma_{\mathfrak{g}})\Phi_D}\!\int_{\mathscr{F}_{\mathfrak{g}}} d{\bf M}_{\mathfrak{g}}\big(4\pi^2\alpha'\,{\det}\,{\rm Im}\Omega_{IJ}\big)^{-D/2}\!\!\!\!\! \sum_{\{N^a_I,M^{'I}_a\}\in \mathbb{Z}}\!\!\!\! \Big|\mathcal{Z}_{\mathfrak{g}}e^{i\frac{\pi\alpha'}{2}\mathbb{Q}^{I}_a(G_{ab})^{-1}\Omega_{IJ}\mathbb{Q}_b^J}\Big|^2.
\end{aligned}
\end{equation}
The first equality is in the {\bf (I,\,I)} picture when the instanton contribution, $\Psi^{\rm cl}|_{j=0}$, is identified with (\ref{eq:Psi^cl original}) evaluated at $\Phi_a^I=\bar{\Phi}_a^I=0$, whereas the second is in the {\bf (I,\,F)} picture. 
For $\mathfrak{g}=1$ however, where $\Omega_{IJ}\rightarrow \tau=\tau_1+i\tau_2$, we should include a further factor of $2\tau_2$ in the denominator in the absence of vertex operators to obtain the vacuum amplitude \cite{Polchinski_v1},
\begin{equation}
\begin{aligned}
Z_1&=\int_{\mathscr{F}_1}\frac{d{\bf M}_{1}}{2\tau_2}\,\mathcal{A}(0),\qquad (\mathfrak{g}=1)\\
&=-i\int d^Dx_0\sqrt{-\textrm{Det}\,g_{\mu\nu}}\,\Lambda_D^{\mathfrak{g=1}},
\end{aligned}
\end{equation}
with the genus-1 dimensionally reduced cosmological constant,
\begin{equation}
\begin{aligned}
\Lambda_D^{\mathfrak{g=1}}&\dfn -V_C\!\!\int_{\mathscr{F}_1}\frac{d^2\tau}{4\tau_2}\,(4\pi^2\alpha'\tau_2)^{-13}|\eta(\tau)|^{-48}\Psi^{\rm cl}|_{j=0}\\
&\,\,=-\int_{\mathscr{F}_{1}} \frac{d^2\tau}{4\tau_2}\big(4\pi^2\alpha'\,\tau_2\big)^{-D/2}|\eta(\tau)|^{-48}\!\!\!\!\! \sum_{\{N^a,M^{'}_a\}\in \mathbb{Z}}\!\!\!\! \Big|e^{i\frac{\pi\alpha'}{2}\mathbb{Q}_a(G_{ab})^{-1}\tau\mathbb{Q}_b}\Big|^2.
\end{aligned}
\end{equation}
For {\it all} amplitudes with at least one vertex operator insertion there is no additional factor of $2\tau_2$ in the denominator. 
The moduli space measure $d{\bf M}_1=\frac{1}{2}d^2\tau$ (the additional factor of $1/2$ here being due to the remaining $\mathbb{Z}_2$ isometry, see Appendix \ref{sec:T2}). We have taken into account that $\int_{T^2}d^2z=2\tau_2$ and made use of the presence of one CKV in order to write all vertex operators in the integrated picture, and then set the number of vertex operators to zero, i.e.~$\prod_{\gamma=1}^n\int d^2z_{\gamma}V_{z_{\gamma}\bar{z}_{\gamma}}\rightarrow 1$. Furthermore, we have defined $\int d^Dx_0\dfn (2\pi)^D\delta^D(0)$, with\footnote{That it is natural to identify the integral over the zero modes, $x_0^{\mu}$, with $(2\pi)^D\delta^D(j_{\mu})|_{j_{\mu}=0}$ (with indices downstairs) follows from the integral representation of the delta function.} $\delta^D(0)=\delta^D(j_{\mu})|_{j_{\mu}=0}$, which has dimensions of $L^D$. 
The above expression for the vacuum amplitude is in precise agreement with standard conventions \cite{GSW2,Polchinski_v1} and serves as a non-trivial check of the normalisation of $\mathcal{A}(j)$ at $\mathfrak{g}=1$.

\ssk
We next discuss correlation functions for generic vertex operator insertions.

\section{Correlation Functions}\label{sec:CF}
Given the result for the generating function (\ref{eq:A(j)full}), whose defining equation is (\ref{eq:Afull}), let us now start to think about generic correlation functions, setting the stage in particular for correlation functions of highly excited strings. We define:
\begin{equation}\label{eq:<exp ijx>}
\mathcal{A}(j)\equiv \Big\langle \exp\Big(i\int d^2z j_M x^M\Big) \Big\rangle,
\end{equation}
where note that this implicitly includes the minimum number of ghost insertions required to make amplitudes not vanish trivially, see (\ref{eq:Afull}). To compute correlation functions of generic operators we take functional derivatives of $\mathcal{A}(j)$ with respect to $j_M(z,\bar{z})$, and subsequently set the source equal to the value of interest, see the footnote on p.~\pageref{foot:j}. For instance, given a set of operators $\{D_i\}$ (that commute with the path integral) we can extract correlation functions from $\mathcal{A}(j)$ as follows:
\begin{equation}\label{eq:D..DeJ}
\prod_{i=1}^{\mathcal{J}}D_i\frac{-i\delta}{\delta j_{M_i}(z_i,\bar{z}_i)}\mathcal{A}(j)\dfn \Big\langle D_1x^{M_1}(z_1,\bar{z}_1)\dots D_{\mathcal{J}}x^{M_{\mathcal{J}}}(z_{\mathcal{J}},\bar{z}_{\mathcal{J}})\exp\Big(i\int d^2z j_M x^M\Big) \Big\rangle. 
\end{equation}
The operators $\{D_i\}$ may denote a set of worldsheet derivatives, e.g.,~$\{\partial_z,\partial_{z}^2,\partial_w,\partial_{\bar{z}},\partial_{\bar{w}},\dots\}$, or (e.g.~in the case of coherent vertex operators \cite{SklirosCopelandSaffin16bb}) they may be more complicated but linear operators \cite{SklirosCopelandSaffin16cc}. 
 In order for this procedure to be useful in the case of composite operators (where multiple $D_ix$'s may be inserted at the same location on the worldsheet), we use the notion of {\it point splitting}, see e.g.~\cite{DabholkarMandalRamadevi98}; that is, we write a normal-ordered operator $:\!O_1O_2(z)\!:$ as $O_1(z_1)O_2(z_2)$, calculate the correlators as specified in (\ref{eq:D..DeJ}), subtract the terms singular in $(z_2-z_1)$, and take the limit $z_1\rightarrow z$, $z_2\rightarrow z$. We refer to the latter step as {\it point merging}.  

\ssk
Carrying out the functional derivatives as specified in (\ref{eq:D..DeJ}) on account of (\ref{eq:A(j)full}), (\ref{eq:HHbar}) and (\ref{eq:BIntPrimeForm}) leads to:\footnote{\label{foot:notation}Notation-wise, $(DD\ln |E|^2)_{\pi(2l-1)\pi(2l)}\equiv D_{\pi(2l-1)}D_{\pi(2l)}\ln |E(z_{\pi(2l-1)},z_{\pi(2l)})|^2$ and $\int j^{M_{\pi(q)}}(D\ln |E|^2)_{\pi(q)}\equiv \int d^2zj^{M_{\pi(q)}}(z,\bar{z})D_{\pi(q)}\ln |E(z_{\pi(q)},z)|^2$.}
\begin{equation}\label{<<V1...Vn>>fixedloop4}
\begin{aligned}
\Big\langle D_1&x^{M_1}(z_1,\bar{z}_1)\dots D_{\mathcal{J}}x^{M_{\mathcal{J}}}(z_{\mathcal{J}},\bar{z}_{\mathcal{J}})\exp\Big(i\int d^2z j_M x^M\Big) \Big\rangle=\\
&=i\bar{\delta}(j\ell_s)
\,g_{\rm eff}^{2\mathfrak{g}-2}\,
\sumint\limits_{(\mathbb{Q},\bar{\mathbb{Q}})}\bigg|\mathcal{Z}_{\mathfrak{g}}\exp \bigg(\frac{\ell_s^2}{4}\int d^2z\int d^2z'\big(j+H\big)\cdot \big(j'+H'\big)\ln E(z,z')\bigg)\bigg|^{2}\\
&\quad\times\sum_{k=0}^{\lfloor\mathcal{J}/2\rfloor}\sum_{\pi\in S_{\mathcal{J}}/\sim}\prod_{l=1}^k\bigg{\{}-\frac{\alpha'}{2}G^{M_{\pi(2l-1)}M_{\pi(2l)}}(DD\ln |E|^2)_{\pi(2l-1)\pi(2l)})\bigg{\}}\\
&\quad\times\prod_{q=2k+1}^{\mathcal{J}}\left\{\pi\alpha' D_{\pi(q)}\Big(\mathbb{Q}_I^{M_{\pi(q)}}\!\!\int^{z_{\pi(q)}}\!\!\!\!\omega_I-\bar{\mathbb{Q}}_I^{M_{\pi(q)}}\!\!\int^{\bar{z}_{\pi(q)}}\!\!\!\!\bar{\omega}_I\Big)-\frac{\alpha'}{2}i\int j^{M_{\pi(q)}}(D\ln |E|^2)_{\pi(q)}\right\},
\end{aligned}
\end{equation}
which on account of (\ref{eq:BIntPrimeForm}) may be viewed as a functional generalisation of,
$$
(-i)^{\mathcal{J}}\frac{\partial^{\mathcal{J}}}{\partial y^{\mathcal{J}}}e^{\tfrac{1}{2}g y^2}=\sum_{k=0}^{\lfloor \mathcal{J}/2\rfloor}\frac{2^{-k}\mathcal{J}!}{k!(\mathcal{J}-2k)!}(-g)^k(-i yg)^{\mathcal{J}-2k}e^{\tfrac{1}{2}gy^2}.
$$
It is also possible to show that for fixed $k$ the number of terms that appear in the sum over permutations in (\ref{<<V1...Vn>>fixedloop4}) before point merging is indeed: $$\frac{2^{-k}\mathcal{J}!}{k!(\mathcal{J}-2k)!},$$ as one would expect from the finite dimensional formula. 
The notation $\lfloor\mathcal{J}/2 \rfloor$ in the sum over $k$ indicates that the maximum value of $k$ is the integer that saturates the inequality $k\leq \mathcal{J}/2$. $S_{\mathcal{J}}$ is the {\it symmetric group} of degree $\mathcal{J}$ \cite{Hamermesh}, the group of all permutations of $\mathcal{J}$ elements, and the equivalence relation `$\sim$' is such that $\pi_i\sim\pi_j$ with $\pi_{i},\pi_j\in S_{\mathcal{J}}$ when they define the same element in (\ref{<<V1...Vn>>fixedloop4}).

\ssk
The point merging procedure will give rise to {\it contact terms}\label{contactterms}, i.e.~terms that only contribute when two or more vertex operators are coincident, e.g.~from contractions of the form $\partial_z\partial_{\bar{w}}\ln |E(z,w)|^2=-2\pi\delta^2(z-w)$. Following a standard argument, in view of the (assumed\footnote{It is not obvious whether analyticity in external momenta is present for generic amplitudes \cite{DHokerPhong95}, and one needs to check this on a case by case basis. In fact, the tachyon and massless tadpoles often cause trouble \cite{DHokerPhong95} in searching for absolute convergence in bosonic string amplitudes, and we one has to adopt a certain prescription in order to extract physical observables. In addition, certain degeneration limits of the worldsheet moduli can lead to trouble when one or more internal lines are forced to be onshell by momentum conservation (such as the separating degeneration of a two-loop two-point amplitude with one vertex operator on either component, or tadpole degeneration limits). These cases require particular care \cite{Polchinski87,Polchinski88,Nelson89,Sen15b}, such as an offshell description \cite{Sen15b}, vertex operators in a larger Hilbert space than the conformally invariant one \cite{Polchinski87,Polchinski88,Nelson89,Sen15b}, and one must introduce a local coordinate dependence \cite{Polchinski88,Sen15b} (i.e.~abandon conformal invariance) that cancels \cite{Sen15b} (see also \cite{PiusRudraSen14b,PiusRudraSen14c}) out of observables. The matter contribution to the generating function introduced here is still applicable in such cases, but one typically needs to consider more general ghost insertions than the minimal number. In this document we assume a region does exist in the complex momentum plane (or complex `Mandelstam variables', or appropriate generalisations thereof for $n$-point amplitudes) of absolute convergence, and physical amplitudes are then obtained by analytic continuation from this region. In sequels \cite{SklirosCopelandSaffin16cc,SklirosCopelandSaffin16dd} tachyon divergences will be carefully identified and some (in particular tadpoles) will be absorbed by background shifts and others dropped by brute force.}) analyticity of string amplitudes in external momenta \cite{DHokerPhong95}, and the fact that the amplitude always contains a factor of the form 
$$
\prod_{i<j}|E(z_i,z_j)|^{\alpha'k_i\cdot k_j},
$$ 
it follows that such terms will not contribute even after the vertex operator positions have been integrated over, and will thus be set to zero  \cite{D'HokerPhong89}.\footnote{\label{ft: analyticity}We mention the argument for completeness. Notice that the exponent of $|E(z_i,z_j)|$ can always be made positive by analytic continuation and that when two vertex insertion points come close together, $E(z_i,z_j)\simeq z_i-z_j$. Therefore, given that (symbolically) $\int d^2z|w-z|^{k_i\cdot k_j}\delta^2(w-z)=0$ when ${\rm Re} \,k_i\cdot k_j>0$ it follows from a famous theorem of complex analysis that the entire expression will vanish for all $k_j$. In amplitudes involving coherent vertex operators one also encounters exponentials of contact terms, and so one also needs to consider multiple delta functions. Similar reasoning to the above leads also to the vanishing of multiple delta functions, e.g. $\int d^2z|w-z|^{k_i\cdot k_j}\delta^2(w-z)\delta^2(w-z)=0$. To see this write this expression as $\lim_{\epsilon\rightarrow0}\int d^2z|w-z+\epsilon|^{k_i\cdot k_j}\delta^2(w-z)\delta^2(w-z+\epsilon)$. Performing the $z$ integration leads to $\lim_{\epsilon\rightarrow0}|\epsilon|^{k_i\cdot k_j}\delta^2(\epsilon)$, which vanishes for the following two reasons: the integral $\int d^2\epsilon|\epsilon|^{k_i\cdot k_j}\delta^2(\epsilon)=0$ and the corresponding integrand is non-negative -- therefore, the integrand must vanish. Extending this reasoning to three or more delta function insertions implies that (unless the momenta under consideration are constrained to vanish identically by momentum conservation, such as in the case of tadpoles) contact terms do not contribute to the amplitudes and will be dropped.}

A very important implication of this latter observation is that of all the permutations that we are to sum over in (\ref{<<V1...Vn>>fixedloop4}), the only ones that will give a non-zero contribution will be those that respect chiral splitting. That is, we can partition the full set of operators $\{D_i\}$ and spacetime indices $\{M_i\}$ into chiral and anti-chiral pieces,
\begin{equation}\label{eq:DjDbarj dfn}
\begin{aligned}
&\{D_1,\dots,D_{\mathcal{J}}\}=\{\mathcal{D}_1,\dots,\mathcal{D}_{\mathcal{I}},\bar{\mathcal{D}}_1,\dots,\bar{\mathcal{D}}_{\bar{\mathcal{I}}}\}\\
&\{M_1,\dots,M_{\mathcal{J}}\}=\{N_1,\dots,N_{\mathcal{I}},\bar{N}_1,\dots,\bar{N}_{\bar{\mathcal{I}}}\},\qquad {\rm with}\qquad \mathcal{J}=\mathcal{I}+\bar{\mathcal{I}},
\end{aligned}
\end{equation}
and then (denoting the worldsheet coordinates where the chiral and anti-chiral operators are inserted by $\{z_j,\bar{z}_j\}$ and $\{w_j,\bar{w}_j\}$ respectively) a careful consideration of the single sum over $k$ in (\ref{<<V1...Vn>>fixedloop4}) shows that it factorises into {\it two} independent sums. In turn, these two independent sums can be extracted from two completely independent correlation functions as follows:
\begin{equation}\label{<<V1...Vn>>fixedloop4chirallysplit}
\begin{aligned}
\Big\langle \mathcal{D}_1&x^{N_1}(z_1,\bar{z}_1)\dots\mathcal{D}_{\mathcal{I}}x^{N_{\mathcal{I}}}(z_{\mathcal{I}},\bar{z}_{\mathcal{I}}) \bar{\mathcal{D}}_1x^{\bar{N}_1}(w_1,\bar{w}_1)\dots\bar{\mathcal{D}}_{\bar{\mathcal{I}}}x^{\bar{N}_{\bar{\mathcal{I}}}}(w_{\mathcal{I}},\bar{w}_{\bar{\mathcal{I}}}) \exp\Big(i\int d^2z j\cdot x(z,\bar{z})\Big) \Big\rangle=\\
&=i\bar{\delta}(j\ell_s)\,g_{\rm eff}^{2\mathfrak{g}-2}\,
\sumint\limits_{(\mathbb{Q},\bar{\mathbb{Q}})}
\mathcal{Z}_{\mathfrak{g}} \Big\langle\mathcal{D}_1x_+^{N_1}(z_1)\dots\mathcal{D}_{\mathcal{I}}x_+^{N_{\mathcal{I}}}(z_{\mathcal{I}}) e^{i\int d^2z (j_L+H)\cdot x_+(z)}\Big\rangle_+\\
&\,\,\,\quad\qquad\qquad\qquad\times\bar{\mathcal{Z}}_{\mathfrak{g}} \Big\langle\bar{\mathcal{D}}_1x_-^{\bar{N}_1}(\bar{w}_1)\dots\bar{\mathcal{D}}_{\bar{\mathcal{I}}}x_-^{\bar{N}_{\bar{\mathcal{I}}}}(\bar{w}_{\bar{\mathcal{I}}})e^{i\int d^2z (\bar{j}_R+\bar{H})\cdot x_-(\bar{z})} \Big\rangle_-,
\end{aligned}
\end{equation}
where $j=j_{\rm L}=j_{\rm R}$ was assumed in the above derivation -- we will discuss the extension to $j_{\rm L}\neq j_{\rm R}$ momentarily. We want to emphasise that on the left-hand side the $x^N(z,\bar{z})=x_0^N+y^N_{\rm cl}(z,\bar{z})+y^N(z,\bar{z})$ appearing contain the zero modes, instanton contributions and quantum fluctuations. Recall the analysis following (\ref{eq:x=x0+ycl+y}). On the right-hand side however, a chiral and anti-chiral field appears, $x_+^N(z)$ and $x_-^N(\bar{w})$ respectively, which does {\it not} contain any zero mode or instanton contributions. These zero mode and instanton contributions are rather contained in $\bar{\delta}(j\ell_s)$ and $H,\bar{H}$ respectively. The relevant correlators on the right-hand side of (\ref{<<V1...Vn>>fixedloop4chirallysplit}) are defined with respect to the ``chiral propagators'' \cite{D'HokerPhong89}, which unlike the full propagator $G(z,w)$ have \cite{Fay,Mumford_v12} non-trivial monodromies around $B_I$ cycles but not around $A_I$ cycles, see (\ref{eq:quasiperiodprimeform}),
\begin{equation}\label{eq:chiral<xx>}
\boxed{\langle x_+^M(z)x_+^N(w)\rangle_+ =-\frac{\alpha'}{2}G^{MN}\ln E(z,w),\qquad \langle x_-^M(\bar{z})x_-^N(\bar{w})\rangle_- =-\frac{\alpha'}{2}G^{MN}\ln \bar{E}(\bar{z},\bar{w})}.
\end{equation}
Note that the non-chirally split contribution in the full propagator (\ref{eq: Greens function dzdbarzG+int}) precisely cancels out when all loop momenta (in both compact and non-compact dimensions) are made manifest. 

\ssk
The chiral correlator in (\ref{<<V1...Vn>>fixedloop4chirallysplit}) reads explicitly:
\begin{equation}\label{eq:<>_+}
\begin{aligned}
\Big\langle& \mathcal{D}_1x_+^{N_1}(z_1)\dots\mathcal{D}_{\mathcal{I}}x_+^{N_{\mathcal{I}}}(z_{\mathcal{I}}) e^{i\int d^2z (j_{\rm L}+H)\cdot x_+(z)}\Big\rangle_+=\\
&
=\exp\bigg(\frac{\alpha'}{4}\int\! \!\!\int j_{{\rm L}M}G^{MN}j'_{{\rm L}N}\ln E(z,z')+i\frac{\pi\alpha'}{2}\mathbb{Q}_{I}^{M}G_{MN}\Omega_{IJ}\mathbb{Q}_{J}^{N}+i\pi\alpha' \mathbb{Q}_{I}^{M}\int\,\!j_{{\rm L}M}\int^z\!\!\omega_I\bigg)\\
&\quad\times\sum_{k=0}^{\lfloor\mathcal{I}/2\rfloor}\sum_{\pi\in S_{\mathcal{J}}/\sim}\prod_{l=1}^k\bigg{\{}-\frac{\alpha'}{2}G^{N_{\pi(2l-1)}N_{\pi(2l)}}(\mathcal{D}\mathcal{D}\ln E)_{\pi(2l-1)\pi(2l)})\bigg{\}}\\
&\quad\times\prod_{q=2k+1}^{\mathcal{I}}\left\{\pi\alpha' \mathbb{Q}_I^{N_{\pi(q)}}\mathcal{D}_{\pi(q)}\!\!\int^{z_{\pi(q)}}\!\!\!\!\omega_I-\frac{\alpha'}{2}i\int j_{\rm L}^{N_{\pi(q)}}(\mathcal{D}\ln E)_{\pi(q)}\right\}
\end{aligned}
\end{equation}
where the argument in the exponential equals $\frac{\ell_s^2}{4}\int d^2z\int d^2z'\big(j_{\rm L}+H\big)\cdot \big(j'_{\rm L}+H'\big)\ln E(z,z')$, and similarly for the anti-chiral half,
\begin{equation}\label{eq:<>_-}
\begin{aligned}
\Big\langle& \bar{\mathcal{D}}_1x_-^{\bar{N}_1}(\bar{w}_1)\dots\bar{\mathcal{D}}_{\bar{\mathcal{I}}}x_-^{\bar{N}_{\bar{\mathcal{I}}}}(\bar{w}_{\bar{\mathcal{I}}})e^{i\int d^2z (j_{\rm R}+\bar{H})\cdot x_-(\bar{z})} \Big\rangle_-=\\
&
=\exp\bigg(\frac{\alpha'}{4}\int\! \!\!\int j_{{\rm R}M}G^{MN}{\bar{j}}'_{{\rm R}N}\ln E(\bar{z},\bar{z}')-i\frac{\pi\alpha'}{2}\bar{\mathbb{Q}}_{I}^{M}G_{MN}\bar{\Omega}_{IJ}\bar{\mathbb{Q}}_{J}^{N}-i\pi\alpha' \bar{\mathbb{Q}}_{I}^{M}\int\,\! j_{{\rm R}M}\int^{\bar{z}}\!\!\bar{\omega}_I\bigg)\\
&\quad\times\sum_{k=0}^{\lfloor\bar{\mathcal{I}}/2\rfloor}\sum_{\pi\in S_{\mathcal{J}}/\sim}\prod_{l=1}^k\bigg{\{}-\frac{\alpha'}{2}G^{\bar{N}_{\pi(2l-1)}\bar{N}_{\pi(2l)}}(\bar{\mathcal{D}}\bar{\mathcal{D}}\ln \bar{E})_{\pi(2l-1)\pi(2l)})\bigg{\}}\\
&\quad\times\prod_{q=2k+1}^{\bar{\mathcal{I}}}\left\{-\pi\alpha' \bar{\mathbb{Q}}_I^{\bar{N}_{\pi(q)}}\bar{\mathcal{D}}_{\pi(q)}\!\!\int^{\bar{w}_{\pi(q)}}\!\!\!\!\bar{\omega}_I-\frac{\alpha'}{2}i\int j^{\bar{N}_{\pi(q)}}_{\rm R}(\bar{\mathcal{D}}\ln \bar{E})_{\pi(q)}\right\}
\end{aligned}
\end{equation}

As mentioned above, in the derivation of (\ref{<<V1...Vn>>fixedloop4chirallysplit}) we assumed $j=j_{\rm L}=j_{\rm R}$, but in fact using the (anti-)chiral representation of amplitudes enables one to consider more general insertions for which asymptotic vertex operators can have non-trivial winding. That is, using the chirally split generating function it is almost obvious how to insert vertex operators of the form:
\begin{equation}\label{eq:ffbareikx+_}
f(\partial x_+,\partial^2x_+,\dots)\bar{f}(\bar{\partial} x_-,\bar{\partial}^2x_-,\dots)e^{i\mathbb{k}_Mx_+^M(z)}e^{i\bar{\mathbb{k}}_Mx_-^M(\bar{z})},
\end{equation}
with $\mathbb{k}_M\neq \bar{\mathbb{k}}_M$, simply by taking $j_{\rm L}$ on the {\it right-hand side} of (\ref{<<V1...Vn>>fixedloop4chirallysplit}) to be independent of $j_{\rm R}$. 
The corresponding insertion on the {\it left-hand} side of (\ref{<<V1...Vn>>fixedloop4chirallysplit}) however is not so obvious, given that here vertex operators associated to (\ref{eq:ffbareikx+_}) should be functionals of the full path integral field, $x^M(z,\bar{z})$. When $\mathbb{k}_M=\bar{\mathbb{k}}_M$, it is clear that to every vertex operator insertion (\ref{eq:ffbareikx+_}) on the right-hand side is associated a vertex operator insertion,
\begin{equation}\label{eq:ffbareikx}
f(\partial x,\partial^2x,\dots)\bar{f}(\bar{\partial} x,\bar{\partial}^2x,\dots)e^{i\frac{1}{2}(\mathbb{k}_M+\bar{\mathbb{k}}_M)x^M(z,\bar{z})},\qquad{\rm with}\qquad x^M=x_0^M+x_{\rm cl}^M(z,\bar{z})+\tilde{x}^M(z,\bar{z}),
\end{equation}
on the left-hand side with total momentum $\frac{1}{2}(\mathbb{k}_M+\bar{\mathbb{k}}_M)$. In order to extend insertions on the left-hand side to vertex operators with non-trivial winding where $\mathbb{k}_M\neq \bar{\mathbb{k}}_M$ we need to integrate over all $x^M(z,\bar{z})$ with source $j_M(z,\bar{z})$, and constrain the integration to fields with non-trivial winding. This may be achieved \cite{DijkgraafVerlindeVerlinde88} by a $j_{\rm L}$-, $j_{\rm R}$-dependent shift in the classical instanton solutions $x_{\rm cl}^M(z,\bar{z})$ of (\ref{eq:xcl sol}). Therefore, with this shift vertex operators of the form (\ref{eq:ffbareikx}) remain valid insertions even in the presence of non-trivial winding. We will not work out the details of this procedure here as there exists a simpler approach. In particular, we will instead enforce chiral splitting of the source, the prescription being the following.\footnote{The authors would like to thank Joe Polchinski for suggesting this alternative procedure.} 
\ssk

Suppose we consider an amplitude with $n$ vertex operator insertions, each of which (in the chiral representation (\ref{eq:ffbareikx+_})) carries an exponential of the form: $e^{i\mathbb{k}^{\gamma}\cdot x_+(z_{\gamma})}e^{i\bar{\mathbb{k}}^{\gamma}\cdot x_-(\bar{z}_{\gamma})}$, with $\gamma=1,\dots,n$, in addition to some polynomial of derivatives of $x_+(z)$ and $x_-(\bar{z})$.  (More generally, every vertex operator will be a superposition of such momentum eigenstates, as is the case for coherent vertex operators for instance.) The statement is that insertions with exponentials of the form:
$$
\prod_{\gamma=1}^ne^{i\mathbb{k}^{\gamma}\cdot x_+(z_{\gamma})}e^{i\bar{\mathbb{k}}^{\gamma}\cdot x_-(\bar{z}_{\gamma})},
$$
on the {\it right-hand side} of (\ref{<<V1...Vn>>fixedloop4chirallysplit}) (with $\mathbb{k}^{\gamma}_M\neq \bar{\mathbb{k}}^{\gamma}_M$ generically) 
correspond to evaluating the source on the {\it left-hand side} of (\ref{<<V1...Vn>>fixedloop4chirallysplit}) at:
\begin{equation}\label{eq:j split}
\begin{aligned}
j_M(z,\bar{z})&=j_{{\rm L}M}(z,\bar{z})+j_{{\rm R}M}(z,\bar{z})\\
&=\sum_{\gamma=1}^n\mathbb{k}_M^{\gamma}\int_{\wp}^{z_{\gamma}} du\delta^2(u-z)\partial_z+\sum_{\gamma=1}^n\bar{\mathbb{k}}_M^{\gamma}\int_{\bar{\wp}}^{\bar{z}_{\gamma}} d\bar{u}\delta^2(u-z)\partial_{\bar{z}},
\end{aligned}
\end{equation}
so that with this choice of source there exists the correspondence:
$$
e^{i\int d^2zj(z,\bar{z})\cdot x(z,\bar{z})}\,\equalhat\,\prod_{\gamma=1}^ne^{i\mathbb{k}^{\gamma}\cdot x_+(z_{\gamma})}e^{i\bar{\mathbb{k}}^{\gamma}\cdot x_-(\bar{z}_{\gamma})},
$$
{\it even though} on the left-hand side the embedding field, $x(z,\bar{z})$, contains (potentially) also instanton or soliton contributions whereas the right-hand side does not.  
So the prescription is to consider $j_M(z,\bar{z})$ on the left-hand side as an operator and act with the derivatives, $\partial_z,\partial_{\bar{z}}$, of (\ref{eq:j split}) before carrying out the line integrals. 
Using the representation for the source (\ref{eq:j split}) makes is obvious that we can simply substitute (\ref{eq:j split}) into the right-hand side of the $j=j_{\rm L}=j_{\rm R}$ expression (\ref{<<V1...Vn>>fixedloop4chirallysplit}), and then the operator nature of the decomposition (\ref{eq:j split}) will ensure that only $j_{\rm L}$ appears in chiral terms and $j_{\rm R}$ in anti-chiral terms, and so we can legitimately extend the result (\ref{<<V1...Vn>>fixedloop4chirallysplit}) to the case where $j\neq j_{\rm L}\neq j_{\rm R}$. This is the desired result. 

\ssk
Having understood how to insert vertex operators with non-trivial winding using either the chiral fixed-loop momenta or the non-chiral integrated-loop momenta representation, a crucial remark is that making use of `{\it chiral vertex operators}' (\ref{eq:ffbareikx+_}) that are constructed out of $x_{\pm}$ and correspondingly the chiral fixed-loop momenta representation of amplitudes (i.e.~working in terms of the right-hand side of (\ref{<<V1...Vn>>fixedloop4chirallysplit})) vastly simplifies computations while preserving complete generality.

\ssk
That the fixed-loop momenta generating function chirally factorises in the critical dimension is in line with the Belavin-Knizhnik theorem \cite{BelavinKnizhnik86,D'HokerPhong89} combined with the chiral splitting theorem \cite{D'HokerPhong89}, although the existing proof of chiral splitting had been established {\it explicitly} only for generic genus-$\mathfrak{g}$ massless and exponential external physical vertex operators. Here we have extended this result to {\it all} correlation functions of operators inserted on generic compact Riemann surfaces. Notice also that this statement is independent of whether the vertex operator insertions are onshell, and given that correlation functions of generic ghost insertions factorise in the same way as above, where $\mathcal{Z}_{\mathfrak{g}}$ ($\bar{\mathcal{Z}_{\mathfrak{g}}}$) may be replaced by more general superpositions of (anti-)chiral ghost correlators, we have shown that generic {\it offshell} amplitudes \cite{Sen15b} also respect chiral splitting. 

\ssk
It is worth re-emphasising that (\ref{<<V1...Vn>>fixedloop4chirallysplit}) is truly a remarkable statement, and it is due to this relation that it is justified to use vertex operators that are constructed out of the chiral fields $x_+(z)$, $x_-(\bar{z})$ (and also $c^z(z)$, $\tilde{c}^{\bar{z}}(\bar{z})$ and $b_{zz}(z),\tilde{b}_{\bar{z}\bar{z}}(\bar{z})$). To compute any string amplitude, for the matter sector we can use either vertex operators constructed out of the full path integral fields, $x(z,\bar{z})$, or the (anti-)chiral fields, $x_+(z)$, $x_-(\bar{z})$, and this choice depends on whether we want to extract correlation functions using the left-hand side of (\ref{<<V1...Vn>>fixedloop4chirallysplit}) or the right-hand side respectively. However, the natural representation for vertex operators that arises from the operator-state correspondence is in terms of the (anti-)chiral fields. Notice that we have {\it not} appealed to any onshell condition in order to split the field in the path integral $x(z,\bar{z})$ into chiral and anti-chiral pieces, $x_+(z)$, $x_-(\bar{z})$. The best way to think of the latter is as fields that arise {\it effectively} after properly taking into account all zero mode contributions (and instantons if they are present) associated to the full field $x(z,\bar{z})$.\footnote{The vertex operator construction in \cite{SklirosCopelandSaffin16bb} makes full use of the {\it (anti-)chiral} fields, $x_+(z)$, $x_-(\bar{z})$, throughout (as opposed to the path integral fields $x(z,\bar{z})$).} 
The above analysis makes it completely manifest when this is justified and why: fixing the loop momenta (in {\it both} compact and non-compact dimensions) is the key to realising these statements. Another point to emphasise is that when vertex operators have winding charges, KK charges and/or polarisations in compact directions we need {\it not} expand the fields $x_+(z)$, $x_-(\bar{z})$ that vertex operators are constructed out of around zero mode or classical instanton contributions, and in addition the simple correlators, 
\begin{equation}\label{eq:OPE'sn}
\begin{aligned}
&b_{zz}(z)\,c^{z'}(z')\sim\frac{1}{z-z'},\qquad b_{zz}(z)\,b_{z'z'}(z')\sim\mathcal{O}(z-z'),\qquad c^{z}(z)\,c^{z'}(z')\sim\mathcal{O}(z-z')\\
&\tilde{b}_{\bar{z}\bar{z}}(\bar{z})\,\tilde{b}_{\bar{z}'\bar{z}'}(\bar{z}')\sim \frac{1}{\bar{z}-\bar{z}'},\qquad \tilde{b}_{\bar{z}\bar{z}}(\bar{z})\,\tilde{b}_{\bar{z}'\bar{z}'}(\bar{z}')\sim\mathcal{O}(\bar{z}-\bar{z}'),\qquad \tilde{c}^{\bar{z}}(\bar{z})\,\tilde{c}^{\bar{z}'}(\bar{z}')\sim\mathcal{O}(\bar{z}-\bar{z}')\\
&x_+^M(z)\,x_+^N(z')\sim -\frac{\alpha'}{2}G^{MN}\ln (z-z'),\qquad x_-^M(\bar{z})\,x_-^N(\bar{z}')\sim -\frac{\alpha'}{2}G^{MN}\ln (\bar{z}-\bar{z}').
\end{aligned}
\end{equation}
are exact in the limit $z\rightarrow z'$ and should be used to carry out the operator product expansions that map states to vertex operators -- this will be discussed in more detail in \cite{SklirosCopelandSaffin16bb}. This appears to be somewhat miraculous, but it is nevertheless true (for arbitrary-genus string amplitudes). 

\ssk
These observations are of course direct generalisations of the classic result of D'Hoker and Phong \cite{D'HokerPhong89}, the differences being that here: 
\begin{itemize}
\item[{\bf (a)}] we consider generic correlation functions (rather than massless asymptotic states) associated to arbitrarily excited string vertex operators (potentially with winding and KK charges and general polarisation tensors and oscillators); 
\item[{\bf (b)}] we explicitly keep a generic constant background, $G_{MN}$, $B_{MN}$ and $\Phi$ (rather than $G_{MN}=\eta_{MN}$ and $B_{MN}=0$), so that these results hold true for generic constant target space K\"ahler and complex structure moduli, torsion and background gauge fields; 
\item[{\bf (c)}] we consider generic target spaces $\mathbb{R}^{D-1,1}\times \mathbb{T}^{D_{\rm cr}-D}$ (rather than $\mathbb{R}^{D_{\rm cr}-1,1}$), implying that there are also instanton contributions (worldsheets that wrap $\mathbb{T}^{D_{\rm cr}-D}$) that are absent in $D=D_{\rm cr}$ and that were hence not made manifest in \cite{D'HokerPhong89}. The latter were discussed in \cite{DijkgraafVerlindeVerlinde88}, building on earlier results \cite{VerlindeVerlinde87}, but the focus there was entirely on exponential insertions, and also there target space moduli were fixed and background gauge fields were absent. 
\item[{\bf (d)}] we derived these results directly without using the ``reverse engineering'' approach, as discussed in the introduction, thus eliminating the potential ambiguity of the type discussed by Sen \cite{Sen16b}.
\end{itemize}

Finally, for completeness let us discuss how to extract connected S-matrix elements.  
Given a set of $n$ external states described by general old covariant quantisation (OCQ) (possibly coherent) vertex operators 
$V_{z_{\gamma}\bar{z}_{\gamma}}$ (with $\gamma=1,2,\dots,n$), 
connected (dimensionless) S-matrix elements are extracted from (for $n\geq2$):
\begin{equation}\label{eq:S-matrix}
\begin{aligned}
\boxed{S_{fi}^C=\delta_{fi}^C+\sum_{\mathfrak{g}=0}^{\infty}\int d{\bf M}_{\mathfrak{g}}\Big\langle V_{z_1\bar{z}_1}\dots V_{z_n,\bar{z}_n}\Big\rangle}
\end{aligned}
\end{equation}
Here it is implied that vertex operators are inserted at $(z_{\gamma},\bar{z}_{\gamma})$ in $\Sigma_{\mathfrak{g}}$ (or the covering space, $\tilde{\Sigma}_{\mathfrak{g}}$, thereof, see e.g.~Fig.~\ref{Fig:CutRiemannSurface} on p.~\pageref{Fig:CutRiemannSurface}), and normalised by the leading singularity of the OPE,
$$
\overline{V_{z_{\gamma}\bar{z}_{\gamma}}}\,V_{z_{\gamma}'\bar{z}_{\gamma}'}\simeq \frac{g_D}{\sqrt{2\mathbb{k}^0V_{D-1}}}\frac{1}{|z_{\gamma}-z_{\gamma}'|^4}+\dots,
$$
where an overline denotes taking the Euclidean adjoint \cite{SklirosCopelandSaffin16bb}. It is conventional to extract out the kinematic factors and define $V_{z\bar{z}}=\frac{1}{\sqrt{2\mathbb{k}^0V_{D-1}}}\mathcal{O}_{z\bar{z}}$, so that invariant amplitudes, $\mathcal{M}_{\rm fi}(1,\dots,n)$, defined below, are most naturally written in terms of $\mathcal{O}_{z\bar{z}}$'s (recall the discussion on p.~\pageref{page:OOnorm}). 
The quantity $\delta_{fi}^C$ represents the interaction-free contribution to the connected S-matrix elements, and given that $S_{fi}^C$ only contains connected contributions $\delta_{fi}^C$ should be non-vanishing only for $n=2$ asymptotic states, because for $n>2$ the interaction-free terms cannot be connected. We have defined the measure:
\begin{equation}\label{eq:intdM}
\int d{\bf M}_{\mathfrak{g}}= \int_{\mathscr{F}_{\mathfrak{g}}}\frac{1}{N_{\mathfrak{g}}} \prod_{j=1}^{\#_{\mathbb{C}}{\rm moduli}}\!\!\!\!\!d^2\tau_j\,\int_{\Sigma_{\mathfrak{g}}}\prod_{\gamma=1}^{n-\#_{\mathbb{C}}{\rm CKVs}}d^2z_{\gamma},
\end{equation}
$N_{\mathfrak{g}}$ being the order of the unfixed global worldsheet diffeomorphisms \cite{Polchinski86}, e.g.~at $\mathfrak{g}=1$ this is $N_1=2$, corresponding to the fact that our gauge choice, $ds^2=|dz|^2$, leaves $z\rightarrow -z$ of ${\rm SL}(2,\mathbb{Z})$ unfixed (the space of global diffeomorphisms being ${\rm SL}(2,\mathbb{Z})/\mathbb{Z}_2$), see Appendix \ref{sec:T2} for further details where also our genus-one conventions are presented.\footnote{One may also extract the full S-matrix elements, $S_{\rm fi}$, (as opposed to just the connected pieces, $S_{fi}^C$) by including a summation over disconnected Riemann surfaces in the definition of the path integral, in which case everything can be cast on equal footing, but this is somewhat impractical and we will not do so here.} 

\ssk
The full S-matrix elements, $S_{\rm fi}$, are in turn extracted from products of these and sums over the various partitions, as explained, e.g., in Sec.~4.3 of \cite{Weinberg_v1}:\footnote{In the superstring (depending on the asymptotic states present) there may also be relative sign differences in the sum (\ref{eq:S=ScSc..}) due to the Grassmann nature of target space fermions.}
\begin{equation}\label{eq:S=ScSc..}
S_{\rm fi}=\sum_{\rm partitions}S_{f_1i_1}^CS_{f_2i_2}^C\dots,
\end{equation}
where the sum is (according to the cluster decomposition principle) over all  distinct partitions $\{\langle f_1|,\langle f_2|,\dots\}$ of $\langle{\rm f}|$ and over distinct partitions $\{|i_1\rangle,|i_2\rangle,\dots\}$ of $|{\rm i}\rangle$, with the ``incoming'' states, $|i_{\gamma}\rangle$ associated to vertex operators $V_{z_{\gamma}\bar{z}_{\gamma}}$, and the ``outgoing'' states $\langle f_{\gamma}|$ associated to Euclidean adjoints, $\overline{V_{z_{\gamma}\bar{z}_{\gamma}}}$. Our conventions are such that $|S_{\rm fi}|^2$ is interpreted as a transition probability associated to going from $|{\rm i}\rangle$ to $\langle{\rm f}|$, 
\begin{equation}\label{eq:Prob|Sfi|^2*}
{\rm Prob}({\rm f}\leftarrow {\rm i})=|S_{\rm fi}|^2,
\end{equation}
whereas S-matrix unitarity corresponds to the statements:
\begin{equation}\label{eq:SSdagger=1xx}
\sum_{\rm h}S^{\dagger}_{\rm hf}S_{\rm hi}=\delta_{\rm fi},\qquad {\rm or}\qquad \sum_hS_{\rm fh}S^{\dagger}_{\rm ih}=\delta_{\rm fi}.
\end{equation}
The precise interpretation of the sum over states, $\sum_{\rm h}$, and also of the delta function, $\delta_{\rm fi}$, requires specifying a basis (and for coherent vertex operators in particular an {\it overcomplete} basis) and will be discussed elsewhere \cite{SklirosCopelandSaffin16bb}. 

Note that (even when both `${\rm f}$' and `${\rm i}$' represent multi-string states) there are generically \cite{Weinberg_v1} also vacuum-to-vacuum contributions in this partitioning \cite{Polchinski94,Green95}, denote these by $S_{00}^C$, as well as explicit tadpole contributions, $S_{0i}^C$ (and/or $S_{f0}^C$) if `$i$' (and/or `$f$') are single string states, in addition to implicit ones (that arise in various regions of the boundary of moduli space where internal lines are forced to lie on the mass shell) that may already be present in $S_{fi}^C$. Summing over distinct partitions in (\ref{eq:S=ScSc..}) shows that the former exponentiate, so there is \cite{Green95} an overall factor $e^{S_{00}^C}$ in $S_{\rm fi}$, and this is analogous to the exponentiation of the D-instanton amplitude in \cite{Polchinski94,Green95}, ultimately suggesting a breakdown of the world-sheet in that context. 

For instance, generic $n=2$-point S-matrix elements are of the form, 
$$
S_{\rm ii}=e^{S_{00}^C}\big(S_{ii}^C+S_{i0}^CS_{0i}^C\big).
$$ 
For $n=2$ (and $n=3$) there is therefore (up to the overall universal factor $e^{S_{0,0}^C}$) no distinction between the two sets of S-matrix elements, $S_{\rm ii}$ and $S_{ii}^C$, in the absence of tadpoles, $S_{i0}^C=0$, as only connected diagrams exist, but for $n>3$ there is a distinction. The tadpole contributions, $S_{0i}^C$, and the vacuum-to-vacuum contribution, $e^{S_{00}^C}$, are pathological in the bosonic string (due to the presence of a tachyon in the spectrum and also massless tadpoles) and these will be absent in the superstring (when the vacuum of interest is stable under quantum corrections).

Finally, let us also note that for momentum eigenstates and when $G_{\mu\nu}=\eta_{\mu\nu}$ it is conventional to extract out the kinematic factors and a momentum conserving delta function and define an invariant amplitude, $\mathcal{M}_{\rm fi}(1,\dots,n)$, as follows,
\begin{equation}\label{eq:SfiM}
\begin{aligned}
S_{\rm fi}
&=\delta_{\rm fi}+i\deltaslash(j)\frac{\mathcal{M}_{\rm fi}(1,\dots,n)}{\sqrt{2\mathbb{k}^0_1V_{D-1}\dots 2\mathbb{k}^0_nV_{D-1}}},
\end{aligned}
\end{equation}
The argument of the delta function, e.g.~$\bar{\deltaslash}(j)\dfn (2\pi)^D\delta^D(\smallint j^{\mu})\delta_{(\smallint j_a\ell_s),0}^{D_{\rm cr}-D}$, see (\ref{eq:bardelta-Smatrix}), enforces momentum conservation, as well as conservation of any other charges (such as KK and winding charges) that may be present in the external states, but note that for coherent vertex operators there will be a sum over such delta function contributions. Factorisation, normalisation and unitarity of string amplitudes (with coherent vertex operator insertions) and related concepts will be discussed in \cite{SklirosCopelandSaffin16cc} where we focus on $n=2$. 

The $n$ external states are assumed to have well-defined energy expectation values\footnote{This is the case for coherent vertex operators as well as for mass eigenstates. Alternatively, we can also switch to light-cone coordinates whereby the kinematic factor $\mathbb{k}^0V_{D-1}$ is replaced by $p^+\mathcal{V}_{D-1}$, and that coherent vertex operators {\it are} eigenstates of $\hat{\mathbb{P}}^+$ but not of $\hat{\mathbb{P}}^0$, making it natural to adopt the latter kinematic factor. More about these details will appear elsewhere \cite{SklirosCopelandSaffin16bb}.} denoted by $\mathbb{k}^0_{\gamma}$, for $\gamma=1,\dots,n$, and $V_{D-1}\dfn (2\pi)^{D-1}\delta^{D-1}(0)$ denotes the formal (infinite) spatial volume of $\mathbb{R}^{D-1}$. 
Another point to emphasise is that (as mentioned above) the formal volume $V_{D-1}$ will always cancel out and does not appear in the observables of interest (cross sections, decay rates, etc.), just as in field theory \cite{LandauLifshitzRQF}. Finally, generically there will be additional delta-function (or Kronecker-delta) constraints (implicit in $\mathcal{M}_{\rm fi}(1,\dots,n)$) in addition to that appearing explicitly in (\ref{eq:SfiM}), associated to the fact that the full invariant amplitude also contains disconnected pieces, i.e.~if $n>3$, depending on context, as exhibited in (\ref{eq:S=ScSc..}).

\section{Discussion}
We have constructed a generating function (and associated correlation functions) for string amplitudes in generic constant string backgrounds, $G_{MN}$, $B_{MN}$, $\Phi$ and $U$, on $\mathbb{R}^{D-1,1}\times \mathbb{T}^{D_{\rm cr}-D}$, so that also {\it all} K\"ahler and complex structure moduli (of the target space torus, $\mathbb{T}^{D_{\rm cr}-D}$) contained in $G_{ab},B_{ab}$, background KK gauge fields, $A_{\mu}^a$ and $B_{\mu a}$, spacetime torsion, $B_{\mu\nu}$ and also spacetime metric, $G_{\mu\nu}$, are allowed to be turned on. In the process, we have derived the chiral splitting theorem of D'Hoker and Phong  \cite{D'HokerPhong89}  for string amplitudes, which we have generalised to the aforementioned background and with arbitrarily excited string vertex operator insertions (with generic KK and winding charges, as well as polarisation tensors associated to generic oscillators and spacetime indices\footnote{And hence the result also applies to generic {\it coherent vertex operators} as we explain in \cite{SklirosCopelandSaffin16bb,SklirosCopelandSaffin16cc}.}).  

\ssk
Our approach differs from that of D'Hoker and Phong \cite{D'HokerPhong89} (and also Sen \cite{Sen16b}), in that we did not make use of the ``reverse engineering'' approach (where the target spacetime embedding fields, $x^M(z,\bar{z})$, are first integrated out and only at a later stage of the computation is it noted that the result can be written as an integral whose integration variables get interpreted as $A_I$-cycle loop momenta). As pointed out in a recent paper by Sen \cite{Sen16b}, such a reverse engineering approach could potentially lead to ambiguities (because the same integrated-loop momenta amplitude can be written in more than one way as an integral over loop momenta \cite{Sen16b}). Sen went on to explain that these ambiguities will not be visible in the final amplitudes after integrating out the loop momenta (while adopting an appropriate analytic continuation for the loop momentum integral contours \cite{PiusSen16}), and that these ambiguities are therefore immaterial. However, as discussed in the Introduction above, it is sometimes desirable to not integrate out the loop momenta, and that this is also of interest for the computation of some physical observables, such as the spectrum of massless radiation associated to a decaying string. Therefore, the reverse engineering method could potentially lead to ambiguous results for observables. In our approach we have resolved this potential ambiguity, in that we introduced loop momenta associated to $A_I$-cycle strings from the outset (by explicit momentum conserving delta function insertions into the original path integral where there is no room for this ambiguity), and have thus shown that the result of D'Hoker and Phong (that one can replace the target space fields, $x^M(z,\bar{z})$, by a set of effective chiral fields, $x_+^M(z)$, $x_-^M(\bar{z})$ for the left- and right-moving degrees of freedom, appropriately modified so as to apply to generic backgrounds, $\mathbb{R}^{D-1,1}\times \mathbb{T}^{D_{\rm cr}-D}$, and vertex operators) is fully justified and leads to the correct un-ambiguous result for the fixed-loop momentum amplitudes.\footnote{Note that there are still expected to be field-redefinition ambiguities that one expects from insight from string field theory, see Sec.~4 in \cite{Sen16b}. The authors thank Ashoke Sen for an extensive discussion of this point.}

\ssk
Let us now zoom in on the statement (\ref{<<V1...Vn>>fixedloop4chirallysplit}). Here it is crucial to note that the left-hand side denotes the usual path integral over matter, $x^M(z,\bar{z})$, and ghost fields, $b,c$, whereas on the right-hand side the matter and ghost fields have been integrated out, and the result has been written in terms of Wick contractions of effective (anti-)chiral fields, $x_+^M(z)$, ($x_-(\bar{z})$), whose correlation functions are determined from the chiral propagators (\ref{eq:chiral<xx>}), with the results given in (\ref{eq:<>_+}) and (\ref{eq:<>_-}). What we want to emphasise here is that on the left-hand side of (\ref{<<V1...Vn>>fixedloop4chirallysplit}) the target space embedding field appearing in vertex operators and the worldsheet action contains (generically) zero modes, instanton (or soliton) contributions, as well as quantum fluctuations, whereas the chiral fields on the right hand side are defined by their correlation functions, so that $x_{\pm}^M$ do {\it not} contain information about zero modes or instanton contributions. The latter have nevertheless been fully taken into account and appear in the overall delta function and loop momenta respectively. Therefore, using the chiral representation of amplitudes significantly simplifies amplitude computations. 
\ssk

Finally, we have also discussed how wave/particle (or rather wave/string) duality is manifested in string theory, and we have shown that the fixed-loop momenta representation can be thought of as the `wave picture', the integrated loop momenta expression yielding the `string picture'. There are also hybrid formulations (or Routhians) whereby the compact and non-compact dimensions are in the wave or string picture, leading to four natural possibilities in total. In a forthcoming article \cite{SklirosCopelandSaffin16dd} we will show that adopting a wave picture leads to significant simplifications and explicit analytic results (a string picture being much less tractable analytically). 
\ssk

The objective here has been to provide a working and efficient handle on computing string amplitudes involving HES vertex operators. In \cite{SklirosCopelandSaffin16bb} we construct chiral HES coherent vertex operators (which is a very natural basis for excited strings) and discuss the notion of Euclidean adjoint vertex operators (which refines the rule of thumb of Polchinski \cite{Polchinski88,Polchinski_v1}, a refinement that is necessary in order for {\it all} vertex operators to have positive norm\footnote{The rule of thumb \cite{Polchinski88,Polchinski_v1} that to obtain the Euclidean adjoint of a vertex operator one is to conjugate all explicit factors of `$i=\sqrt{-1}$' is not sufficient when vertex operators have winding $N-\bar{N}\in 2\mathbb{Z}+1$, in that there are some additional phases (here $N,\bar{N}$ are level numbers).}). These vertex operators are then \cite{SklirosCopelandSaffin16cc} used to derive a generic expression for two-point amplitudes (where we keep the genus of the worldsheet generic in order to study generic properties), whose imaginary part at one loop \cite{SklirosCopelandSaffin16dd} yields decay rates and power emitted into massless and massive radiation (including radiative backreaction and in particular $\alpha'$ corrections), the real part giving mass shifts (relevant for black hole physics \cite{DamourVeneziano00}). In \cite{SklirosCopelandHindmarshSaffin16dd2} we discuss decay rates associated to gravitational radiation in particular and in \cite{SklirosCopelandSaffin16ee} we make the connection to low energy effective field theory. 

\section*{Acknowledgements}
DPS would like to first and foremost thank Joseph Polchinski for numerous insightful discussions and guidance that ultimately made this work possible. In addition, he would like to thank Mark Hindmarsh, Eric D'Hoker, Erik Verlinde, for various discussions and insights, and Jorge Russo and especially Ashoke Sen for extensive correspondence and comments on the draft. Last but not least DPS would like to thank Lara Callegari for technical help with the figures. 
The research of EC, PMS and DPS was supported partly by the STFC consolidated grant No.~ST/L000393/1, and DPS was further supported by an Advanced Research Fellowship of Anastasios Avgoustidis at the University of Nottingham, the London Centre for Terauniverse Studies (LCTS), using funding from the European Research Council via the Advanced Investigator Grant 26732, and a visiting fellowship at Institut des Hautes \'Etudes Scientifiques (IH\'ES). 

\appendix

\section{Conventions}\label{sec:TFFFPF}

\subsection{Complex Tensors and Riemann Surfaces}\label{sec:RS}
In this subsection we collect some useful formulas and conventions used in the main text, on the local and global properties of compact Riemann surfaces, $\Sigma_{\mathfrak{g}}$. We will be completely explicit, because although we largely adopt the Polchinski conventions \cite{Polchinski_v1}, we follow the approach of D'Hoker and Phong \cite{DHokerPhong} who use slightly different conventions.

Focusing on a local patch of the worldsheet, for a given set of real coordinates $(x,y)$ we define a complex set $(z,\bar{z})$ by $z=x+iy$, $\bar{z}=x-iy$, with $\partial_z=\frac{1}{2}(\partial_x-i\partial_y)$, $\partial_{\bar{z}}=\frac{1}{2}(\partial_x+i\partial_y)$. We use the following convention throughout, 
$$
d^2z \equiv idz\wedge d\bar{z}=2dx\wedge dy,
$$ 
Two-dimensional Riemannian manifolds are conformally flat, $g=g_{z\bar{z}}(dz\otimes d\bar{z}+d{\bar z}\otimes dz)$, see e.g.~\cite{Nakahara03}, and it is useful to note that $\sqrt{g}g^{z\bar{z}}=1$. The corresponding Ricci scalar in our conventions ($+\!+\!+$ in the classification of Misner, Thorne and Wheeler \cite{MisnerThorneWheeler74}) reads:
\begin{equation}\label{eq:R}
\begin{aligned}
R_{(2)} &= g^{\alpha\beta}R_{\alpha\beta}
=2g^{z\bar{z}}R_{z\bar{z}}
=2g^{z\bar{z}}R^{\alpha}_{\phantom{a}z\alpha \bar{z}}=2g^{z\bar{z}}R^{z}_{\phantom{a}zz \bar{z}},
\end{aligned}
\end{equation}
where the components of Riemann curvature tensor in terms of the Christofel symbol read,
$$
R^{\alpha}_{\phantom{a}\beta\gamma\delta} = \partial_{\gamma}\Gamma^{\alpha}_{\beta\delta}-\partial_{\delta}\Gamma^{\alpha}_{\beta\gamma}+\Gamma^{\alpha}_{\sigma\gamma}\Gamma^{\sigma}_{\beta\delta}-\Gamma^{\alpha}_{\sigma\delta}\Gamma^{\sigma}_{\beta\gamma},
$$
and in the above coordinate system the only non-vanishing Christofel symbols are $\Gamma^z_{zz}=\partial_z\ln g_{z\bar{z}}$ and $\Gamma^{\bar{z}}_{\bar{z}\bar{z}}=\partial_{\bar{z}}\ln g_{z\bar{z}}$,
\begin{equation}\label{eq:Rzzzzbar}
\begin{aligned}
R^{z}_{\phantom{a}zz \bar{z}}& = -\partial_{\bar{z}}\Gamma^z_{zz}\\
&=-\partial_{\bar{z}}\partial_z\ln g_{z\bar{z}},
\end{aligned}
\end{equation}
so that:
$$
R_{(2)} = 2g^{z\bar{z}}(-\partial_{\bar{z}}\partial_z\ln g_{z\bar{z}}).
$$

A tensor $V$ of conformal weight $(h,\bar{h})$ is of the form:
\begin{equation}\label{eq: tensor V}
V=V_{ z\dots z\bar{z}\dots\bar{z}}(dz)^{h}(d\bar{z})^{\bar{h}}\in K^{(h,\bar{h})}
\end{equation}
so that $K^{(h,\bar{h})}$ is the space of tensors of weight $(h,\bar{h})$ and spin $h-\bar{h}=\frac{1}{2}\mathbb{Z}$. The components of $V$  are sometimes referred to as {\it conformal primary operators}. 
Examples used in the main text are:
\begin{equation}
\begin{aligned}
&x(z,\bar{z})\in K^{(0,0)}\\
&c=c^z(dz)^{-1}\in K^{(-1,0)}\\
&b=b_{zz}(dz)^2\in K^{(2,0)}\\
&\mu=\mu_{\bar{z}}^{\phantom{a}z}(dz)^{-1}d\bar{z}\in K^{(-1,1)}
\end{aligned}
\end{equation}
Define $K^{(n,0)}\equiv K^{n}$ (and $K^{(0,n)}\equiv \bar{K}^n$). Using the metric $g_{z\bar{z}}$ to raise and lower indices there is an isomorphism $(n-m,0)\sim (n,m)\sim (0,m-n)$, and one may therefore express all tensors in terms of holomorphic indices, e.g. we write, $g^{z\bar{z}}V_{\bar{z}}=V^{z}$, with $g^{z\bar{z}}g_{z\bar{z}}=1$. Covariant derivatives satisfy\footnote{We occasionally drop the index `$(n)$' from covariant derivatives when there is no ambiguity about the type of tensor it acts upon.} $\nabla^{(n)}_z: K^{n}\rightarrow K^{n+1}$,
\begin{equation}\label{eq: nabla V}
\begin{aligned}
\nabla^{(n)}_zV &= (\partial_z-n\Gamma_{zz}^z)V\otimes dz.
\end{aligned}
\end{equation}
It is straightforward to show, using the explicit expression for the Christoffel symbols above (\ref{eq:Rzzzzbar}), that (\ref{eq: nabla V}) is equivalent to:
$$
\nabla_z^{(n)}V=g_{z\bar{z}}^n\partial_z\big(g_{z\bar{z}}^{-n}V\big)\otimes dz.
$$
In addition, there is the Cauchy-Riemann operator $\partial_{\bar{z}}$; formally $\nabla_{\bar{z}}^n: K^{n}\rightarrow K^{n,1}$,
\begin{equation}\label{eq: Cauchy-Riemann operator}
\nabla^{(n)}_{\bar{z}}V = \partial_{\bar{z}}V\otimes d\bar{z}.
\end{equation}
According to the above identification we could also have written the Cauchy-Riemann operator as $\nabla^z_{(n)}: K^{n}\rightarrow K^{n-1}$,
\begin{equation}\label{eq: Cauchy-Riemann operator}
\nabla^z_{(n)}V = g^{z\bar{z}}\partial_{\bar{z}}V\otimes (dz)^{-1}.
\end{equation}
We shall not always display the differentials $dz$ ($d\bar{z}$) in $\nabla_z$ ($\nabla_{\bar{z}}$) but include them in the definitions in order to make their transformation properties clear.

The natural inner product between tensors $V_{1,2}\in K^n$ with respect to the metric $g$ is
\begin{equation}\label{eq: (V1,V2)}
\langle V_1,V_2\rangle = \int_{\Sigma}d^2z\sqrt{g}\,(g^{z\bar{z}})^n\,V_1^*\,V_2,
\end{equation}
and we define the adjoint operators $\nabla^{(n)\dagger}_z$ and $\nabla^{z\dagger}_{(n)}$ with respect to this, $\langle V_1,\nabla^{(n)\dagger}_zV_2\rangle \equiv \langle \nabla^{(n)}_zV_1, V_2\rangle$. When $V_1=V_2$ we also write $\| V\|^2=\langle V,V\rangle $. Using the definitions it follows that
\begin{equation}
\nabla^{(n)\dagger}_z=-\nabla_{(n+1)}^z,\qquad \nabla_{(n)}^{z\dagger}=-\nabla^{(n-1)}_z.
\end{equation}
We can construct two, in general distinct, Laplacians using the differential operators (\ref{eq: nabla V}) and (\ref{eq: Cauchy-Riemann operator})
\begin{equation}\label{eq: Laplacians}
\begin{aligned}
\Delta_{(n)}^{+}&=-2\nabla_{(n+1)}^{z}\nabla_z^{(n)}\\
\Delta_{(n)}^{-}&=-2\nabla^{(n-1)}_{z}\nabla^z_{(n)},
\end{aligned}
\end{equation}
and so from $[\nabla_z,\nabla_{\bar{z}}]V_{zz\dots}=nR^z_{\phantom{a}zz\bar{z}}V_{zz\dots}$ (for $V\in K^n$) and (\ref{eq:R}) it follows that $\Delta_{(n)}^{+}-\Delta_{(n)}^{-}=nR_{(2)}$. 
Therefore, these two Laplacians are equal when acting on scalars (where $n=0$) or when $R_{(2)}=0$. In the former case we define $\Delta_{(0)}\equiv\Delta_{(0)}^{+}=\Delta_{(0)}^{-}$. The factor of $-2$ in the definitions (\ref{eq: Laplacians}) is conventional and is included so as to agree with the definition of the conventional Laplacian $\Delta_{(0)}=-\frac{1}{\sqrt{g}}\partial_{\alpha}\sqrt{g}g^{\alpha\beta}\partial_{\beta}$. In particular, in the $z,\bar{z}$ coordinates $\Delta_{(0)}=-2g^{z\bar{z}}\partial_z\partial_{\bar{z}}$, in agreement with both $\Delta_{(0)}^+$ and $\Delta_{(0)}^-$. 

\ssk
The string embedding $x^M(z,\bar{z})$ is a scalar from the 2-dimensional point of view. Its derivatives are tensor fields in the sense of (\ref{eq: tensor V}). In accordance with (\ref{eq: nabla V}) and (\ref{eq: Cauchy-Riemann operator}) we write: 
$$
\nabla^{(0)}_z x\equiv \partial_zx\,dz\equiv \partial x, \qquad \nabla^{(0)}_{\bar{z}}x\equiv \partial_{\bar{z}}x\,d\bar{z}\equiv \bar{\partial}x,\qquad{\rm and}\qquad dx\equiv \partial x+\bar{\partial}x.
$$
In particular, $\partial x=\partial_zxdz$ is a tensor of weight $(1,0)$, and using the derivatives (\ref{eq: nabla V}) one can form tensors of weight $(\ell,0)$ as follows $\nabla_z^{(\ell-1)}\dots\nabla_z^{(1)}(\partial x)$. In practice we write this as $\nabla_z^{\ell-1}\partial_z x$ and may not in general (as mentioned above) display the differentials. In the main text, the $\Gamma_{zz}^z$ dependence will always drop out (due to Weyl invariance) and we shall write instead $\partial^l_zx$ when there is no ambiguity, and likewise for the anti-holomorphic counterpart. 

\ssk
We now move on to discuss certain global topological aspects of Riemann surfaces. 
A key relation is the Atiyah-Singer-Riemann-Roch index theorem:
\begin{equation}\label{eq:Atiyah-Singer}
{\rm dim}_{\mathbb{C}}\,{\rm ker}\,\nabla_z^{(n)}-{\rm dim}_{\mathbb{C}}\,{\rm ker}\,\nabla_{(n+1)}^z = \frac{1}{2}(2n+1)\chi(\Sigma_{\mathfrak{g}}),
\end{equation}
and this relates the number of zero modes of tensors in $K^{n}$, tensors in $K^{n+1}$, (for $n\in \frac{1}{2}\mathbb{Z}$) and the Euler characteristic, $\chi(\Sigma_{\mathfrak{g}})$, of the Riemann surface. For compact Riemann surfaces the latter reads:
\begin{equation}\label{eq:chi}
\chi(\Sigma_{\mathfrak{g}}) =\frac{1}{2\pi}\int_{\Sigma_{\mathfrak{g}}} d^2zR_{z\bar{z}}= 2-2\mathfrak{g}.
\end{equation}

Following D'Hoker and Phong \cite{DHokerPhong} (see also \cite{AlvarezGaumeNelson}), we parametrise the genus-$\mathfrak{g}$ compact Riemann surface, $\Sigma_{\mathfrak{g}}$, by choosing a canonical intersection basis for the $2\mathfrak{g}$ homology cycles of the associated {\it first homology group}, $H^1(\Sigma_{\mathfrak g},\mathbb{Z})=\mathbb{Z}^{2\mathfrak{g}}$,
\begin{equation}\label{eq:intersectionbasis}
\# (A_I,A_J) = \#(B_I,B_J)=0,\qquad \# (A_I,B_J) = -\#(B_I,A_J)=\delta_{I,J},\qquad I,J=1,\dots,\mathfrak{g},
\end{equation}
and denote the dual $2\mathfrak{g}$ holomorphic 1-forms by, $\omega_I=\omega_I(z)dz$, $\bar{\omega}_I=\bar{\omega}_I(\bar{z})d\bar{z}$, whose existence is guaranteed by the index theorem (\ref{eq:Atiyah-Singer}): taking $n=0$ and noting that ${\rm dim}_{\mathbb{C}}\,{\rm ker}\,\nabla_z^{(0)}=1$ yields the result of interest, 
$
{\rm dim}_{\mathbb{C}}\,{\rm ker}\,\nabla_z^{(1)}=\mathfrak{g}.
$ 
We normalise these by the duality relation in the usual manner,
\begin{equation}\label{eq:omegaA}
\oint_{A_I}\omega_J = \delta_{IJ},
\end{equation}
and define the period matrix, $\Omega_{IJ}$, by:
\begin{equation}\label{eq:omegaB}
\oint_{B_I}\omega_J \equiv \Omega_{IJ}.
\end{equation}
This has the properties $\Omega_{IJ}=\Omega_{JI}$, ${\rm Im}\,\Omega_{IJ}> 0$, which in turn follow from the Riemann bilinear identity, 
\begin{equation}\label{eq:RiemannBilinear}
\int_{\Sigma_{\mathfrak{g}}} \omega\wedge \eta = \sum_{I=1}^{\mathfrak{g}} \oint_{A_I}\omega\oint_{B_I}\eta-\oint_{B_I}\omega\oint_{A_I}\eta,
\end{equation}
for any closed 1-forms $\omega,\eta$ (in the absence of poles \cite{LugoRusso89}), 
a useful corollary of which is,
\begin{equation}\label{eq:oo}
i\int_{\Sigma_{\mathfrak{g}}}\omega_I\wedge \bar{\omega}_J=2({\rm Im}\Omega)_{IJ}.
\end{equation}
The space $\mathcal{H}_{\mathfrak{g}}=\{\Omega\in \mathbb{C}^{\mathfrak{g}}\,|\,\Omega_{IJ}=\Omega_{JI},\,\,{\rm Im}\Omega>0\}$ is the Siegel upper half space.

Fixing the loop momenta in amplitudes breaks {\it manifest} modular invariance, but of course integrating out the loop momenta restores it. In order to keep track of this, let us briefly mention how modular transformations act on the various ingredients that appear in amplitudes \cite{VerlindeVerlinde87}. 

Modular transformations act on the canonical basis $A_I$, $B_I$ as follows:
\begin{equation}\label{eq:A_I'B_I'}
A'_I=D_{IJ}A_J+C_{IJ}B_J,\qquad B'_I=B_{IJ}A_J+A_{IJ}B_J.
\end{equation}
The primed quantities satisfy (\ref{eq:intersectionbasis}) provided the $2\mathfrak{g}\times2\mathfrak{g}$ matrix $\bigl(\begin{smallmatrix} A&B\\ C&D \end{smallmatrix}\bigr)$ is an element of the symplectic (or modular) group ${\rm Sp}(2\mathfrak{g},\mathbb{Z})$: 
$$
\biggl(\begin{matrix} A\,&B\\ C\,&D \end{matrix}\biggr)^T\biggl(\begin{matrix} 0&1\\ -1&0 \end{matrix}\biggr)\biggl(\begin{matrix} A\,&B\\ C\,&D \end{matrix}\biggr)=\biggl(\begin{matrix} 0&1\\ -1&0 \end{matrix}\biggr),
$$
as can be explicitly verified. These transformations generate $2\pi$ twists (or Dehn twists) around the $A_I$ and $B_I$ cycles and generate the group ${\rm Diff}_{\rm gl}(\Sigma_{\mathfrak{g}})$ of global diffeomorphisms that are not connected to the identity. The abelian differentials and period matrix, see (\ref{eq:omegaA}) and (\ref{eq:omegaB}), in turn transform under modular transformations (\ref{eq:A_I'B_I'}) according to:
\begin{equation}\label{eq:o'oO'O}
\begin{aligned}
& \omega_I'=\omega_J\big(C\Omega+D\big)^{-1}_{JI},\\
&\Omega_{IJ}'=\big(A\Omega+B\big)_{IK}\big(C\Omega+D\big)_{KJ}^{-1}.
\end{aligned}
\end{equation}
The first of these follows from requiring that $\oint_{A_I} \omega_J=\delta_{IJ}$ remains invariant, whereas the second follows from the first and the definition $\oint_{B_I'}\omega_J'\equiv \Omega_{IJ}'$, but see also \cite{AlvarezGaumeNelson,HamidiVafa87}. Note that $\Omega_{IJ}'$ is also an element of $\mathcal{H}_{\mathfrak{g}}$ when $\Omega_{IJ}$ is. 

Period matrices related as in (\ref{eq:o'oO'O}) refer to the {\it same} Riemann surface, but in fact restricting to the quotient $\mathcal{H}_{\mathfrak{g}}/{\rm Sp}(2\mathfrak{g},\mathbb{Z})$ is still a redundant description of the moduli space,  $\mathscr{F}_{\mathfrak{g}}$, which is contained in $\mathcal{H}_{\mathfrak{g}}/{\rm Sp}(2\mathfrak{g},\mathbb{Z})$ in a rather complicated manner for generic genus $\mathfrak{g}$ surfaces, see e.g.~\cite{Mulase83,Mulase84} for a detailed discussion and \cite{HarrisMorrison} for a broader overview, and also \cite{AlvarezGaumeNelson,DHokerPhong} for discussions with a physics-motivated approach. A detailed discussion of the moduli space would take us far afield, but it is useful to always keep in mind the physical picture whereby different points in $\mathscr{F}_{\mathfrak{g}}$ correspond to distinct deformations of the Riemann surface (i.e.~that cannot be undone by using a symmetry transformation, namely global and local diffeomorphisms and Weyl transformations of $\Sigma_{\mathfrak{g}}$), whereas the boundary of moduli space (upon compactification, $\mathscr{F}_{\mathfrak{g}}\rightarrow \bar{\mathscr{F}}_{\mathfrak{g}}$) can be identified with the set of degenerations whereby one or more isotopically distinct cycles in $\Sigma_{\mathfrak{g}}$ (with cycles encircling vertex operator insertions considered non-trivial) are shrunk to points. 

Given any base point $\wp_0$ we may associate to every point $\wp$ on $\Sigma$ a complex $\mathfrak{g}$-component vector  ${\bf z}$ by the {\it Jacobi map} (referred to also as the {\it Abel map})\index{Jacobi map}\index{Abel map}:
\begin{equation}\label{eq: Abel map}
\mathbb{I}:\wp\rightarrow {\bf z}(\wp)=\left(\int_{\wp_0}^{\wp}\omega_1,\dots,\int_{\wp_0}^{\wp}\omega_{\mathfrak{g}}\right).
\end{equation}
This vector is unique up to periods (\ref{eq:omegaA}), (\ref{eq:omegaB}). We associate to $\Omega$ a lattice $L_{\Omega}\subset \mathbb{C}^{\mathfrak{g}}$, such that $L_{\Omega}\equiv \mathbb{Z}^{\mathfrak{g}}+\Omega\mathbb{Z}^{\mathfrak{g}}$. The vector ${\bf z}$ is an element of the complex torus $J(\Sigma)$, also known as the {\it Jacobian variety} of $\Sigma$,
\begin{equation}\label{eq:Jacobian variety}
J(\Sigma)\equiv  \mathbb{C}^{\mathfrak{g}}/L_{\Omega}=\mathbb{C}^{\mathfrak{g}}/(\mathbb{Z}^{\mathfrak{g}}+\Omega\mathbb{Z}^{\mathfrak{g}}).
\end{equation}

We next discuss Riemann theta functions and the prime form, both of which are fundamental in the construction of correlation functions on Riemann surfaces. We will again present only the essential material required to follow the main text, given that all of this material is very lucidly explained in \cite{Mumford_v12,Fay} and we refer the reader to these references for detailed proofs; see also \cite{DHokerPhong,VerlindeVerlinde87} for a more concise overview. The {\it Riemann theta function}\index{theta function,Riemann theta function}, associated to $\Omega$\footnote{$\Omega$ need not be identified with the Riemann surface period matrix in the definition of $\theta({\bf z},\Omega)$ but we shall do so.} is then defined for $({\bf z},\Omega)\in \mathbb{C}^{\mathfrak{g}}\times\mathcal{H}_{\mathfrak{g}}$ by,
\begin{equation}\label{eq: Riemanntheta}
\vartheta\left({\bf z},\Omega\right) \equiv  \sum_{n\in \mathbb{Z}^{\mathfrak{g}}}\exp \bigg{\{}2\pi i\left(\frac{1}{2}n^T\Omega n+n^T{\bf z}\right)\bigg{\}}.\qquad{\rm (Riemann\,\,theta\,\,function)}
\end{equation}
We first note that $\vartheta\left({\bf z},\Omega\right)$ defines a \cite{Mumford_v12} {\it holomorphic} function on $\mathbb{C}^{\mathfrak{g}}\times\mathcal{H}_{\mathfrak{g}}$. Secondly, it is quasi-periodic (periodic up to a multiplicative factor) with respect to lattice translations, ${\bf z}\rightarrow {\bf z}+c$, with $c\in L_{\Omega}$, and is invariant under parity ${\bf z}\rightarrow-{\bf z}$:
\begin{subequations}\label{eq: Riemanntheta properties}
\begin{align}
\vartheta\left({\bf z} + m+\Omega n,\Omega\right)&= \exp \bigg{\{}2\pi i\left(-\frac{1}{2}n^T\Omega n-n^T{\bf z}\right)\bigg{\}}\vartheta\left({\bf z},\Omega\right)\qquad ({\rm translations})\label{eq: Riemanntheta translations}\\
\vartheta\left({\bf z},\Omega\right) &= \vartheta\left(-{\bf z},\Omega\right)\qquad ({\rm parity})\label{eq: Riemanntheta parity}
\end{align}
\end{subequations}
where $n,m\in \mathbb{Z}^{\mathfrak{g}}$. Notice that the RHS of (\ref{eq: Riemanntheta translations}) is independent of $m$, thus implying that the Riemann theta function is invariant under integer shifts ${\bf z}\rightarrow{\bf z} + m$. 

The theta function satisfies a ``heat equation'',
\begin{equation}
\frac{\partial\vartheta({\bf z},\Omega)}{\partial \Omega_{IJ}}=\frac{1}{2\pi i}\frac{\partial^2\vartheta({\bf z},\Omega)}{\partial z_I\partial z_J}\times\left\{ \begin{array}{l}
1\,\,\quad{\rm for}\quad I\neq J\\
\frac{1}{2}\,\,\quad{\rm \texttt{"}}\,\,\quad I=J
\end{array} \right. \qquad({\rm heat\,\,equation})
\end{equation}
where $z_I$, for $I=1,\dots,\mathfrak{g}$, denote the components of the vector ${\bf z}$.

We also need the notion of a {\it Riemann theta function with (rational) characteristics}, $[\begin{smallmatrix}a\\ b\end{smallmatrix}]$, defined by:
\begin{equation}\label{eq: Riemannthetacharact}
\vartheta[\begin{smallmatrix}a\\ b\end{smallmatrix}]\left({\bf z},\Omega\right) \equiv  \sum_{n\in \mathbb{Z}^{\mathfrak{g}}}\exp \bigg{\{}2\pi i\left(\frac{1}{2}(n+a)^T\Omega (n+a)+(n+a)^T({\bf z}+b)\right)\bigg{\}},\quad\forall\,a,b\in \mathbb{Q}^{\mathfrak{g}}.
\end{equation}
This is also quasiperiodic with respect to lattice translations ${\bf z}\rightarrow {\bf z}+c$, with $c\in L_{\Omega}$,
\begin{equation}
\vartheta[\begin{smallmatrix}a\\ b\end{smallmatrix}]({\bf z} + m+\Omega n,\Omega)= e^{2\pi i \left(a^Tm-b^Tn\right)}
\exp \bigg{\{}2\pi i\left(-\frac{1}{2}n^T\Omega n-n^T{\bf z}\right)\bigg{\}}\vartheta[\begin{smallmatrix}a\\ b\end{smallmatrix}]\left({\bf z},\Omega\right).
\end{equation}
In terms of the Riemann theta function,
\begin{equation}\label{eq: Riemannthetacharact_intermsof}
\vartheta[\begin{smallmatrix}a\\ b\end{smallmatrix}]\left({\bf z},\Omega\right) = \exp \bigg{\{}2\pi i\left(\frac{1}{2}a^T\Omega a+a^T({\bf z}+b)\right)\bigg{\}}\vartheta\left({\bf z} + b+\Omega a,\Omega\right),
\end{equation}
and so the original theta function is just $\vartheta\left({\bf z},\Omega\right)=\vartheta[\begin{smallmatrix}0\\ 0\end{smallmatrix}]\left({\bf z},\Omega\right)$. The theta function with characteristics is invariant under parity, ${\bf z}\rightarrow-{\bf z}$, provided we also take $a,b\rightarrow-a,-b$, so that $\vartheta[\begin{smallmatrix}a\\ b\end{smallmatrix}]\left({\bf z},\Omega\right)=\vartheta[\begin{smallmatrix}-a\\ -b\end{smallmatrix}]\left(-{\bf z},\Omega\right)$. This follows from (\ref{eq: Riemanntheta parity}) and (\ref{eq: Riemannthetacharact_intermsof}). In the case of integer or half-integer characteristics, $a,b\in (\tfrac{1}{2}\mathbb{Z}/\mathbb{Z})^{\mathfrak{g}}$, this relation simplifies further, in that it can be either even or odd under ${\bf z}\rightarrow -{\bf z}$,
$$
\vartheta[\begin{smallmatrix}a\\ b\end{smallmatrix}]\left(-{\bf z},\Omega\right)=(-)^{4a^Tb}\vartheta[\begin{smallmatrix}a\\ b\end{smallmatrix}]\left({\bf z},\Omega\right),\qquad \forall a,b\in (\tfrac{1}{2}\mathbb{Z}/\mathbb{Z})^{\mathfrak{g}}.
$$
By induction one can show that there are $2^{\mathfrak{g}-1}(2^{\mathfrak{g}}-1)$ choices of $[\begin{smallmatrix}a\\ b\end{smallmatrix}]$ for which $4a^Tb\in 2\mathbb{Z}-1$, and $2^{\mathfrak{g}-1}(2^{\mathfrak{g}}+1)$ choices of $[\begin{smallmatrix}a\\ b\end{smallmatrix}]$ for which $4a^Tb\in 2\mathbb{Z}$, leading to a total of $2^{2\mathfrak{g}}$ distinct choices. The corresponding characteristics $[\begin{smallmatrix}a\\ b\end{smallmatrix}]$ are referred to as odd or even respectively. For example, at genus $\mathfrak{g}=1$ there is 1 odd characteristic, $[\begin{smallmatrix}1/2\\ 1/2\end{smallmatrix}]$, and 3 even characteristics, $[\begin{smallmatrix}1/2\\ 0\end{smallmatrix}]$, $[\begin{smallmatrix}0\\ 1/2\end{smallmatrix}]$, and $[\begin{smallmatrix}0\\ 0\end{smallmatrix}]$. 

Consider the case of odd characteristics, which is of particular relevance for our purposes, and consider the function: 
$
f(z,w)= \vartheta[\begin{smallmatrix}a\\ b\end{smallmatrix}]\big(\int_w^z\omega,\Omega\big).
$ 
From the above, note primarily that for odd characteristics $[\begin{smallmatrix}a\\ b\end{smallmatrix}]$ it must be that $f(z,w)$ has a single zero for $z=w$. 
In addition, according to {\it Riemann's vanishing theorem} \cite{Mumford_v12} we know that there will be additional single zeros for $z=r_i$ and $w=r_i$, for $i=1,\dots,\mathfrak{g}-1$, so that when both $z$ and $w$ are close to one of the $r_i$ $f(z,w)$ will be of the form $f(z,w)\simeq {\rm const.}(z-w)(z-r_i)(w-r_i)$. Therefore, differentiating with respect to $w$ at $w=z$ implies the one form, $\omega_I(z)\partial_I\vartheta[\begin{smallmatrix}a\\ b\end{smallmatrix}](0,\Omega)$, has $\mathfrak{g}-1$ double zeroes for $z=r_i$, with an analogous reasoning (upon replacing $z\leftrightarrow w$) implying $\mathfrak{g}-1$ double zeros at $w=r_i$ also. Therefore, taking an appropriate ratio of $f(z,w)$ over two square roots of these one-forms will lead to a quantity that has only one (simple) zero at $z=w$, and these observations lead one to define a very useful quantity known as the prime form \cite{Fay,Mumford_v12,DHokerPhong}.

\ssk
The prime form generalises the notion of distance between two points, $z-w$, on $\mathbb{C}$ to higher genus surfaces. In terms of the Riemann theta function it reads \cite{Fay,Mumford_v12,DHokerPhong}:
\begin{equation}\label{eq: primeform}
E(z,w) = \frac{\vartheta[\begin{smallmatrix}a\\ b\end{smallmatrix}]\left(\int_w^z\omega,\Omega\right)}{h_{[\begin{smallmatrix}a\\ b\end{smallmatrix}]}(z)h_{[\begin{smallmatrix}a\\ b\end{smallmatrix}]}(w)},
\end{equation}
where the characteristics $[\begin{smallmatrix}a\\ b\end{smallmatrix}]$ are odd (although actually $E(z,w)$ is independent of the precise choice), and the holomorphic half-differentials, $h_{[\begin{smallmatrix}a\\ b\end{smallmatrix}]}(z)$, are defined (according to the above discussion) by:
$$
h_{[\begin{smallmatrix}a\\ b\end{smallmatrix}]}(z)\dfn \sqrt{\omega_I(z)\partial_I\vartheta[\begin{smallmatrix}a\\ b\end{smallmatrix}](0,\Omega)},
$$
and correspond to a spin bundle associated to $[\begin{smallmatrix}a\\ b\end{smallmatrix}]$. The prime form is a (or rather the components of a) holomorphic differential form of weight $(-\frac{1}{2},-\frac{1}{2})$ on $\tilde{\Sigma}_{\mathfrak{g}}\times\tilde{\Sigma}_{\mathfrak{g}}$, with $\tilde{\Sigma}_{\mathfrak{g}}$ the universal cover of $\Sigma_{\mathfrak{g}}$. In the notation of (\ref{eq: tensor V}),  locally, $E(z,w)dz^{-\frac{1}{2}}dw^{-\frac{1}{2}}\in K^{(-\frac{1}{2},0)}\times K^{(-\frac{1}{2},0)}$. Note that $E(z,w)$ is quasi-periodic around the $A_I$ and $B_I$ cycles,
\begin{subequations}\label{eq: monodromiesE}
\begin{align}
&E(z+A_I,w)= E(z,w),\\
&E(z+B_I,w)=E(z,w)\exp\left(-\pi i\Omega_{II}-2\pi i\int_w^z\omega_I\right),
\end{align}
\end{subequations}
so that when we transport $z$ around a generic homology cycle $\gamma = n_IA_I+m_IB_I$, 
\begin{equation}\label{eq:quasiperiodprimeform}
E(z+n_IA_I+m_IB_I,w)=E(z,w)\exp\Big[2\pi i\Big(-\frac{1}{2}m_I \Omega_{IJ}m_J-m_I\int_w^z\omega_I\Big)\Big],
\end{equation}
which is again to be understood in the sense explained in the footnote on p.~\pageref{foot:z+A}.

The prime form in turn transforms under the modular group, ${\rm Sp}(2\mathfrak{g},\mathbb{Z})$, as follows \cite{VerlindeVerlinde87},
\begin{equation}
E(z,w)\rightarrow \exp\Big(\pi i\int_w^z\omega_I\big(C\Omega+D\big)^{-1}_{IJ}C_{JK}\int_w^z\omega_K\Big)E(z,w).
\end{equation}

Finally, it is useful to also have at hand Green's (or 2-d Stoke's) theorem, 
$$
\oint_{\partial D}A=\int_DdA,
$$
which in the $(z,\bar{z})$ coordinate system reads (using the above conventions, displayed explicitly in the beginning of this section):
$$
\oint_{\partial D} dz A_z+d\bar{z}A_{\bar{z}} =\int_D dz\wedge d\bar{z}\big(\partial_zA_{\bar{z}}-\partial_{\bar{z}}A_z\big)
$$
with the boundary integral in a counterclockwise direction if $D$ is ``inside'' the contour (i.e.~with $D$ to the left of the contour ``arrow''), and $A_z=A_x-iA_y$, $A_{\bar{z}}=A_x+iA_y$. 

\subsection{Green Function}\label{sec:GF}
Our convention for the genus-$\mathfrak{g}$ torus Green function \cite{DHokerPhong} is,
\begin{subequations}\label{eq: Greens function dzdbarzG+int}
\begin{align}
&\partial_z\partial_{\bar{z}}G(z,w) = -\pi\alpha'\delta^2(z-w)+\frac{\pi\alpha' g_{z\bar{z}}}{\int_{\Sigma_{\mathfrak{g}}}d^2z\sqrt{g}},\label{eq: Greens function dzdbarzG}\\
&\partial_z\partial_{\bar{w}}G(z,w) = \pi\alpha'\delta^2(z-w)-\frac{\pi \alpha'}{2}\omega_I(z)\left({\rm Im}\Omega\right)_{IJ}^{-1}\bar{\omega}_J(\bar{w}),\label{eq: Greens function dzdbarzG2},
\end{align}
\end{subequations}
satisfying $\int_{\Sigma_{\mathfrak{g}}}d^2z\sqrt{g}\,G(z,w) = 0$, which is concisely expressed in terms of Fay's prime form  \cite{Fay}, $E(z,w)$, see (\ref{eq: primeform}), the period matrix $\Omega_{IJ}$, and abelian differentials $\omega_I,\bar{\omega}_I$. 
For compact and oriented genus-$\mathfrak{g}$ Riemann surfaces \cite{HamidiVafa87,VerlindeVerlinde87,DHokerPhong}:
\begin{equation}\label{eq: Green Function+regular*}
G(z,w) = -\frac{\alpha'}{2}\ln \left|E(z,w)\right|^2 + \pi\alpha'\, {\rm Im}\int\limits_{z}^{w}\omega_I\left({\rm Im}\Omega\right)_{IJ}^{-1}{\rm Im}\int\limits_{z}^{w}\omega_J.
\end{equation}
which, up to terms of the form $f(z,\bar{z})+g(w,\bar{w})$ which do not contribute to amplitudes (in spacetimes for which charge and momentum is conserved), is determined uniquely by the requirement that it be single-valued around $A_I$ and $B_I$ cycles, and that it have the correct singular behaviour as $z\rightarrow w$, $G(z,w)\simeq -\ln |z-w|^2+\dots$. The prime form is quasi-periodic on $\Sigma_{\mathfrak{g}}$, see (\ref{eq:quasiperiodprimeform}).

\section{The Torus, $T^2$}\label{sec:T2}
\subsection{Coordinates and Moduli Space}
To specify a point on the torus we need two coordinates, $\sigma^1,\sigma^2$, chosen conveniently such that $\sigma^1\in[0,1)$, 
$\sigma^2\in [0,1)$, with identifications, $\sigma^1\sim \sigma^1+1$ and $\sigma^2\sim \sigma^2+1$. Locally, we can always express the metric in the form $ds^2=2g_{z\bar{z}}dzd\bar{z}$, in terms of which the Ricci scalar $R_{(2)}=-g^{z\bar{z}}\partial_z\partial_{\bar{z}}\ln g_{z\bar{z}}$. 
The uniformization theorem \cite{DHokerPhong} then enables us to choose a gauge slice that is tangent to zero curvature metrics, e.g.~$g_{z\bar{z}}=1/2$, although observables do not depend on this choice. As there is one (complex) conformal Killing vector (CKV) on the torus, i.e.~${\rm dim}_{\mathbb{C}}\,{\rm ker}\,\nabla_{(-1)}^z={\rm dim}_{\mathbb{C}}\,{\rm ker}\,\nabla^{(1)}_z=1$, the Euler characteristic being $\chi(T^2) = 0$, the Atiyah-Singer-Riemann-Roch index theorem (\ref{eq:Atiyah-Singer}) 
implies there is one (complex) modulus, ${\rm dim}_{\mathbb{C}}\,{\rm ker}\,\nabla^z_{(2)}={\rm dim}_{\mathbb{C}}\,{\rm ker}\,\nabla^{(-2)}_{z}=1$, call it $\tau=\tau_1+i\tau_2$. 
A useful (global) coordinate system is then $z=\sigma^1+\tau\sigma^2$, $\bar{z}=\sigma^1+\bar{\tau}\sigma^2$, with:
\begin{equation}\label{eq:metric}
ds^2 =|dz|^2= \big|d\sigma^1+\tau d\sigma^2\big|^2. 
\end{equation}
At genus one the (anti-)holomorphic abelian differentials (\ref{eq:omegaA}) reduce to $\omega=dz$, $\bar{\omega}=d\bar{z}$, with 
$$
\oint_Adz=1,\qquad {\rm and} \qquad \oint_Bdz=\tau,
$$
so that the period matrix, $\Omega_{11}$, is identified with $\tau$. 

\ssk
Starting from a metric (\ref{eq:metric}), we can deform the complex structure moduli by turning on Beltrami differentials, $(\mu,\bar{\mu})=(\mu_{\bar{z}}^{\phantom{z}z}(dz)^{-1}d\bar{z},\mu_{z}^{\phantom{z}{\bar{z}}}dz(d\bar{z})^{-1})$, so that any other metric is (up to a conformal rescalling) of the form $d\tilde{s}^2 = |dz+\mu d\bar{z}|^2$. These therefore provide a parametrisation of the space of metrics on the Riemann surface. There is a single insertion of, $|\langle\mu,b\rangle|^2$, in the amplitude, reflecting the presence of a single complex modulus. The pairing, $\langle\mu,b\rangle$, is defined with respect to the natural inner product of the space, see (\ref{eq: (V1,V2)}), and is independent of a metric, $\langle\mu,b\rangle=\int_{\Sigma}d^2z\,\mu_{\bar{z}}^{\phantom{z}z}b_{zz}$. When we compare the variation $\delta g_{\bar{z}\bar{z}}\equiv\delta\tau g_{z\bar{z}}\mu_{\bar{z}}^{\phantom{z}z}$, with the infinitesimal deformation $\tau\rightarrow \tau+\delta \tau$ of the flat metric (\ref{eq:metric}), we find $\mu_{\bar{z}}^{\phantom{z}z}=i/\tau_2$.\label{Beltrami} 
\begin{figure}
\begin{center}
\includegraphics[width=0.48\textwidth]{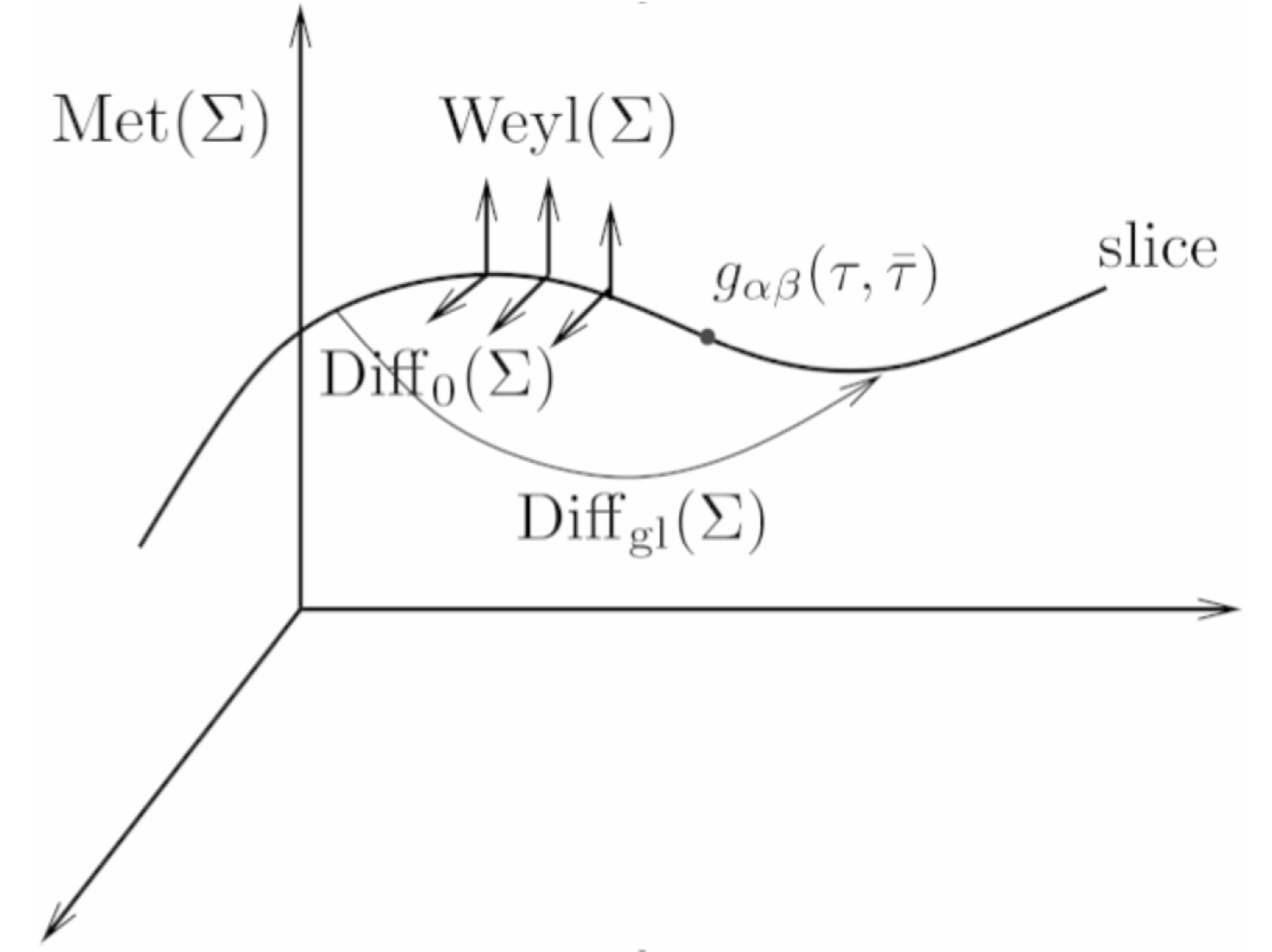}
\end{center}
\vspace{-0.3cm}
\caption{\small An illustration of a gauge slice in the space of worldsheet metrics. The ghost path integral ensures that the gauge slice, parametrized by $\tau,\bar{\tau}$, is orthogonal to the local worldsheet symmetries, whereas the restriction to a fundamental domain of integration, $\mathscr{F}_1$, ensures that we do not integrate over worldsheet deformations that are related by global diffeomorphisms. The gauge slice is specified by the choice of Beltrami differentials.}\label{fig:gaugeslice}
\end{figure}

Most of the invariance under global diffeomorphisms (see Fig.~\ref{fig:gaugeslice}), ${\rm Diff}_{\rm gl}(\Sigma)={\rm SL}(2,\mathbb{Z})/\mathbb{Z}_2$,  \cite{Polchinski86}, $\sigma^1\rightarrow a\sigma^1+b\sigma^2$, and $\sigma^2\rightarrow c\sigma^1+d\sigma^2$, with $a,b,c,d\in\mathbb{Z}$ and $ \det \Big({ \footnotesize\begin{matrix} 
      a & b \\
      c & d \\
   \end{matrix}}\Big)=1$, (equivalently $\tau\rightarrow \frac{d\tau+b}{c\tau+a}$), as well as the periodicities $\sigma^1\sim \sigma^1+1$, $\sigma^2\sim\sigma^2+1$ can be fixed by restricting the integration regions, respectively, to fundamental domains,
$$
\mathscr{F}_1 = \Big\{\tau_1,\tau_2\,\Big|\, -\tfrac{1}{2}\leq \tau_1\leq \tfrac{1}{2},\quad \sqrt{1-\tau_1^2}\leq \tau_2<\infty\Big\},\qquad \textrm{with}\qquad d^2\tau \equiv 2d\tau_1\wedge d\tau_2,
$$
$$
\Sigma_1 = \Big\{\sigma^1,\sigma^2\,\Big|\, 0\leq \sigma^1<1,\quad 0\leq \sigma^2\leq 1\Big\},\qquad \textrm{with}\qquad d^2z= 2\tau_2d\sigma^1\wedge d\sigma^2.
$$
This fixes most of the global diffeomorphisms, namely ${\rm SL}(2,\mathbb{Z})$, leaving a remaining $\mathbb{Z}_2$ isometry, $z\rightarrow -z$, and the latter leads to the factor $N_{1}=2$ in (\ref{eq:S-matrix}).

\subsection{Ghost Contributions}\label{sec:ghosts}
The ghost path integral, $\mathcal{A}_{\rm gh}$, evaluated on a genus-1 surface is a standard computation for which we provide some details for completeness:
\begin{equation}\label{eq:ghost_apx}
\begin{aligned}
\mathcal{A}_{\rm gh}&=\int \mathcal{D}(b,\tilde{b},c,\tilde{c})\!\!\prod_{j=1}^{\#_{\mathbb{C}}{\rm moduli}}\!\!\!\!\!|\langle \mu_j,b\rangle|^2\,\prod_{s=1}^{\#_{\mathbb{C}}{\rm CKVs}}\!\!\!|c(w_s)|^2\,\,e^{-I_{\rm gh}},
\end{aligned}
\end{equation}
with
\begin{equation}\label{eq:GhostAction_apx}
I_{\rm gh}= \frac{1}{2\pi} \int_{\Sigma_{\mathfrak{g}}}d^2z\sqrt{g} \big(b\nabla^z_{(-1)}c+\tilde{b}\nabla^{\bar{z}}_{(1)}\tilde{c}\big).
\end{equation}
Following \cite{DHoker}, we expand $b\in K^{2}$ and $c\in K^{-1}$, in an orthonormal (with respect to the natural inner products (\ref{eq: (V1,V2)})) complex basis of eigenfunctions of $\Delta_{(-1)}^-$, $\Delta_{(2)}^{-}$. There exist two complex zero modes, call them $\psi\in {\rm ker}\,\nabla^z_{(-1)}$ and $\phi\in {\rm ker}\,\nabla^{(-2)}_z$, so that recalling the discussion above (\ref{eq:metric}),
\begin{equation}\label{eq:numb of mod+CKVs}
\begin{aligned}
\#_{\mathbb{C}}{\rm CKV}&\dfn {\rm dim}_{\mathbb{C}}\,{\rm ker}\,\nabla_{(-1)}^z\big|_{\mathfrak{g}=1}=1\\
\#_{\mathbb{C}}{\rm moduli}&\dfn {\rm dim}_{\mathbb{C}}\,{\rm ker}\,\nabla^{(-2)}_{z}\big|_{\mathfrak{g}=1}=1.
\end{aligned}
\end{equation}
The corresponding (real, non-zero) eigenvalues, $\lambda_{\alpha}$, and corresponding eigenfunctions turn out to be related, $\Delta_{(-1)}^-\Psi_{\alpha}=\lambda_{\alpha}^2 \Psi_{\alpha}$, and $\Delta_{(2)}^-(g_{z\bar{z}})^2\Phi^*_{\alpha} = \lambda_{\alpha}(g_{z\bar{z}})^2\Phi^*_{\alpha}$, where $\sqrt{2}\nabla^z_{(-1)}\Psi_{\alpha}\equiv\lambda_{\alpha}\Phi_{\alpha}$. 
We thus have the orthogonal expansions, $c=c_0\psi+\sum_{\alpha}c_{\alpha}\Psi_{\alpha}$, $b=b_0\phi+\sum_{\alpha}b_{\alpha}(g_{z\bar{z}})^2\Phi^*_{\alpha}$ (with Grassmann-valued coefficients $c_0,c_{\alpha},b_0,b_{\alpha}$), so that plugging these into $\mathcal{A}_{\rm gh}$ above and integrating out the ghosts yields:  
\begin{equation}\label{eq:ghostpathint1}
\begin{aligned}
\mathcal{A}_{\rm gh}\big|_{\mathfrak{g}=1}={\rm det}'\Delta_{(-1)}^-|\langle\mu,\phi\rangle|^2|\psi(w)|^2.
\end{aligned}
\end{equation}
The prime indicates omission of zero modes, and we have normalised the fields by their natural inner products  (\ref{eq: (V1,V2)}), such that $\langle\Phi_{\alpha},\Phi_{\beta}\rangle=\langle\Psi_{\alpha},\Psi_{\beta}\rangle=\delta_{\alpha,\beta}$ and $\langle\phi,\phi\rangle=\langle\psi,\psi\rangle=1$. (Note that we could have also derived the right-hand side of (\ref{eq:ghostpathint1}) directly from the decomposition of the path integral measure associated to worldsheet metrics after decomposing it into a gauge and a moduli contribution and reading off the appropriate Jacobian \cite{DHokerPhong}.) We next evaluate (\ref{eq:ghostpathint1}) explicitly, in terms of the complex structure moduli of the worldsheet, $\tau,\bar{\tau}$, see (\ref{eq:metric}).

The $b$-ghost zero modes, $\phi=\phi_{zz}(dz)^2$, are the (normalizable) {\it holomorphic quadratic differentials}, and span the space orthogonal to Weyl transformations and diffeomorphisms, the non-trivial constraint (or defining relation) being, $\langle\delta_{\rm Diff_0} g_{zz},\phi_{zz}\rangle\equiv 0$, with $\delta_{\rm Diff_0} g_{zz}=2\nabla^{(1)}_z\delta v_z$). Correspondingly, the $c$-ghost zero modes, $\psi$, are {\it conformal Killing vectors} (CKV), and generate isometries associated to rigid shifts along the A- and B-cycles of the torus. The choice of metric (\ref{eq:metric}) admits one globally defined CKV and one quadratic differential. To solve the equations, $\nabla^z_{(-1)}\psi^{z}=0$, $\nabla^{(2)}_{\bar{z}}\phi_{zz}=0$, note that the only doubly periodic holomorphic functions on a torus are the constants. Thus, using the aforementioned normalization conditions, 
$ 
\psi=\frac{1}{\sqrt{\tau_2}}(dz)^{-1}$, and $\phi=\frac{1}{2\sqrt{\tau_2}}(dz)^2.
$ 
Notice that the components are independent of $z,\bar{z}$. Above we picked a gauge slice that is parametrized by the complex number $\tau$, see p.~\pageref{Beltrami}, and this determined the Beltrami differential $\mu_{\bar{z}}^{\phantom{z}z}=i/\tau_2$. Hence $|\langle\mu,\phi\rangle|^2=|\int d^2z\mu_{\bar{z}}^{\phantom{z}z}\phi_{zz}|^2=1/\tau_2$. The remaining quantity to evaluate in (\ref{eq:ghostpathint1}) is the determinant of the Laplace operator.

To compute the determinant of the Laplace operator, note primarily that the various Laplacians are equal for flat metrics, $\Delta_{(0)}=\Delta^-_{(-1)}=-2g^{z\bar{z}}\partial_z\partial_{\bar{z}}=-4\partial_z\partial_{\bar{z}}$. Here one starts from a complete set of eigenfunctions, $\psi_{n,m}(z,\bar{z})$, of $\Delta^-_{(-1)}$ which satisfy the torus periodicities, $z\sim z+1$ and $z\sim z+\tau$, so that: $\Delta^-_{(-1)}\psi_{n,m}=\lambda_{n,m}\psi_{n,m}$. If the basis vectors, $\psi_{n,m}(z,\bar{z})$, are orthogonal, ${\rm det}'\Delta_{(-1)}^-={\prod}'_{n,m}\lambda_{n,m}=(\prod_{n\neq0}\lambda_{n,0})(\prod_{m\neq0}\lambda_{0,m})(\prod_{n\neq0}\prod_{m\neq0}\lambda_{n,m})$. For example, a complete basis that has the correct periodicities is $\psi_{n,m}=e^{2\pi i n\sigma^2+2\pi i m \sigma^2}$, with $z=\sigma^1+\tau\sigma^2$, in which case $\Delta_{(0)}=-\frac{1}{\tau_2}|\tau\partial_1-\partial_2|^2$, and $\lambda_{n,m}=\big(\frac{2\pi}{\tau_2}\big)^2|m+\tau n|^2$. The resulting infinite products can be determined by zeta-function regularisation (to show that $\prod_{n>0}a=1/\sqrt{a}$ for a constant $a$) and the product representation for the eta function \cite{DHokerPhong}, or by making use of properties of the Eisenstein series \cite{Nakahara03}; the result is (up to an immaterial constant) ${\rm det}'\Delta_{(0)}={\rm det}'\Delta_{(-1)}^-=\tau_2^2|\eta(\tau)|^{4}$. Collecting the above,
\begin{equation}\label{eq:ghostpathint}
\mathcal{A}_{\rm gh}\big|_{\mathfrak{g}=1} =|\eta(\tau)|^{4},
\end{equation}
and this is independent of $w,\bar{w}$, i.e.~of where we place the $c,\tilde{c}$ ghosts on the worldsheet. (We are working with the critical string where the non-chiral Liouville action \cite{VerlindeVerlinde87} is absent.) This is the standard result for the ghost contribution at genus $\mathfrak{g}=1$, but in the main text we are rather interested in the quantity $\mathcal{Z}_1$, defined in (\ref{eq:ghostb}), and so from the above, on account of ${\det}\,{\rm Im}\Omega_{IJ}\big|_{\mathfrak{g}=1}=\tau_2$ and $\int_{\Sigma_{1}}d^2z\sqrt{g}=\tau_2$, it follows immediately that we can also write (\ref{eq:ghostpathint}) as follows,
\begin{equation}\label{eq:ghost_apx}
\begin{aligned}
\mathcal{A}_{\rm gh}\big|_{\mathfrak{g}=1}&=
 \bigg(\frac{{\det}'\Delta_{(0)}}{{\det}\,{\rm Im}\Omega_{IJ}\int_{\Sigma_{1}}d^2z\sqrt{g}}\bigg)^{13}\big|\eta(\tau)^{-24}\big|^2,
\end{aligned}
\end{equation}
allowing us to conclude that (up to an immaterial phase):
\begin{equation}\label{eq:Z1=eta..}
\mathcal{Z}_1=\eta(\tau)^{-24},\qquad \bar{\mathcal{Z}}_1=\eta(\bar{\tau})^{-24}.
\end{equation}

\subsection{Dedekind $\eta$ Function}\label{Apx:Dedekiind}
Writing $v=e^{2\pi i\tau}$, with ${\rm Im}\tau>0$, the Dedekind $\eta$-function is defined as:
\begin{equation}
\eta(\tau)=v^{1/24}\prod_{n>0}\big{(}1-v^n\big{)},
\end{equation}
with the property,
$$
\eta(-1/\tau)=\sqrt{-i\tau}\eta(\tau).
$$
The following explicit expansion is useful when focusing on the contribution of the lightest decay channels to the amplitude:
\begin{equation}\label{eq:eta-24}
\begin{aligned}
\eta(\tau)^{-24}&= v^{-1} \prod_{n>0}(1-v^n)^{-24}\\
&=v^{-1}+24+324v+3200\,v^2+25650v^3+\mathcal{O}(v^4).\phantom{\big|}
\end{aligned}
\end{equation}
\subsection{Jacobi Theta Functions}
The Jacobi theta function is defined as \cite{Mumford_v12}:
\begin{equation}\label{eq:vartheta}
\begin{aligned}
\vartheta(z|\tau)&=\sum_{n\in\mathbb{Z}}\exp\big{(}\pi i n^2\tau+2\pi inz\big{)}
\end{aligned}
\end{equation}
with $\tau_2={\rm Im}\tau>0$. 
The variant $\vartheta_1(z|\tau)$ is of particular relevance and has the following representations,
\begin{subequations}\label{eq:vartheta_1}
\begin{align}
\vartheta_1(z|\tau)&=-\sum_{n\in\mathbb{Z}}\exp\big{(}\pi i (n+\tfrac{1}{2})^2\tau+2\pi i(n+{\tfrac{1}{2}})(z+\tfrac{1}{2})\big{)},\label{eq:vartheta_1a}\\
&=2\sum_{n=0}^{\infty}(-)^{n}v^{\frac{1}{2}(n+\frac{1}{2})^2}\sin(2n+1)\pi z\label{eq:vartheta_1b}\\
&=2v^{1/8}\sin\pi z\prod_{n>0}\big{(}1-2v^n\cos2\pi z+v^{2n}\big{)}\big{(}1-v^n\big{)}\label{eq:vartheta_1c}
\end{align}
\end{subequations}
A useful quantity that appears in the definition of the prime form, $E(z,z')$, is $\vartheta_1'(0|\tau)\equiv \partial_z\vartheta_1(z|\tau)|_{z=0}$, an explicit expression for which follows directly from (\ref{eq:vartheta_1c}):
\begin{equation}\label{eq:theta'(0|t)}
\vartheta'(0|\tau) = 2\pi v^{1/8}\prod_{n>0}(1-v^n)^3.
\end{equation}

The quantity $\vartheta_1(z|\tau)$ is odd under parity, $\vartheta_1(z|\tau)=-\vartheta_1(-z|\tau)$, 
and hence $\vartheta_1(0|\tau)=0$. In fact, the zeros of $\vartheta_1(z|\tau)$ are located at: 
\begin{equation}\label{eq: zeros}
\begin{aligned}
z=m+n\tau\quad&\Leftrightarrow\quad\vartheta_1(z|\tau)=0,\qquad {\rm with }\qquad n,m\in \mathbb{Z}.
\end{aligned}
\end{equation}

\subsection{Prime Form}\label{sec:FPF}
The one loop (genus $\mathfrak{g}=1$) expression for Fay's prime form (\ref{eq: primeform}) \cite{Fay,Mumford_v12,DHokerPhong} is given in terms of Jacobi theta functions,
\begin{equation}\label{eq:primeform}
E(z) = \frac{\vartheta_1(z|\tau)}{\vartheta_1'(0|\tau)},
\end{equation}
with $\vartheta_1'(0|\tau)\equiv \partial_z\vartheta_1(z|\tau)|_{z=0}$, see (\ref{eq:vartheta_1}) and (\ref{eq:theta'(0|t)}). Writing,
$$
u\equiv e^{2\pi iz},\qquad{\rm and}\qquad v\equiv e^{2\pi i \tau},
$$ 
from (\ref{eq:vartheta_1c}) and (\ref{eq:theta'(0|t)}) it follows that an explicit product representation is,
\begin{subequations}\label{eq:E(z)}
\begin{align}
E(z)& = \frac{\sin \pi z}{\pi}\prod_{n>0}\left(\frac{1-2v^n\cos 2\pi z+v^{2n}}{1-2v^n+v^{2n}}\right),\label{E(z)a}\\
&=\frac{1}{2\pi i}\Big(u^{1/2}-u^{-1/2}\Big)\prod_{n>0}\left(\frac{(1-v^nu)(1-v^nu^{-1})}{(1-v^n)^2}\right).\label{eq:E(z)b}
\end{align}
\end{subequations}
It is clear that $E(z)$ has a simple zero at $z=0$. In fact, for generic $v$ it follows immediately from (\ref{E(z)a}) that:
$$
2\pi iE(z)\big|_{z\rightarrow 0}\simeq 2\pi iz+\Big(\frac{1}{24}-\sum_{n>0}\frac{v^n}{1-2v^n+v^{2n}}\Big)(2\pi iz)^3+\mathcal{O}(z^5),
$$
and, in fact, the prime form is the unique holomorphic object on a Riemann surface that has a simple zero at $z=0$ and is non-vanishing elsewhere (modulo lattice periodicities, see below). $E(z)$ therefore generalises the notion of distance on topologically non-trivial Riemann surfaces. In addition, the prime form has the following monodromies,
\begin{subequations}\label{eq: monodromiesE}
\begin{align}
&E(z+1)=-E(z),\phantom{\bigg|}\\
&E(z+\tau)=-\exp\left(-\pi i\tau-2\pi iz\right)E(z),
\end{align}
\end{subequations}
around the $A$ and $B$ cycles of the torus respectively. 

\ssk
If we exponentiate the infinite product in (\ref{eq:E(z)b}), expand the resulting logarithms, perform the geometric sums and make use of the identity $(2\sin\theta)^2=2-2\cos 2\theta$, we may equivalently write:
\begin{subequations}\label{eq:primeform}
\begin{align}
&2\pi iE(z) = (2i\sin \pi z)\exp\left\{\sum_{n>0}\frac{1}{n}\frac{v^n}{1-v^n}\,(2\sin\pi nz)^2\right\},\label{eq:2pi iE}\\
&E(z)^2\partial_z^2\ln E(z) = \bigg\{-1+(2\sin\pi z)^2\sum_{n>0}\frac{2nv^n}{1-v^n}\,\cos 2\pi nz\bigg\}\exp\bigg\{\sum_{n>0}\frac{2}{n}\frac{v^n}{1-v^n}\,(2\sin\pi nz)^2\bigg\},\label{eq:E2dlnE}
\end{align}
\end{subequations}
where in the second line we have exhibited another combination that appears in string amplitudes. 
It is convenient to consider these expressions as a series expansion in $v$, which is useful in discussing the $\tau_2\rightarrow \infty$ boundary of moduli space (with $\sigma^1,\sigma^2$ generic). Defining $S(u)\equiv u^{1/2}-u^{-1/2}$, $C(u)\equiv u^{1/2}+u^{-1/2}$,
\begin{subequations}\label{eq:prime form apx}
\begin{align}
&2\pi iE(z) =S(u)-vS(u)^3-3v^2S(u)^3
+\mathcal{O}(v^3)\phantom{\Bigg|}\label{eq:2pi iE qexp}\\
&E(z)^2\partial_z^2\ln E(z) = -1-v\,S(u)^4-6v^2S(u)^4+\mathcal{O}(v^3)\label{eq:E2dlnE q exp}.
\end{align}
\end{subequations}



\begin{thebibliography}{10}

\bibitem{GiveonItzhaki12}
A.~Giveon and N.~Itzhaki, {\it {String Theory Versus Black Hole
  Complementarity}},  {\em JHEP} {\bf 12} (2012) 094,
  [\href{http://arxiv.org/abs/1208.3930}{{\tt arXiv:1208.3930}}].

\bibitem{ShenkerStanford14b}
S.~H. Shenker and D.~Stanford, {\it {Black holes and the butterfly effect}},
  {\em JHEP} {\bf 03} (2014) 067, [\href{http://arxiv.org/abs/1306.0622}{{\tt
  arXiv:1306.0622}}].

\bibitem{ShenkerStanford14a}
S.~H. Shenker and D.~Stanford, {\it {Multiple Shocks}},  {\em JHEP} {\bf 12}
  (2014) 046, [\href{http://arxiv.org/abs/1312.3296}{{\tt arXiv:1312.3296}}].

\bibitem{GiveonItzhaki13}
A.~Giveon and N.~Itzhaki, {\it {String theory at the tip of the cigar}},  {\em
  JHEP} {\bf 09} (2013) 079, [\href{http://arxiv.org/abs/1305.4799}{{\tt
  arXiv:1305.4799}}].

\bibitem{Silverstein14}
E.~Silverstein, {\it {Backdraft: String Creation in an Old Schwarzschild Black
  Hole}},  \href{http://arxiv.org/abs/1402.1486}{{\tt arXiv:1402.1486}}.

\bibitem{MertensVerscheldeZakharov15}
T.~G. Mertens, H.~Verschelde, and V.~I. Zakharov, {\it {Perturbative String
  Thermodynamics near Black Hole Horizons}},  {\em JHEP} {\bf 06} (2015) 167,
  [\href{http://arxiv.org/abs/1410.8009}{{\tt arXiv:1410.8009}}].

\bibitem{ShenkerStanford15}
S.~H. Shenker and D.~Stanford, {\it {Stringy effects in scrambling}},  {\em
  JHEP} {\bf 05} (2015) 132, [\href{http://arxiv.org/abs/1412.6087}{{\tt
  arXiv:1412.6087}}].

\bibitem{Martinec15}
E.~J. Martinec, {\it {The Cheshire Cap}},  {\em JHEP} {\bf 03} (2015) 112,
  [\href{http://arxiv.org/abs/1409.6017}{{\tt arXiv:1409.6017}}].

\bibitem{GiveonItzhakiKutasov15}
A.~Giveon, N.~Itzhaki, and D.~Kutasov, {\it {Stringy Horizons}},  {\em JHEP}
  {\bf 06} (2015) 064, [\href{http://arxiv.org/abs/1502.03633}{{\tt
  arXiv:1502.03633}}].

\bibitem{MertensVerscheldeZakharov16}
T.~G. Mertens, H.~Verschelde, and V.~I. Zakharov, {\it {The long string at the
  stretched horizon and the entropy of large non-extremal black holes}},  {\em
  JHEP} {\bf 02} (2016) 041, [\href{http://arxiv.org/abs/1505.04025}{{\tt
  arXiv:1505.04025}}].

\bibitem{DodelsonSilverstein15b}
M.~Dodelson and E.~Silverstein, {\it {String-theoretic breakdown of effective
  field theory near black hole horizons}},
  \href{http://arxiv.org/abs/1504.05536}{{\tt arXiv:1504.05536}}.

\bibitem{DodelsonSilverstein15a}
M.~Dodelson and E.~Silverstein, {\it {Longitudinal nonlocality in the string
  S-matrix}},  \href{http://arxiv.org/abs/1504.05537}{{\tt arXiv:1504.05537}}.

\bibitem{Ben-IsraelGiveonItzhakiLiram15}
R.~Ben-Israel, A.~Giveon, N.~Itzhaki, and L.~Liram, {\it {Stringy Horizons and
  UV/IR Mixing}},  {\em JHEP} {\bf 11} (2015) 164,
  [\href{http://arxiv.org/abs/1506.07323}{{\tt arXiv:1506.07323}}].

\bibitem{Ben-IsraelGiveonItzhakiLiram16}
R.~Ben-Israel, A.~Giveon, N.~Itzhaki, and L.~Liram, {\it {On the Stringy
  Hartle-Hawking State}},  {\em JHEP} {\bf 03} (2016) 019,
  [\href{http://arxiv.org/abs/1512.01554}{{\tt arXiv:1512.01554}}].

\bibitem{BianchiMoralesPieri16}
M.~Bianchi, J.~F. Morales, and L.~Pieri, {\it {Stringy Origin of 4d Black Hole
  Microstates}},  {\em JHEP} {\bf 06} (2016) 003,
  [\href{http://arxiv.org/abs/1603.05169}{{\tt arXiv:1603.05169}}].

\bibitem{GiveonItzhakiKutasov15b}
A.~Giveon, N.~Itzhaki, and D.~Kutasov, {\it {Stringy Horizons II}},
  \href{http://arxiv.org/abs/1603.05822}{{\tt arXiv:1603.05822}}.

\bibitem{MertensVerscheldeZakharov16b}
T.~G. Mertens, H.~Verschelde, and V.~I. Zakharov, {\it {String Theory in Polar
  Coordinates and the Vanishing of the One-Loop Rindler Entropy}},  {\em JHEP}
  {\bf 08} (2016) 113, [\href{http://arxiv.org/abs/1606.06632}{{\tt
  arXiv:1606.06632}}].

\bibitem{PuhmRojasUgajin16}
A.~Puhm, F.~Rojas, and T.~Ugajin, {\it {(Non-adiabatic) string creation on nice
  slices in Schwarzschild black holes}},
  \href{http://arxiv.org/abs/1609.09510}{{\tt arXiv:1609.09510}}.

\bibitem{Mathur05}
S.~D. Mathur, {\it {The Quantum structure of black holes}},  {\em Class. Quant.
  Grav.} {\bf 23} (2006) R115, [\href{http://arxiv.org/abs/hep-th/0510180}{{\tt
  hep-th/0510180}}].

\bibitem{Mathur09}
S.~D. Mathur, {\it {The Information paradox: A Pedagogical introduction}},
  {\em Class. Quant. Grav.} {\bf 26} (2009) 224001,
  [\href{http://arxiv.org/abs/0909.1038}{{\tt arXiv:0909.1038}}].

\bibitem{MathurTurton14}
S.~D. Mathur and D.~Turton, {\it {Oscillating supertubes and neutral rotating
  black hole microstates}},  {\em JHEP} {\bf 04} (2014) 072,
  [\href{http://arxiv.org/abs/1310.1354}{{\tt arXiv:1310.1354}}].

\bibitem{BenaMartinecTurtonWarner16}
I.~Bena, E.~Martinec, D.~Turton, and N.~P. Warner, {\it {Momentum Fractionation
  on Superstrata}},  {\em JHEP} {\bf 05} (2016) 064,
  [\href{http://arxiv.org/abs/1601.05805}{{\tt arXiv:1601.05805}}].

\bibitem{BenaGiustoMartinecRussoShigemoriTurtonWarner16}
I.~Bena, S.~Giusto, E.~J. Martinec, R.~Russo, M.~Shigemori, D.~Turton, and
  N.~P. Warner, {\it {Smooth horizonless geometries deep inside the black-hole
  regime}},  \href{http://arxiv.org/abs/1607.03908}{{\tt arXiv:1607.03908}}.

\bibitem{'tHooft90}
G.~'t~Hooft, {\it {The Black Hole Interpretation of String Theory}},  {\em
  Nucl. Phys.} {\bf B335} (1990) 138--154.

\bibitem{SusskindThorlaciusUglum93}
L.~Susskind, L.~Thorlacius, and J.~Uglum, {\it {The Stretched horizon and black
  hole complementarity}},  {\em Phys. Rev.} {\bf D48} (1993) 3743--3761,
  [\href{http://arxiv.org/abs/hep-th/9306069}{{\tt hep-th/9306069}}].

\bibitem{Susskind93a}
L.~Susskind, {\it {String theory and the principles of black hole
  complementarity}},  {\em Phys. Rev. Lett.} {\bf 71} (1993) 2367--2368,
  [\href{http://arxiv.org/abs/hep-th/9307168}{{\tt hep-th/9307168}}].

\bibitem{Susskind93}
L.~Susskind, {\it {Strings, black holes and Lorentz contraction}},  {\em Phys.
  Rev.} {\bf D49} (1994) 6606--6611,
  [\href{http://arxiv.org/abs/hep-th/9308139}{{\tt hep-th/9308139}}].

\bibitem{LowePolchinskiSusskindThorlaciusUglum95}
D.~A. Lowe, J.~Polchinski, L.~Susskind, L.~Thorlacius, and J.~Uglum, {\it
  {Black hole complementarity versus locality}},  {\em Phys.Rev.} {\bf D52}
  (1995) 6997--7010.

\bibitem{HorowitzPolchinski97}
G.~T. Horowitz and J.~Polchinski, {\it {A Correspondence principle for black
  holes and strings}},  {\em Phys.Rev.} {\bf D55} (1997) 6189--6197.

\bibitem{HorowitzPolchinski98}
G.~T. Horowitz and J.~Polchinski, {\it {Selfgravitating fundamental strings}},
  {\em Phys.Rev.} {\bf D57} (1998) 2557--2563.

\bibitem{AmatiRusso99}
D.~Amati and J.~Russo, {\it {Fundamental strings as black bodies}},  {\em
  Phys.Lett.} {\bf B454} (1999) 207--212.

\bibitem{DamourVeneziano00}
T.~Damour and G.~Veneziano, {\it {Self-gravitating fundamental strings and
  black holes}},  {\em Nucl. Phys.} {\bf B568} (2000) 93--119.

\bibitem{CornalbaCostaPenedonesVieira06}
L.~Cornalba, M.~S. Costa, J.~Penedones, and P.~Vieira, {\it {From fundamental
  strings to small black holes}},  {\em JHEP} {\bf 12} (2006) 023.

\bibitem{Polchinski16}
J.~Polchinski, {\it {The Black Hole Information Problem}},  in {\em
  Theoretical Advanced Study Institute in Elementary Particle Physics: New
  Frontiers in Fields and Strings (TASI 2015) Boulder, CO, USA, June 1-26,
  2015}, 2016.
\newblock \href{http://arxiv.org/abs/1609.04036}{{\tt arXiv:1609.04036}}.

\bibitem{FlorakisRizos16}
I.~Florakis and J.~Rizos, {\it {Chiral Heterotic Strings with Positive
  Cosmological Constant}},  {\em Nucl. Phys.} {\bf B913} (2016) 495--533,
  [\href{http://arxiv.org/abs/1608.04582}{{\tt arXiv:1608.04582}}].

\bibitem{CopelandMyersPolchinski04}
E.~J. Copeland, R.~C. Myers, and J.~Polchinski, {\it {Cosmic F- and
  D-strings}},  {\em JHEP} {\bf 06} (2004) 013.

\bibitem{DvaliVilenkin04}
G.~Dvali and A.~Vilenkin, {\it {Formation and evolution of cosmic D-strings}},
  {\em JCAP} {\bf 0403} (2004) 010.

\bibitem{Polchinski04}
J.~Polchinski, {\it {Introduction to Cosmic F- and D-Strings}},
  \href{http://arxiv.org/abs/hep-th/0412244}{{\tt hep-th/0412244}}.

\bibitem{BeckerBeckerKrause06}
K.~Becker, M.~Becker, and A.~Krause, {\it {Heterotic cosmic strings}},  {\em
  Phys. Rev.} {\bf D74} (2006) 045023.

\bibitem{CopelandPogosianVachaspati11}
E.~J. Copeland, L.~Pogosian, and T.~Vachaspati, {\it {Seeking String Theory in
  the Cosmos}},  {\em Class. Quant. Grav.} {\bf 28} (2011) 204009,
  [\href{http://arxiv.org/abs/1105.0207}{{\tt arXiv:1105.0207}}].

\bibitem{CopelandKibble10}
E.~J. Copeland and T.~W.~B. Kibble, {\it {Cosmic Strings and Superstrings}},
  {\em Proc. Roy. Soc. Lond.} {\bf A466} (2010) 623--657.

\bibitem{Hindmarsh11}
M.~Hindmarsh, {\it {Signals of Inflationary Models with Cosmic Strings}},  {\em
  Prog.Theor.Phys.Suppl.} {\bf 190} (2011) 197--228.

\bibitem{SarangiTye02}
S.~Sarangi and S.~H.~H. Tye, {\it {Cosmic String Production Towards the End of
  Brane Inflation}},  {\em Phys. Lett.} {\bf B536} (2002) 185--192.

\bibitem{JonesStoicaTye02}
N.~T. Jones, H.~Stoica, and S.~H.~H. Tye, {\it {Brane Interaction as the Origin
  of Inflation}},  {\em JHEP} {\bf 07} (2002) 051.

\bibitem{MajumdarDavis02}
M.~Majumdar and A.~Christine-Davis, {\it {Cosmological creation of D-branes and
  anti-D-branes}},  {\em JHEP} {\bf 03} (2002) 056.

\bibitem{BarnabyBerndsenCLineStoica05}
N.~Barnaby, A.~Berndsen, J.~M. Cline, and H.~Stoica, {\it {Overproduction of
  Cosmic Superstrings}},  {\em JHEP} {\bf 06} (2005) 075.

\bibitem{AvgoustidisCopelandMossSkliros12}
A.~Avgoustidis, E.~J. Copeland, A.~Moss, and D.~Skliros, {\it {Fast Analytic
  Computation of Cosmic String Power Spectra}},  {\em Phys. Rev.} {\bf D86}
  (2012) 123513, [\href{http://arxiv.org/abs/1209.2461}{{\tt
  arXiv:1209.2461}}].

\bibitem{CharnockAvgoustidisCopelandMoss16}
T.~Charnock, A.~Avgoustidis, E.~J. Copeland, and A.~Moss, {\it {CMB constraints
  on cosmic strings and superstrings}},  {\em Phys. Rev.} {\bf D93} (2016),
  no.~12 123503, [\href{http://arxiv.org/abs/1603.01275}{{\tt
  arXiv:1603.01275}}].

\bibitem{BinetruyBoheCapriniDufaux12}
P.~Binetruy, A.~Bohe, C.~Caprini, and J.-F. Dufaux, {\it {Cosmological
  Backgrounds of Gravitational Waves and eLISA/NGO: Phase Transitions, Cosmic
  Strings and Other Sources}},  {\em JCAP} {\bf 1206} (2012) 027,
  [\href{http://arxiv.org/abs/1201.0983}{{\tt arXiv:1201.0983}}].

\bibitem{Dufaux12b}
J.-F. Dufaux, {\it {Cosmological Backgrounds of Gravitational Waves and
  eLISA}},  {\em ASP Conf. Ser.} {\bf 467} (2013) 91--102,
  [\href{http://arxiv.org/abs/1209.4024}{{\tt arXiv:1209.4024}}].

\bibitem{Burden85}
C.~J. Burden, {\it {Gravitational Radiation from a Particular Class of Cosmic
  Strings}},  {\em Phys. Lett.} {\bf B164} (1985) 277.

\bibitem{VachaspatiVilenkin85}
T.~Vachaspati and A.~Vilenkin, {\it {Gravitational Radiation from Cosmic
  Strings}},  {\em Phys. Rev.} {\bf D31} (1985) 3052.

\bibitem{DamourVilenkin00}
T.~Damour and A.~Vilenkin, {\it {Gravitational wave bursts from cosmic
  strings}},  {\em Phys. Rev. Lett.} {\bf 85} (2000) 3761--3764.

\bibitem{DamourVilenkin01}
T.~Damour and A.~Vilenkin, {\it {Gravitational wave bursts from cusps and kinks
  on cosmic strings}},  {\em Phys. Rev.} {\bf D64} (2001) 064008.

\bibitem{DamourVilenkin05}
T.~Damour and A.~Vilenkin, {\it {Gravitational radiation from cosmic
  (super)strings: Bursts, stochastic background, and observational windows}},
  {\em Phys. Rev.} {\bf D71} (2005) 063510.

\bibitem{SklirosCopelandSaffin13}
D.~P. Skliros, E.~J. Copeland, and P.~M. Saffin, {\it {Duality and Decay of
  Macroscopic F-Strings}},  {\em Phys. Rev. Lett.} {\bf 111} (2013) 041601,
  [\href{http://arxiv.org/abs/1304.1155}{{\tt arXiv:1304.1155}}].

\bibitem{HindmarshSkliros10}
M.~Hindmarsh and D.~Skliros, {\it {Covariant Closed String Coherent States}},
  {\em Phys. Rev. Lett.} {\bf 106} (2011) 1602,
  [\href{http://arxiv.org/abs/1006.2559}{{\tt arXiv:1006.2559}}].

\bibitem{SklirosHindmarsh11}
D.~Skliros and M.~Hindmarsh, {\it {String Vertex Operators and Cosmic
  Strings}},  {\em Phys.Rev.} {\bf D84} (2011) 126001,
  [\href{http://arxiv.org/abs/1107.0730}{{\tt arXiv:1107.0730}}].

\bibitem{SklirosCopelandSaffin16bb}
D.~P. Skliros, E.~J. Copeland, and P.~M. Saffin, {\it {Highly Excited Strings
  II: Vertex Operators}},  {\em (to appear)}.

\bibitem{SklirosCopelandSaffin16cc}
D.~P. Skliros, E.~J. Copeland, and P.~M. Saffin, {\it {Highly Excited Strings
  III: Two-Point Amplitudes}},  {\em (to appear)}.

\bibitem{SklirosCopelandSaffin16dd}
D.~P. Skliros, E.~J. Copeland, and P.~M. Saffin, {\it {Highly Excited Strings
  IV: Decay Rates}},  {\em (to appear)}.

\bibitem{SklirosCopelandHindmarshSaffin16dd2}
D.~P. Skliros, E.~J. Copeland, M.~B. Hindmarsh, and P.~M. Saffin, {\it {Highly
  Excited Strings IV: Gravitational Radiation}},  {\em (to appear)}.

\bibitem{SklirosCopelandSaffin16ee}
D.~P. Skliros, E.~J. Copeland, and P.~M. Saffin, {\it {Highly Excited Strings
  V: Effective Theory}},  {\em (to appear)}.

\bibitem{D'HokerPhong89}
E.~D'Hoker and D.~H. Phong, {\it {Conformal Scalar Fields and Chiral Splitting
  on Super-Riemann Surfaces}},  {\em Commun. Math. Phys.} {\bf 125} (1989) 469.

\bibitem{PiusSen16}
R.~Pius and A.~Sen, {\it {Cutkosky Rules for Superstring Field Theory}},  {\em
  JHEP} {\bf 10} (2016) 024, [\href{http://arxiv.org/abs/1604.01783}{{\tt
  arXiv:1604.01783}}].

\bibitem{Sen16}
A.~Sen, {\it {Equivalence of Two Contour Prescriptions in Superstring
  Perturbation Theory}},  \href{http://arxiv.org/abs/1610.00443}{{\tt
  arXiv:1610.00443}}.

\bibitem{Sen16b}
A.~Sen, {\it {One Loop Mass Renormalization of Unstable Particles in
  Superstring Theory}},  \href{http://arxiv.org/abs/1607.06500}{{\tt
  arXiv:1607.06500}}.

\bibitem{Sen16c}
A.~Sen, {\it {Wilsonian Effective Action of Superstring Theory}},
  \href{http://arxiv.org/abs/1609.00459}{{\tt arXiv:1609.00459}}.

\bibitem{Sen16d}
A.~Sen, {\it {Unitarity of Superstring Field Theory}},
  \href{http://arxiv.org/abs/1607.08244}{{\tt arXiv:1607.08244}}.

\bibitem{PiusRudraSen14}
R.~Pius, A.~Rudra, and A.~Sen, {\it {String Perturbation Theory Around
  Dynamically Shifted Vacuum}},  {\em JHEP} {\bf 10} (2014) 70,
  [\href{http://arxiv.org/abs/1404.6254}{{\tt arXiv:1404.6254}}].

\bibitem{PiusRudraSen14b}
R.~Pius, A.~Rudra, and A.~Sen, {\it {Mass Renormalization in String Theory:
  General States}},  {\em JHEP} {\bf 07} (2014) 062,
  [\href{http://arxiv.org/abs/1401.7014}{{\tt arXiv:1401.7014}}].

\bibitem{PiusRudraSen14c}
R.~Pius, A.~Rudra, and A.~Sen, {\it {Mass Renormalization in String Theory:
  Special States}},  {\em JHEP} {\bf 07} (2014) 058,
  [\href{http://arxiv.org/abs/1311.1257}{{\tt arXiv:1311.1257}}].

\bibitem{Sen15b}
A.~Sen, {\it {Off-shell Amplitudes in Superstring Theory}},  {\em Fortsch.
  Phys.} {\bf 63} (2015) 149--188, [\href{http://arxiv.org/abs/1408.0571}{{\tt
  arXiv:1408.0571}}].

\bibitem{Berera94}
A.~Berera, {\it {Unitary string amplitudes}},  {\em Nucl. Phys.} {\bf B411}
  (1994) 157--180.

\bibitem{Witten13b}
E.~Witten, {\it {The Feynman $i \epsilon$ in String Theory}},  {\em JHEP} {\bf
  04} (2015) 055, [\href{http://arxiv.org/abs/1307.5124}{{\tt
  arXiv:1307.5124}}].

\bibitem{Polchinski88b}
J.~Polchinski, {\it {Collision of Macroscopic Fundamental Strings}},  {\em
  Phys. Lett.} {\bf B209} (1988) 252.

\bibitem{DaiPolchinski89}
J.~Dai and J.~Polchinski, {\it {The Decay of Macroscopic Fundamental Strings}},
   {\em Phys. Lett.} {\bf B220} (1989) 387.

\bibitem{MitchellTurokWilkinsonJetzer89}
D.~Mitchell, N.~Turok, R.~Wilkinson, and P.~Jetzer, {\it {The Decay of Highly
  Excited Open Strings}},  {\em Nucl. Phys.} {\bf B315} (1989) 1.

\bibitem{WilkinsonTurokMitchell90}
R.~B. Wilkinson, N.~Turok, and D.~Mitchell, {\it {The Decay of Highly Excited
  Closed Strings}},  {\em Nucl. Phys.} {\bf B332} (1990) 131.

\bibitem{IengoRusso02}
R.~Iengo and J.~G. Russo, {\it {The Decay of Massive Closed Superstrings with
  Maximum Angular Momentum}},  {\em JHEP} {\bf 11} (2002) 045.

\bibitem{DHokerPhong95}
E.~D'Hoker and D.~Phong, {\it {The Box Graph in Superstring Theory}},  {\em
  Nucl.Phys.} {\bf B440} (1995) 24--94.

\bibitem{Schwarz82}
J.~H. Schwarz, {\it {Superstring Theory}},  {\em Phys. Rept.} {\bf 89} (1982)
  223--322.

\bibitem{DijkgraafVerlindeVerlinde88}
R.~Dijkgraaf, E.~P. Verlinde, and H.~L. Verlinde, {\it {$c = 1$ Conformal Field
  Theories on Riemann Surfaces}},  {\em Commun. Math. Phys.} {\bf 115} (1988)
  649--690.

\bibitem{VerlindeVerlinde87}
E.~P. Verlinde and H.~L. Verlinde, {\it {Chiral Bosonization, Determinants and
  the String Partition Function}},  {\em Nucl. Phys.} {\bf B288} (1987) 357.

\bibitem{BelavinKnizhnik86}
A.~Belavin and V.~Knizhnik, {\it {Algebraic Geometry and the Geometry of
  Quantum Strings}},  {\em Phys.Lett.} {\bf B168} (1986) 201--206.

\bibitem{Polchinski_v1}
J.~Polchinski, {\em String Theory. Vol. 1: An Introduction to the Bosonic
  String}.
\newblock Cambridge Univ. Pr., UK, 1998.

\bibitem{Fay}
D.~Fay, John, {\em {Theta Functions on Riemann Surfaces}}.
\newblock 1973.
\newblock Springer-Verlag New Germany.

\bibitem{Mumford_v12}
D.~Mumford, {\em {Tata lectures on Theta I,II}}.
\newblock Modern Birkhauser Classics, 1984.

\bibitem{DHokerPhong}
E.~D'Hoker and D.~H. Phong, {\it {The Geometry of String Perturbation Theory}},
   {\em Rev. Mod. Phys.} {\bf 60} (1988) 917.

\bibitem{AokiD'HokerPhong04}
K.~Aoki, E.~D'Hoker, and D.~H. Phong, {\it {Two loop superstrings on orbifold
  compactifications}},  {\em Nucl. Phys.} {\bf B688} (2004) 3--69,
  [\href{http://arxiv.org/abs/hep-th/0312181}{{\tt hep-th/0312181}}].

\bibitem{DabholkarMandalRamadevi98}
A.~Dabholkar, G.~Mandal, and P.~Ramadevi, {\it {Nonrenormalization of mass of
  some nonsupersymmetric string states}},  {\em Nucl. Phys.} {\bf B520} (1998)
  117--131.

\bibitem{IengoKalkkinen00}
R.~Iengo and J.~Kalkkinen, {\it {Decay Modes of Highly Excited String States
  and Kerr Black Holes}},  {\em JHEP} {\bf 11} (2000) 025.

\bibitem{IengoRusso03}
R.~Iengo and J.~G. Russo, {\it {Semiclassical decay of strings with maximum
  angular momentum}},  {\em JHEP} {\bf 03} (2003) 030.

\bibitem{ChialvaIengoRusso03}
D.~Chialva, R.~Iengo, and J.~G. Russo, {\it {Decay of long-lived massive closed
  superstring states: Exact results}},  {\em JHEP} {\bf 12} (2003) 014.

\bibitem{ChialvaIengo04}
D.~Chialva and R.~Iengo, {\it {Long lived large type II strings: Decay within
  compactification}},  {\em JHEP} {\bf 07} (2004) 054.

\bibitem{ChialvaIengoRusso05}
D.~Chialva, R.~Iengo, and J.~G. Russo, {\it {Search for the Most Stable Massive
  State in Superstring Theory}},  {\em JHEP} {\bf 01} (2005) 001.

\bibitem{GutperleKrym06}
M.~Gutperle and D.~Krym, {\it {Decays of near BPS heterotic strings}},  {\em
  Phys. Rev.} {\bf D74} (2006) 086007.

\bibitem{IengoRusso06}
R.~Iengo and J.~G. Russo, {\it {Handbook on String Decay}},  {\em JHEP} {\bf
  0602} (2006) 041.

\bibitem{Polchinski88}
J.~Polchinski, {\it {Factorization of Bosonic String Amplitudes}},  {\em Nucl.
  Phys.} {\bf B307} (1988) 61.

\bibitem{Wilson74}
K.~G. Wilson, {\it {Confinement of Quarks}},  {\em Phys. Rev.} {\bf D10} (1974)
  2445--2459. [,45(1974)].

\bibitem{WilsonKogut74}
K.~G. Wilson and J.~B. Kogut, {\it {The Renormalization group and the epsilon
  expansion}},  {\em Phys. Rept.} {\bf 12} (1974) 75--200.

\bibitem{Polchinski84}
J.~Polchinski, {\it {Renormalization and Effective Lagrangians}},  {\em Nucl.
  Phys.} {\bf B231} (1984) 269--295.

\bibitem{DabholkarHarvey89}
A.~Dabholkar and J.~A. Harvey, {\it {Nonrenormalization of the Superstring
  Tension}},  {\em Phys.Rev.Lett.} {\bf 63} (1989) 478.

\bibitem{DabholkarGibbonsHarveyRuiz90}
A.~Dabholkar, G.~W. Gibbons, J.~A. Harvey, and F.~Ruiz~Ruiz, {\it {Superstrings
  and Solitons}},  {\em Nucl.Phys.} {\bf B340} (1990) 33--55.

\bibitem{Tseytlin90}
A.~A. Tseytlin, {\it {On 'Macroscopic String' Approximation in String Theory}},
   {\em Phys.Lett.} {\bf B251} (1990) 530--534.

\bibitem{Polchinski87}
J.~Polchinski, {\it {Vertex Operators in the Polyakov Path Integral}},  {\em
  Nucl. Phys.} {\bf B289} (1987) 465.

\bibitem{Witten12c}
E.~Witten, {\it {Superstring Perturbation Theory Revisited}},
  \href{http://arxiv.org/abs/1209.5461}{{\tt arXiv:1209.5461}}.

\bibitem{Witten13}
E.~Witten, {\it {More On Superstring Perturbation Theory}},
  \href{http://arxiv.org/abs/1304.2832}{{\tt arXiv:1304.2832}}.

\bibitem{Witten15}
E.~Witten, {\it {The Super Period Matrix With Ramond Punctures}},  {\em J.
  Geom. Phys.} {\bf 92} (2015) 210--239,
  [\href{http://arxiv.org/abs/1501.02499}{{\tt arXiv:1501.02499}}].

\bibitem{D'HokerPhong15}
E.~D'Hoker and D.~H. Phong, {\it {The Super Period Matrix with Ramond Punctures
  in the supergravity formulation}},  {\em Nucl. Phys.} {\bf B899} (2015)
  772--809, [\href{http://arxiv.org/abs/1501.02675}{{\tt arXiv:1501.02675}}].

\bibitem{D'HokerPhong15b}
E.~D'Hoker and D.~H. Phong, {\it {Higher Order Deformations of Complex
  Structures}},  {\em SIGMA} {\bf 11} (2015) 047,
  [\href{http://arxiv.org/abs/1502.03673}{{\tt arXiv:1502.03673}}].

\bibitem{SenWitten15}
A.~Sen and E.~Witten, {\it {Filling the gaps with PCO?s}},  {\em JHEP} {\bf 09}
  (2015) 004, [\href{http://arxiv.org/abs/1504.00609}{{\tt arXiv:1504.00609}}].

\bibitem{NarainSarmadiWitten87}
K.~S. Narain, M.~H. Sarmadi, and E.~Witten, {\it {A Note on Toroidal
  Compactification of Heterotic String Theory}},  {\em Nucl. Phys.} {\bf B279}
  (1987) 369.

\bibitem{MaharanaSchwarz93}
J.~Maharana and J.~H. Schwarz, {\it {Noncompact symmetries in string theory}},
  {\em Nucl. Phys.} {\bf B390} (1993) 3--32,
  [\href{http://arxiv.org/abs/hep-th/9207016}{{\tt hep-th/9207016}}].

\bibitem{Hamermesh}
M.~Hamermesh, {\it {Group Theory and Its Application to Physical Problems}}, .
  Dover Publications.

\bibitem{Nelson89}
P.~C. Nelson, {\it {Covariant Insertion of General Vertex Operators}},  {\em
  Phys. Rev. Lett.} {\bf 62} (1989) 993.

\bibitem{BeckerBeckerRobbins12}
K.~Becker, G.-Y. Guo, and D.~Robbins, {\it {Disc amplitudes, picture changing
  and space-time actions}},  {\em JHEP} {\bf 01} (2012) 127,
  [\href{http://arxiv.org/abs/1106.3307}{{\tt arXiv:1106.3307}}].

\bibitem{Minahan87}
J.~A. Minahan, {\it {One Loop Amplitudes on Orbifolds and the Renormalization
  of Coupling Constants}},  {\em Nucl. Phys.} {\bf B298} (1988) 36--74.

\bibitem{BergBuchbergerSchlotterer16}
M.~Berg, I.~Buchberger, and O.~Schlotterer, {\it {From maximal to minimal
  supersymmetry in string loop amplitudes}},
  \href{http://arxiv.org/abs/1603.05262}{{\tt arXiv:1603.05262}}.

\bibitem{LugoRusso89}
A.~Lugo and J.~Russo, {\it {Hamiltonian Formulation and Scattering Amplitudes
  in String Theory at Genus $g$}},  {\em Nucl. Phys.} {\bf B322} (1989) 210.

\bibitem{DHoker}
E.~D'Hoker, {\em {Quantum Fields and Strings: A Course for Mathematicians. Vol.
  2}}.
\newblock AMS, USA, 1999.

\bibitem{BakasLust15}
I.~Bakas and D.~Luest, {\it {T-duality, Quotients and Currents for
  Non-Geometric Closed Strings}},  {\em Fortsch. Phys.} {\bf 63} (2015)
  543--570, [\href{http://arxiv.org/abs/1505.04004}{{\tt arXiv:1505.04004}}].

\bibitem{LandauLifshitzRQF}
E.~M. Berestetskii, V.~B.~Lifshitz and L.~P. Pitaevskii, {\em Quantum
  Electrodynamics}, vol.~4 of {\em Landau and Lifshitz Course of Theoretical
  Physics}.
\newblock Butterworth-Heinmann, United Kingdom, 1982.

\bibitem{Kiritsis}
E.~Kiritsis, {\em {String Theory in a Nutshell}}.
\newblock Princeton University Press, United Kingdom, 2007.

\bibitem{EllisMavromatosSkliros15}
J.~Ellis, N.~E. Mavromatos, and D.~P. Skliros, {\it {Complete Normal Ordering
  1: Foundations}},  {\em Nucl. Phys.} {\bf B909} (2016) 840--879,
  [\href{http://arxiv.org/abs/1512.02604}{{\tt arXiv:1512.02604}}].

\bibitem{GSW2}
M.~B. Green, J.~H. Schwarz, and E.~Witten, {\em {Superstring Theory. Vol. 2:
  Loop Amplitudes, Anomalies and Phenomenology}}.
\newblock Cambridge, UK: Univ. Pr., UK, 1987.

\bibitem{Polchinski86}
J.~Polchinski, {\it {Evaluation of the One Loop String Path Integral}},  {\em
  Commun. Math. Phys.} {\bf 104} (1986) 37.

\bibitem{Weinberg_v1}
S.~Weinberg, {\em {The Quantum theory of fields. Vol. 1: Foundations}}.
\newblock Cambridge, UK: Univ. Pr., UK, 1995.

\bibitem{Polchinski94}
J.~Polchinski, {\it {Combinatorics of Boundaries in String Theory}},  {\em
  Phys. Rev.} {\bf D50} (1994) 6041--6045.

\bibitem{Green95}
M.~B. Green, {\it {A Gas of D instantons}},  {\em Phys. Lett.} {\bf B354}
  (1995) 271--278, [\href{http://arxiv.org/abs/hep-th/9504108}{{\tt
  hep-th/9504108}}].

\bibitem{Nakahara03}
M.~Nakahara, {\it {Geometry, topology and physics}}, . Boca Raton, USA: Taylor
  \& Francis (2003) 573 p.

\bibitem{MisnerThorneWheeler74}
C.~W. Misner, K.~Thorne, and J.~Wheeler, {\em {Gravitation}}.
\newblock 1974.

\bibitem{AlvarezGaumeNelson}
L.~Alvarez-Gaume and P.~C. Nelson, {\it {Riemann Surfaces and String
  Theories}}, . CERN-TH-4615/86.

\bibitem{HamidiVafa87}
S.~Hamidi and C.~Vafa, {\it {Interactions on Orbifolds}},  {\em Nucl. Phys.}
  {\bf B279} (1987) 465.

\bibitem{Mulase83}
M.~Mulase, {\it {Algebraic Geometry of Soliton Equations}},  {\em Proc. Japan
  Acad.} {\bf 59} (1983) 285.

\bibitem{Mulase84}
M.~Mulase, {\it {Cohomological Structure in Soliton Equations and Jacobian
  Varieties}},  {\em J. Differential Geometry} {\bf 19} (1984) 403--430.

\bibitem{HarrisMorrison}
J.~Harris and I.~Morrison, {\em Moduli of Curves}.
\newblock Springer-Verlag, New York, 1998.


\end{thebibliography}
\end{document}